\documentclass[prx,twocolumn,superscriptaddress,tightenlines,longbibliography]{revtex4-1}

\usepackage{color}
\usepackage{amssymb}
\usepackage{amsmath}
\usepackage{graphicx}
\usepackage{xcolor}
\usepackage[normalem]{ulem}
\usepackage[unicode=true,pdfusetitle,hypertexnames=false,
 bookmarks=true,bookmarksnumbered=false,bookmarksopen=false,
  breaklinks=true,pdfborder={0 0 0},pdfborderstyle={},backref=false,colorlinks=true]
   {hyperref}
\usepackage{subcaption}

\definecolor{darkgreen}{RGB}{89, 175, 99}

\definecolor{grey}{rgb}{.6,.6,.6}
\definecolor{eggplant}{RGB}{180,33,147}

\usepackage{cancel}

\begin{document}

\title{A density-matrix renormalization group algorithm for simulating quantum circuits with a finite fidelity}

\author{Thomas Ayral}
\affiliation{Atos Quantum Laboratory, Les Clayes-sous-Bois, France}

\author{Thibaud Louvet}
\affiliation{PHELIQS, Université  Grenoble Alpes, CEA, Grenoble INP, IRIG, Grenoble 38000, France}

\author{Yiqing Zhou}
\affiliation{Department of Physics, Cornell University, Ithaca, NY 14853, USA }

\author{Cyprien Lambert}
\affiliation{Atos Quantum Laboratory, Les Clayes-sous-Bois, France}

\author{E. Miles Stoudenmire}
\affiliation{Center for Computational Quantum Physics, Flatiron Institute, New York, NY 10010, USA}

\author{Xavier Waintal}
\affiliation{PHELIQS, Université Grenoble Alpes, CEA, Grenoble INP, IRIG, Grenoble 38000, France}
\date{\today}

\begin{abstract}

We develop a density-matrix renormalization group (DMRG) algorithm for the simulation of quantum circuits. This algorithm can be seen as the extension of time-dependent DMRG from the usual situation of hermitian Hamiltonian matrices to quantum circuits defined by unitary matrices. For small circuit depths, the technique is exact and equivalent to other matrix product state (MPS) based techniques. For larger depths, it becomes approximate in exchange for an exponential speed up in computational time. Like an actual quantum computer, the quality of the DMRG results is characterized by a finite fidelity. However, unlike a quantum computer, the fidelity depends strongly on the quantum circuit considered. For the most difficult possible circuit for this technique, the so-called ``quantum supremacy" benchmark of Google Inc.\ \cite{Arute2019}, we find that the DMRG algorithm can generate bit strings of the same quality as the seminal Google experiment on a single computing core. For a more structured circuit used for combinatorial optimization (Quantum Approximate Optimization Algorithm or QAOA), we find a drastic improvement of the DMRG results with error rates dropping by a factor of $100$ compared with random quantum circuits. Our results suggest that the current bottleneck of quantum computers is their fidelities rather than the number of qubits.
\end{abstract}

\maketitle


\section{Introduction}
Quantum computers and the quantum many-body problem are intimately connected.
On the one hand, a quantum computer is essentially an instance of the quantum many-body problem on which one has a high level of control.
On the other hand, the most promising applications that are foreseen for quantum computers  correspond to solving other instances of the quantum many-body problem such as calculating the properties of new materials \cite{Bauer2020}, of  new molecules for medicine, or of new catalysts for important chemical reactions \cite{Reiher2016}.

A common misconception of the field of quantum computing is that all quantum many-body problems are exponentially difficult to solve by classical computers because the size of the Hilbert space grows exponentially as $2^N$ when the system size $N$ increases. This supposedly dooms many-body simulations on classical computers to failure, and therefore calls for computers with quantum physics inside.
This ``large Hilbert space fallacy'' is however contradicted by the success of classical many-body methods for tackling many of these hard problems. At heart, these methods use the fact that physical problems are \emph{structured}. Physicists take advantage of the separation of time, energy or length scales, of statistical (mean field) behavior, or of symmetries, to design methods to solve seemingly exponentially hard problems.
Even for genuine strongly correlated systems, there exist very powerful many-body techniques that can solve them in a variety of situations, taking advantage of an underlying feature.
Without these features---namely had the physical world been a random point in the Hilbert space---there would indeed be nothing to understand and the problem would be exponentially difficult. However, since the physical world actually {\it makes sense}, one can  argue that simulating a many-body physical problem is not as hopeless as one could naively think.

The subject of this article is to discuss the problem of simulability in the context of quantum computers that use quantum circuits, that is discrete sequences of quantum gates. This questioning is at the core of the possibility for a ``quantum advantage", for if a quantum computer can be easily simulated, one might as well use the (classical) simulator instead of developing a genuine quantum computer. The quantum circuit model of quantum computing
ignores many structures of the underlying many-body problem. For instance, it does not contain the basic concepts of space, time or energy. Nor does it contain the concept of fermionic or bosonic statistics nor the representation of symmetries (spin, relativity). As a result, it might seem that such a quantum computer must be much harder to simulate than a physics-based many-body problem. An extreme version of a quantum circuit designed to be as featureless as possible is the  seminal ``quantum supremacy experiment" \cite{Arute2019} of Google Inc. Google initially claimed
that simulating their supremacy experiment would require 10,000 years on the largest existing supercomputer. Subsequent studies (that we shall review in Section \ref{sec:review}) \cite{Gray2021hyperoptimized,Zhang2021simulating,Pan2021solving} showed that this initial surmise was exaggerated and that the task could be executed in a few hundred seconds. This progress in classical simulations could be obtained through a precise analysis of the structure of the quantum circuit. Yet, the computational cost of these simulations remains exponential in the number $N$ of qubits. 

In this article, we use a different class of algorithms, borrowed from quantum many-body theory, whose complexity increases only as a power of $N$, making the simulation of hundreds of qubits possible. This exponential gain in computational complexity is obtained in exchange for a \emph{quantum state compression} that implies a finite fidelity of the calculation. In this sense, these algorithms share some characteristics with actual quantum computers, which also suffer from a finite fidelity. A first step in that direction was taken in \cite{Zhou2020}, where some of us  developed a quantum circuit version of the time-evolving bond decimation \cite{Daley_2004,Vidal2004efficient,White2004realtime} (TEBD) algorithm.
It was found that surprisingly good fidelities could be obained at a relatively low computational cost.
Here, we develop a generalization of the density-matrix renormalization group \cite{White:1993,Schollwoeck2011} (DMRG) algorithm to quantum circuits. Although technically more complex, DMRG allows one to improve on the TEBD algorithm in a systematic way.  

We benchmark our quantum-circuit-DMRG algorithm on three different quantum circuits: The ``quantum supremacy" sequence of \cite{Arute2019}---optimized to be as difficult to simulate as
possible, another slightly easier random circuit and a ``useful" circuit used in the Quantum Approximate Optimization Algorithm (QAOA) \cite{Farhi2014} for combinatorial optimization. We find that even with the most difficult ``quantum supremacy'' sequence, DMRG can produce bitstrings that have the same quality as the one demonstrated in \cite{Arute2019} on a single computing core. More importantly, we find that for the 
QAOA sequence, we reach a fidelity per gate much higher than the one found in the Sycamore processor.
Our numerical experiments provide important benchmarks of the fidelities that can be reached on a classical computer, and therefore of what the quantum hardware must fulfill to claim genuine quantum supremacy or advantage. Since our DMRG algorithm scales only polynomially with $N$, it implies that quantum computers must improve their fidelities to access regimes that cannot be simulated.

This article is organized as follows: in section \ref{sec:review}, we review the current status of quantum supremacy and of quantum circuit simulation techniques. Section \ref{sec:results} contains a summary of the main findings of this article.
The DMRG technique is developed in section \ref{sec:dmrg}. Section \ref{sec:details} showcases how DMRG works in practice
and discusses some implementation details.
Section \ref{sec:optimum} discusses in which regime the DMRG algorithm provides an optimum solution.
Section \ref{sec:FvsFB} shows the relation between the fidelity and the cross-entropy benchmarking obtained in our simulations. Finally, we conclude in section \ref{sec:conclusions}.
This articles relies heavily on tensor network techniques. Appendix \ref{app:tensor_networks}
contains a short self-contained introduction to tensor networks for readers unfamiliar with these techniques.


\section{A critical review of quantum supremacy}
\label{sec:review}

Almost three years ago, the annoucement by Google of having reached the milestone of ``quantum supremacy"
dazzled both the academic community and the general public \cite{Arute2019}. Google managed to control
$N=53$ transmon qubits and to perform a circuit comprising $430$ two-qubit gates. They obtained a quantum state that had
a small ($\approx 0.002$), yet measurable overlap with the ideal state that they were supposed to get.
While this state did not permit any useful computation, Google surmised that producing something similar using classical simulations
would be prohibitive (10,000 years on the largest supercomputer) and hence claimed to have reached quantum supremacy.

Since Ref.~\cite{Arute2019}, another group has produced an almost identical experiment using a very similar technology \cite{Wu2021}.
There has also been other claims of quantum supremacy, most notably using ``boson sampling" \cite{Zhong2020}. 
Here, we shall not discuss these more recent claims for two reasons. First, the hardness of these tasks is strongly debated \cite{Popova2021,Villalonga2021,Oh2022}.
Second, and more importantly: while the Google experiment represents a milestone on the path
towards building a quantum computer, these most recent claims corresponds to very specific tasks and the devices are not programmable.

Here, we review the status of the classical simulation challenges to these ``quantum supremacy" claims, namely we review the various works that have attempted to simulate
the experiments in a reasonable classical computing time \cite{Gray2021hyperoptimized,Zhang2021simulating,Pan2021solving,Yong2021closingsupremacygap}.
In particular, we shall try to explain in simple terms why the initial claim of ``10,000 years" was challenged a few days later to be only two days and why it has eventually been shown that a simulation of quantum supremacy could be performed in a few hundred seconds. 

\subsection{An exponentially difficult experiment}
A quantum computer with $N$ qubits has an internal state that can be written as
\begin{equation}
|\Psi\rangle = \sum_{i_1i_2i_3\dots i_N} \Psi_{i_1i_2i_3\dots i_N} 
|{i_1i_2i_3\dots i_{N}}\rangle,
\end{equation}
where $i_1\in\{0,1\}$, $i_2\in\{0,1\}$,$\dots i_N\in\{0,1\}$, correspond to the different qubits.
The vector $\Psi$, whose components are the complex numbers $\Psi_{i_1i_2i_3\dots i_N}$, can be considered as a large vector
of dimension $2^N$. One initializes the state in $|\Psi (0)\rangle$ (typically all qubits in state $0$, i.e. ${\Psi_{i_1i_2i_3\dots i_N} =\prod_p \delta_{i_p,0}}$) and then applies a sequence of unitary gates. Formally, these gates transform the state of the quantum computer into
\begin{equation}
\label{eq:perfect}
|\Psi (D)\rangle = U^{(D)} U^{(D-1)} \cdots\  U^{(2)} U^{(1)}  |\Psi (0) \rangle, 
\end{equation}
where the $U^{(p)}$ are unitary matrices (two-qubit gates or combination thereof). A direct simulation of Eq.~(\ref{eq:perfect}) by a series of (sparse) matrix-vector multiplications is referred to as a ``Schr\"odinger approach". In an experiment, however, one does not have access to the many-qubit wavefunction. Instead, one measures the different qubits
and obtains a bitstring $x=i_1i_2\dots i_N$ with probability $Q(x)$.

In Ref.~\cite{Arute2019}, the authors used $N=53$ qubits and a highly unstructured set of $N_{\rm 2g} = 430$ two-qubit gates spread over $D=20$
layers. The experiment was repeated $N_\# \approx 10^6$ times to produce a sequence of bitstrings $x_1,\dots,x_{N_\#}$. Since the
quantum sequence was highly unstructured, the distribution $Q(x)$ was expected to be fairly chaotic so that all bitstrings $x$
would have a probability to be obtained of order $\propto 1/2^N$, i.e. the experiment essentially outputs random bitstrings.
Ref.~\cite{Arute2019} is primarily a global system validation experiment. The authors performed exact numerical simulations of
Eq.~(\ref{eq:perfect}) to obtain the exact distribution $P(x)= |\langle x| \Psi (D)\rangle|^2 = |\Psi_{i_1i_2i_3\dots i_N}^{(D)} |^2$
that should have been obtained from the experiment. By estimating how the perfect distribution $P(x)$ correlates with the
distribution $Q(x)$ obtained experimentally, one is able to assert to which degree the quantum computer has performed the requested task.
The metric used to analyse this correlation is the ``cross-entropy benchmarking" 
\begin{equation}
\label{eq:FB-sup}
{\cal F}_{\cal B} \equiv 2^N \sum_x P(x) Q(x)   \ - 1,
\end{equation}
which can be estimated experimentally as
\begin{equation}
\label{eq:FB-estim}
{\cal F}_{\cal B} \approx \frac{2^N }{N_\#} \sum_{\alpha=1}^{N_{\#}} P(x_\alpha) \ - 1.
\end{equation}
It is expected theoretically, and observed experimentally, that due to decoherence and imprecisions in the gates and measurements,
the cross-entropy benchmarking shall decay exponentially:
\begin{equation}
\label{eq:FB-error}
{\cal F}_{\cal B} \propto e^{-\epsilon_{\cal B} N D /2},
\end{equation}
with an error rate $\epsilon_{\cal B}$. Ref.~\cite{Arute2019} was able to verify Eq.~(\ref{eq:FB-error}) with an error
$\epsilon_{\cal B}\approx 1\%$. For the largest depth $D=20$ where the exact distribution $P(x)$ was too computationally costly to be calculated, 
they extrapolated that ${\cal F}_{\cal B}\approx 0.2\%$. The ``quantum supremacy" claim was that obtaining a set of $N_\#$ bitstrings
with the same fidelity ${\cal F}_{\cal B}\approx 0.2\%$ by simulations would require 10,000 years on the largest supercomputer.

It should be  noted that this experiment is exponentially difficult. Indeed, in order for the set of bitstrings to be distinguishable
from just plain random bitstrings distributed uniformly, one needs the statistical error in the estimation of  Eq.~(\ref{eq:FB-estim}) to be smaller than what is estimated, i.e. ${\cal F}_{\cal B}$. Since the statistical error decreases as $1/\sqrt{N_\#}$, it implies an exponentially large number of samples,
\begin{equation}
N_\# \propto e^{\epsilon_{\cal B} N D}.
\end{equation}
Ref.~\cite{Arute2019} pushed the quantum chip to the extreme limit where there remained just enough fidelity for the signal to be measured. For instance going to $D=40$ would have implied an increase of the measurement time by a factor $10^6$. The authors also had to give up a little on the universality or programmability of the chip to maintain a low enough error rate $\epsilon_{\cal B}$: they chose, for each pair of qubit, the two-qubit gate that had the best fidelity
by optimizing the microwave pulses. Subsequent experiments that used the same chip but focused on ``useful" quantum circuits used only 10-20 qubits to retain accurate enough results~\cite{Arute2020}.

\subsection{Exchanging a smaller memory footprint for an exponential increase of computational time}
Immediately following Google's supremacy claim, a team from IBM proposed an algorithm that, according to their estimation, would require
only $2.5$ days to complete the supremacy task instead of $10,000$ years~\cite{Pednault2019}. Such a speed up (a factor $10^5$) begs for an explanation. A direct naive ``Schr\"odinger" evaluation of Eq.~(\ref{eq:perfect}) would require of the order of $N_{\rm 2g} 2^N$ floating operations by holding the vector $\Psi$ in memory and applying the two-qubit gates one by one. Such an algorithm would require $10^{18}$ operations. Hence, since large supercomputers can perform more than $10^{17}$ floating operations per second, the supremacy task could, according to this naive estimation, be performed in at most a few minutes, not thousands of years. This however requires one to hold a vector of size $2^N$ in memory, i.e. $10^5$ TBytes of RAM which is more than what supercomputers have (typically by more than a factor 10). To get around this difficulty, one designs algorithms that require exponentially more operations in exchange for a smaller memory footprint.

To illustrate how the tradeoff between memory footprint and computational time can be implemented in practice, 
 imagine that we group the qubits into 2 groups of $N_1$ and $N_2$ qubits ($N_1+N_2=N$). A first index $\alpha$ labels the first group $i_1i_2\dots i_{N_1}$ and a second index $\beta$ labels the second group $i_{N_1+1}\dots i_{N}$. An arbitrary gate $U^{(p)}$ has matrix elements
$U^{(p)}_{\alpha\beta;\alpha'\beta'}$. Such a tensor can be considered as a matrix where the two indices $(\alpha,\alpha')$ are considered as a meta-index that index the lines and the two other indices $(\beta,\beta')$ index the columns. Using singular value decomposition, such a matrix can be factorized into a sum of $\chi_p$ terms of the form
\begin{equation}
U^{(p)}_{\alpha\beta;\alpha'\beta'} = \sum_{a=1}^{\chi_p} V^{(p)}_{a\alpha\alpha'}W^{(p)}_{a\beta\beta'}
\label{eq:SVD}
\end{equation}
where $V^{(p)}$ and $W^{(p)}$ act separately on the first and second group of qubits, respectively (see Appendix~\ref{app:tensor_networks} for details on the SVD operation). Since the wave function initially factorizes, 
$\Psi_{\alpha\beta}^{(0)} = \Psi_{1\alpha}^{(0)}\Psi_{2\beta}^{(0)}$, one can rewrite Eq.~(\ref{eq:perfect}) as,
\begin{equation}
\label{eq:schrofeynman}
\Psi^{(D)}_{\alpha\beta} = \sum_{a_1, \dots, a_D} \Psi_{1\alpha}^{(D)}\Psi_{2\beta}^{(D)}
\end{equation}
with
\begin{eqnarray}
\label{eq:schrofeynman1}
\Psi_{1}^{(D)} &= V^{(D)}_{a_D} V^{(D-1)}_{a_{D-1}}\cdots V^{(1)}_{a_1}\Psi_{1}^{(0)} \\
\label{eq:schrofeynman2}
\Psi_{2}^{(D)} &= W^{(D)}_{a_D} W^{(D-1)}_{a_{D-1}}\cdots W^{(1)}_{a_1}\Psi_{2}^{(0)} 
\end{eqnarray}
Now, we need only to perform matrix vector products of much smaller sizes, $2^{N_1}\ll 2^N$ and $2^{N_2}\ll 2^N$. In return for this much smaller memory footprint, $\Psi_{1}^{(D)}$ and $\Psi_{2}^{(D)}$ depend explicitly on $a_1\dots a_D$. 
One has to repeat the calculation for each $a_1, \dots, a_D$ to perform the sum, which has an exponential computational cost $\propto \prod_p \chi_p$. 
Simulations that use Eqs.~(\ref{eq:schrofeynman},\ref{eq:schrofeynman1},\ref{eq:schrofeynman2}) are known as "Schr\"odinger-Feynman" simulations.
In a Schr\"odinger-Feynman simulation, the only problematic gates are the two-qubit gates that couple the two groups. 
These gates have $\chi_p=2$ (Control-NOT or Control-Z) or at most $\chi_p=4$ (arbitrary two-qubit gates). 
For all the gates that do not couple the two groups of qubits, $\chi_p=1$ and there is no increase of computational time. 
Another aspect is that one can calculate the amplitude $\Psi_{\alpha,\beta}$ for as many configurations $\alpha,\beta$ as needed with no additional cost except for the one of storing these amplitudes in memory. 
The initial statement of Google of 10,000 years was associated with an estimation of the computational cost of a  Schr\"odinger-Feynman simulation. 
We see that this computational estimation is strongly tied to the available memory. 
More memory would allow one to perform an optimized splitting of the qubits into two groups or no splitting at all, resulting in a much smaller computational cost. 
The IBM proposal~\cite{Pednault2019}, which was not implemented, was to take advantage of the hard drives of the supercomputer as temporary storage so that the full $N$ qubit wave-function could be held in memory, thereby considerably reducing the computational time. 
The drawback of this approach, besides the obvious difficulty of performing an actual implementation, is the fact that adding just one extra qubit would require a doubling of the memory footprint hence making the simulation out of reach.

\subsection{The hierarchy of ``open" versus ``closed" versus ``weak" simulations } 

The final blow on the supremacy claim came from a combination of works Refs~\onlinecite{Gray2021hyperoptimized,Zhang2021simulating,Pan2021solving,Yong2021closingsupremacygap}, in which the authors found a route to perform the simulation of the ``quantum supremacy" experiment in a few hundred of seconds and demonstrated that the solution could be implemented in an actual very large supercomputer.
This series of works essentially closed the gap between the simulations and the experiments.
This corresponds to a drop by a factor $10^9$ with respect to the initial estimate of 10,000 years. This new gain comes from the combination of two new ingredients.

The first important point is that there are several simulation modes of decreasing power. The Schr\"odinger (Schr\"odinger-Feynman) simulation provides the full $N$-qubit wavefunction (as many
amplitudes as one can store). We refer to this simulation mode as ``open" in the sense that they do not target a specific bitstring $x$. Open simulations produce much more information than what the experiment outputs. Another simulation mode, that we refer to as a ``closed" simulation, targets a single bitstring $x$ and computes the amplitude
\begin{equation}
\label{eq:close}
\Psi_x^{(D)} = \langle x | \Psi (D) \rangle = \langle x | U^{(D)}\cdots U^{(2)} U^{(1)} |0\rangle.
\end{equation}
Closed simulations are generically much easier than open ones. A last type of simulation, ``weak" simulations, would produce the same output as an actual quantum computer, namely random bitstrings distributed according to 
the probability $|\Psi_x^{(D)}|^2$. There exists a clear hierarchy between these different simulation modes: an open simulation provides more information than a closed one which in turn provides more information than a weak simulation (or an actual quantum computer). One of the key steps in speeding up our simulations was to go from the open mode to the closed one.

The fact that closed simulations are superior to weak ones is not totally obvious. It follows from a simple algorithm recently proposed in Ref.~\cite{Bravyi2021} that allows one to sample $|\Psi_x^{(D)}|^2$ from
the calculation of a polynomial number of individual amplitudes $\Psi_x^{(D')}$ at smaller depth $D'\le D$. 
The algorithm of Ref.~\cite{Bravyi2021} constructs a bitstring $x^D$ iteratively, starting from the initial bitstring $x^0=00\dots 0$ and taking into account the two-qubit gates one by one. 
$x^{D'+1}$ is identical to $x^{D'}$ except for the two qubits that are affected by the two-qubit gate.  
There are only four such bitstrings $x_1,x_2,x_3,x_4$. One computes the four  probabilities $p_i = |\Psi_{x_i}^{D'+1}|^2/\sum_j|\Psi_{x_j}^{D'+1}|^2$ and samples from this conditional distribution, 
i.e.  $x^{D+1} = x_i$ with probability $p_i$. It is straightfoward to verify that this scheme indeed provides a bitstring distributed according to $|\Psi_x^{(D)}|^2$. Note that in the context of the supremacy experiment where all $\Psi_x^{(D)}$ have similar orders of magnitude, this algorithm is not necessary and a simple Metropolis-Hastings sampling could be used instead (see the discussion in Appendix \ref{sec:metropolis}). 

\subsection{Optimized contraction strategies of tensor networks}

In the closed simulation mode, the challenge of the calculation lies in the summation over all the internal
indices of the tensors in Eq.~(\ref{eq:close}). Finding the optimum order for these summation is a hard (NP complete)
problem, but there exist good heuristic algorithms for finding close to optimum contraction strategies \cite{Gray2021hyperoptimized}. These optimum
orders are in general very different from a simple summation from right to left, i.e.\ from the contraction done in a Schr\"odinger or Schr\"odinger-Feynman simulation. The memory/CPU tradeoff is implemented using a ``slicing" approach: 
one carefully selects a few indices that are frozen in order to lower the cost of contracting the tensor network. The different values taken by these indices (the equivalent of the $a_1\dots a_D$ in the Schr\"odinger-Feynman simulations)
are distributed over different computing nodes or GPU cards as these tasks are embarrassingly parallel. For the reader not familiar with the concept of tensor contractions and slicing, a small introduction is given in appendix~\ref{app:tensor_networks}.

Using these techniques (i.e. closed simulations and good contraction strategies), the authors in Ref.~\cite{Gray2021hyperoptimized} estimated that the time to compute a single amplitude on a graphics card (GPU) could be reduced down to $3,088$ years with perfect fidelity. The same authors estimated that it would take $197$ days to match the supremacy experiment, i.e. produce one million samples with the $0.2 \%$ cross-entropy benchmarking fidelity. To arrive at this estimate, they took advantage of (i) the computing ressources of the large supercomputer ``Summit", (ii) the fact that the computing time to compute a few amplitudes that differ only by  the value of a few qubits is not significantly higher than computing a single amplitude and (iii) that a fidelity of $0.2\%$ can be obtained by mixing a few ($2,000$) high amplitude (large $|\Psi_x|^2$) samples with $998,000$ bitstrings sampled from a uniform random distribution.

A few months later, the estimated time to sample one million bitstrings with $0.2 \%$ fidelity was further reduced to $19.3$ days~\cite{Huang2020}, based on similar ideas and refinements by a team at Alibaba. These authors estimated the time to compute one perfect sample on Summit to $833$ seconds. Like the  previously mentioned study~\cite{Gray2021hyperoptimized}, they proposed to sample Sycamore by computing batches of $64$ amplitudes at no significant cost increase in a partly ``closed", partly ``open" mode.

The approach was further optimized by Pan and Zhang~\cite{Zhang2021simulating} in the so-called ``big-batch method" that optimized the choice of the qubits left in open mode. They managed to compute two million bitstrings with a large $73.9 \%$ cross-entropy benchmarking in 5 days on a small cluster of 60 GPUs~\cite{Zhang2021simulating}. However these bitstrings had many qubits in common, hence were strongly correlated. In a second study~\cite{Pan2021solving}, they have proposed a new  "sparse state method" that lowers the contraction cost of the supremacy circuit tensor network by cleverly introducing a few errors at specific locations. They have produced $2^{20}$ independent batches of $64$ correlated bitstrings in $15$ hours on a cluster of $512$ GPUs, and via importance sampling finally obtained one million uncorrelated samples with $0.37 \%$ fidelity. In the same work, they estimated that the sampling time for Sycamore could be reduced to a few dozen of seconds on a large supercomputer.

Ref.~\cite{Yong2021closingsupremacygap} produced  two million correlated bitstrings from Sycamore with $0.2 \%$ cross-entropy benchmarking fidelity in $304$ seconds. This calculation was performed on the Sunway TaihuLight supercomputer with 42 million effective cores, with an algorithm inspired by the work of~\cite{Zhang2021simulating}, and taking advantage of a new heuristic for slicing and contraction path optimization. It demonstrated that the parallelization of these algorithms could be effectively implemented on a very large supercomputer. Together Ref.\cite{Pan2021solving} and Ref.\cite{Yong2021closingsupremacygap} convincingly show that a supercomputer can match the Sycamore chip, even for a task whose only interest lies in having been optimized to be difficult to simulate.

Hence, the initial claim of ``quantum supremacy" has been, to a large extent, deflated. However,
the classical simulations reviewed above required colossal resources to achieve this goal. Since the problem is exponentially hard, a marginal improvement of the qubit fidelity (which would allow the experiment to go to larger depth) would have made the problem inaccessible to simulations. A second aspect is that problems that are impossible to simulate are easy to find, and the quantum supremacy experiment is to a large extent an artificial problem constructed for the sole purpose of being difficult to simulate. The question remains of what the quantum supremacy experiment taught us in terms of where one stands in the route to building genuine quantum computing capabilities for useful problems.
In this article, we also use tensor network simulations to benchmark the performance of quantum computers.
However, our focus is very different from the above high-performance computing calculations. While previous simulations are essentially exact with a small linear speed-up coming from the finite targeted fidelity,
we borrow quantum state compression techniques from many-body theory that exchange a finite fidelity for an exponential gain in computing time. As we shall see, these complementary techniques provide strong insights in the influence of a finite fidelity on computing capabilities and what it would take for a quantum computer to reach a regime that is both interesting and out of reach of simulations.

\section{Summary of the main results}
\label{sec:results}

\begin{figure}
	    \centering
	    	\includegraphics[trim=80 20 80 20,clip,width=0.8\columnwidth]{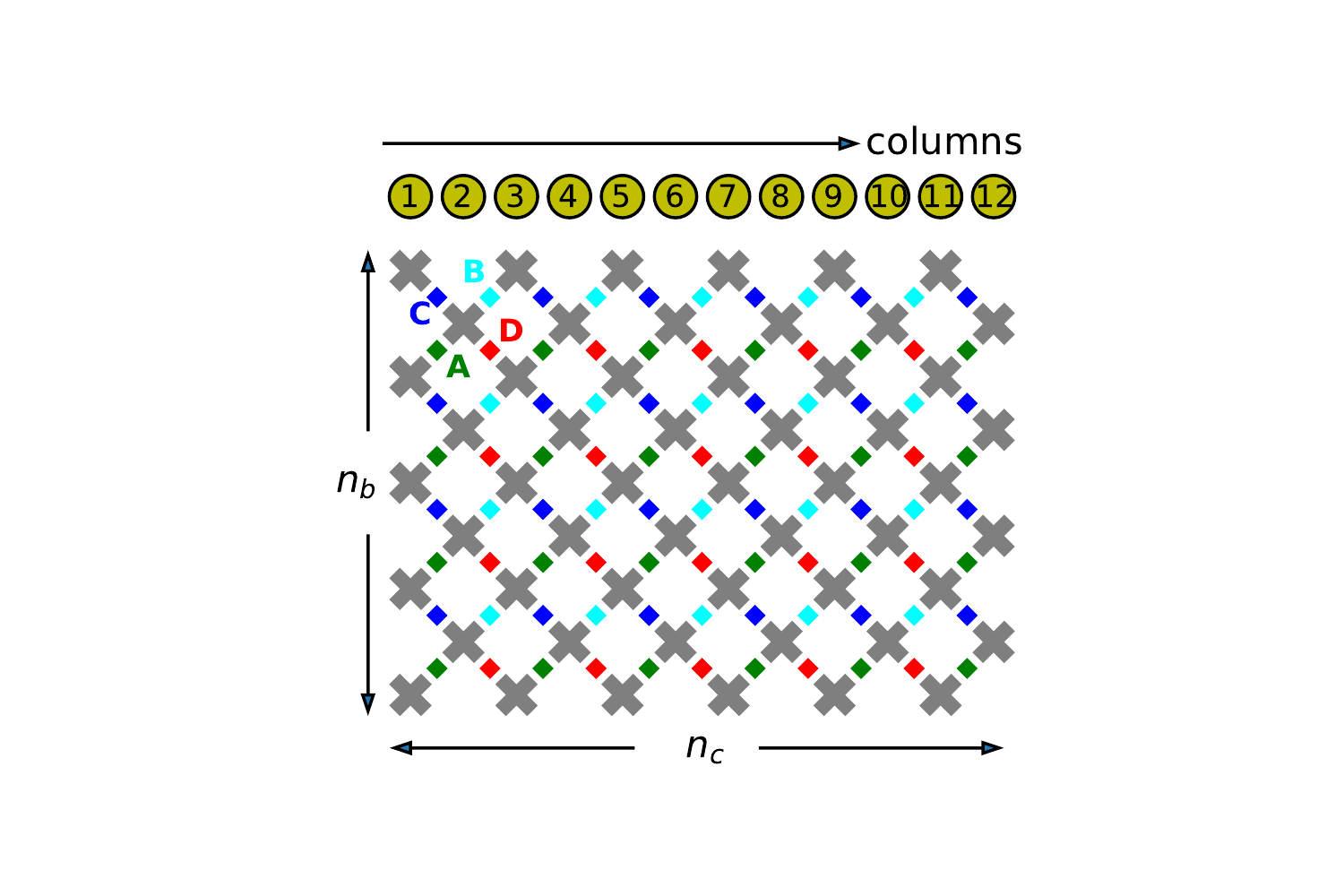}
	    	\caption{\emph{Topology of the ``quantum processor" simulated in this work.} The system has $n_c$ columns containing alternatively $n_b$ and $n_b-1$ qubits. $n_c=12$ with $n_b=5$ corresponds to a 54-qubit planar processor that has the same topology as the Sycamore chip. The qubits have nearest-neighbor connectivity. In the quantum circuits corresponding to sequence I and II (see text), the circuit is split in layers. In each layer, one applies a two-qubit gate between all the pairs of qubits coupled by a green rectangle (A layers), a light blue rectangle (B layers), a dark blue rectangle (C layers) or a red rectangle (D layers).}
	    	\label{fig:qubits}
\end{figure}

In this article, we develop an approximate DMRG algorithm for the simulation of quantum circuits. Before going into the mathematical details of how the technique works, we report on the results of our simulations for a few relevant circuits. Denoting by $| \Psi_P \rangle$ the perfect state that one should obtain and by $| \Psi_Q \rangle$ the actual approximate state obtained in the simulation, the main quantity of interest in this article is the fidelity ${\cal F}$ of the simulation, defined as
\begin{equation}
\label{eq:F-sup}
{\cal F} = |\langle \Psi_Q | \Psi_P \rangle|^2.
\end{equation}
Since ${\cal F}$ decreases exponentially with the number $N_{\rm 2g}$ of two-qubit gates (${\cal F} \approx \exp(-\epsilon N_{\rm 2g})$) in these simulations, we define the error rate per two-qubit gate $\epsilon$ as
\begin{equation}
\label{eq:epsilon}
\epsilon = 1 - {\cal F}^{1/N_\mathrm{2g}}\approx -\frac{1}{N_{\rm 2g}} \log {\cal F}.
\end{equation}
This quantity can be directly compared to experiments. For instance, assuming ${\cal F_B}={\cal F}$,
then the ${\cal F_B} = 0.2\%$ of the quantum supremacy experiment translates into an error $\epsilon= 1.4\%$
per two-qubit gate, for each of the $N_{2g}=430$ two-qubit gates. This effective value accounts for the actual two-qubit gate errors (around $1\%$), the one-qubit gate errors (around $0.1\%$) and the measurement errors (around $3 \%$ ). 

We perform most of our simulations on a $N=54$ qubit system very close to the $53$-qubit Sycamore chip of Ref.~\onlinecite{Arute2019}, see Fig.~\ref{fig:qubits}. All the simulations presented here have been performed with limited computational ressources: one to few computing processes (fewer than 12) that have lasted at most a few hours. 
We will consider three different quantum circuits.

\emph{Sequence I} is essentially the quantum supremacy sequence of Google. $D=20$ layers are applied. For each layer, one applies a random one-qubit gate on each qubit, followed by the so-called fsim gate on all pairs of qubits according to the pattern ABCD-CDBA-ABCD-... (see Fig.~\ref{fig:qubits}). Sequence I has been designed to entangle the qubits as quickly as possible and therefore be as hard to simulate as possible.

\emph{Sequence II} is identical to sequence I but with a pattern rotated by 90 degrees, i.e. the sequence is CDBA-BACD-CDBA...

\emph{Sequence III} is, in contrast to sequence I and II, designed to perform a supposedly useful task. It implements the Quantum Approximate Optimization Algorithm (QAOA). QAOA is attracting a lot of attraction as a candidate algorithm to solve combinatorial optimization problems on a (noisy) quantum computer \cite{Harrigan2021}. It can be viewed as a discrete variational version of the adiabatic quantum computing paradigm.

Our main findings are summarized in the next three subsections.

\begin{figure*}
    \begin{subfigure}{\columnwidth}
    	\includegraphics[width=\textwidth]{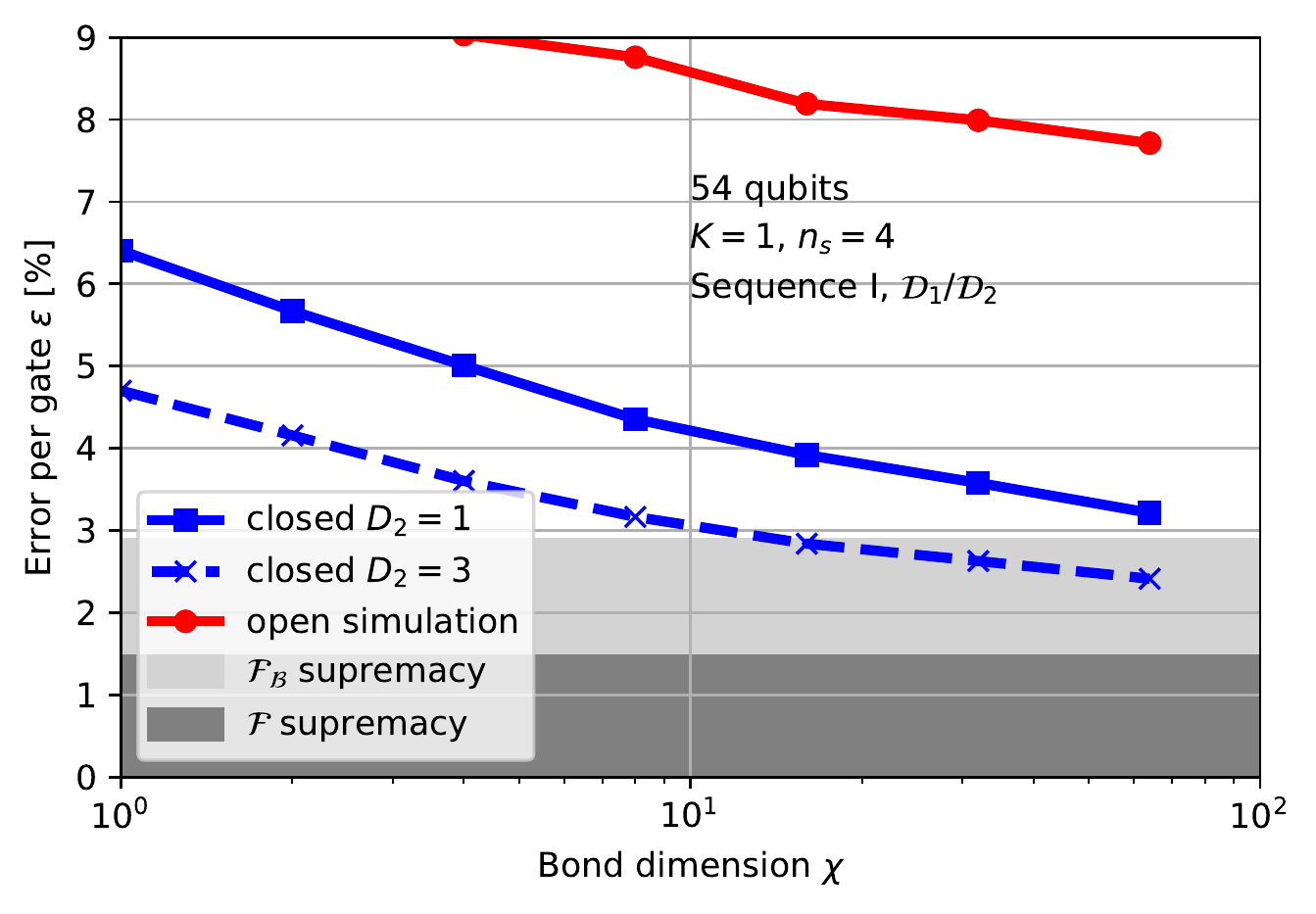}
    	\caption{}
    	\label{fig:strong_vs_chi_seqI}
    \end{subfigure}
    \begin{subfigure}{\columnwidth}
    	\includegraphics[width=\textwidth]{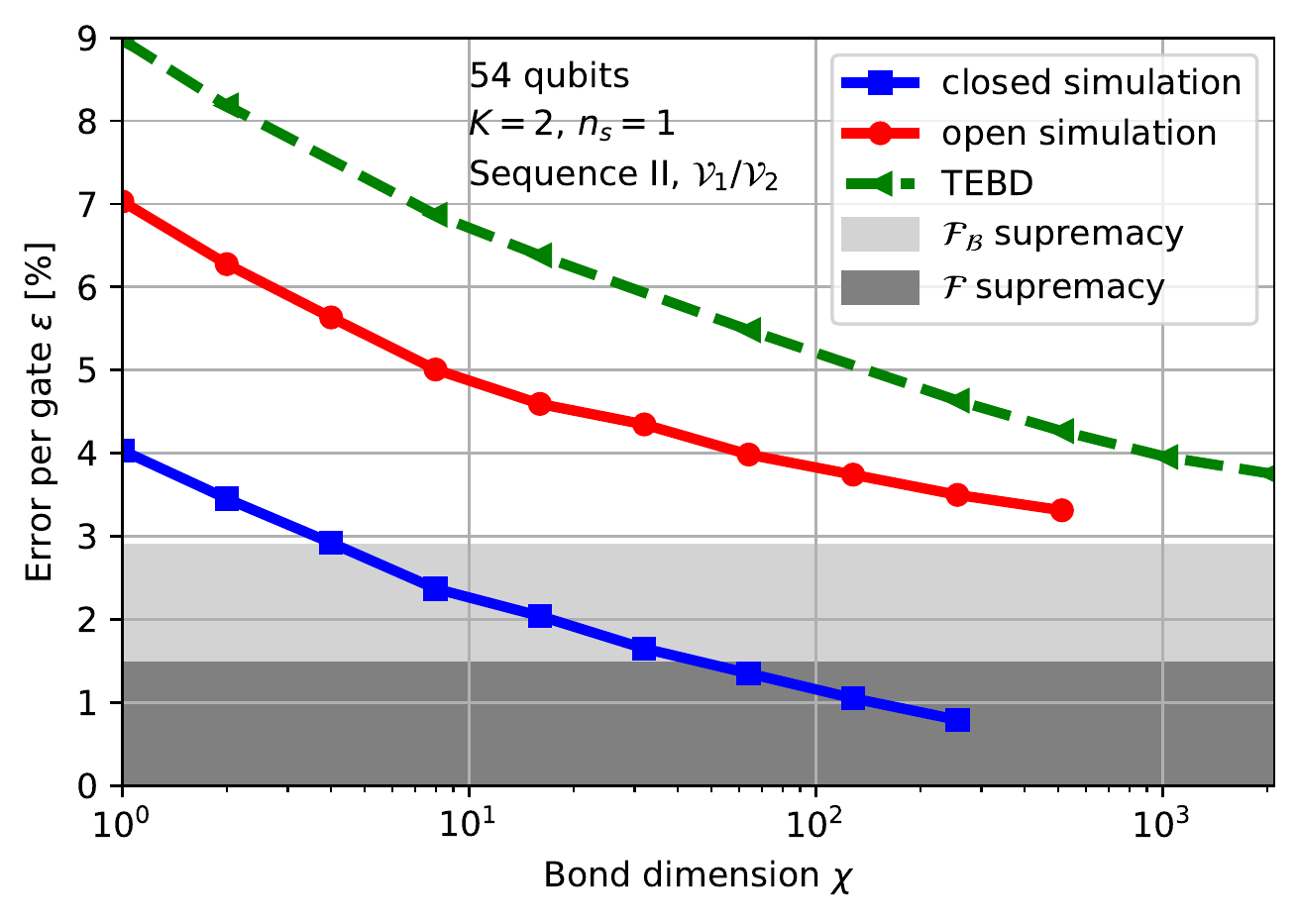}
    	\caption{}
    	\label{fig:strong_vs_chi_seqII}
    \end{subfigure}
    \begin{subfigure}{\columnwidth}
    \centering
    \includegraphics[width=\textwidth]{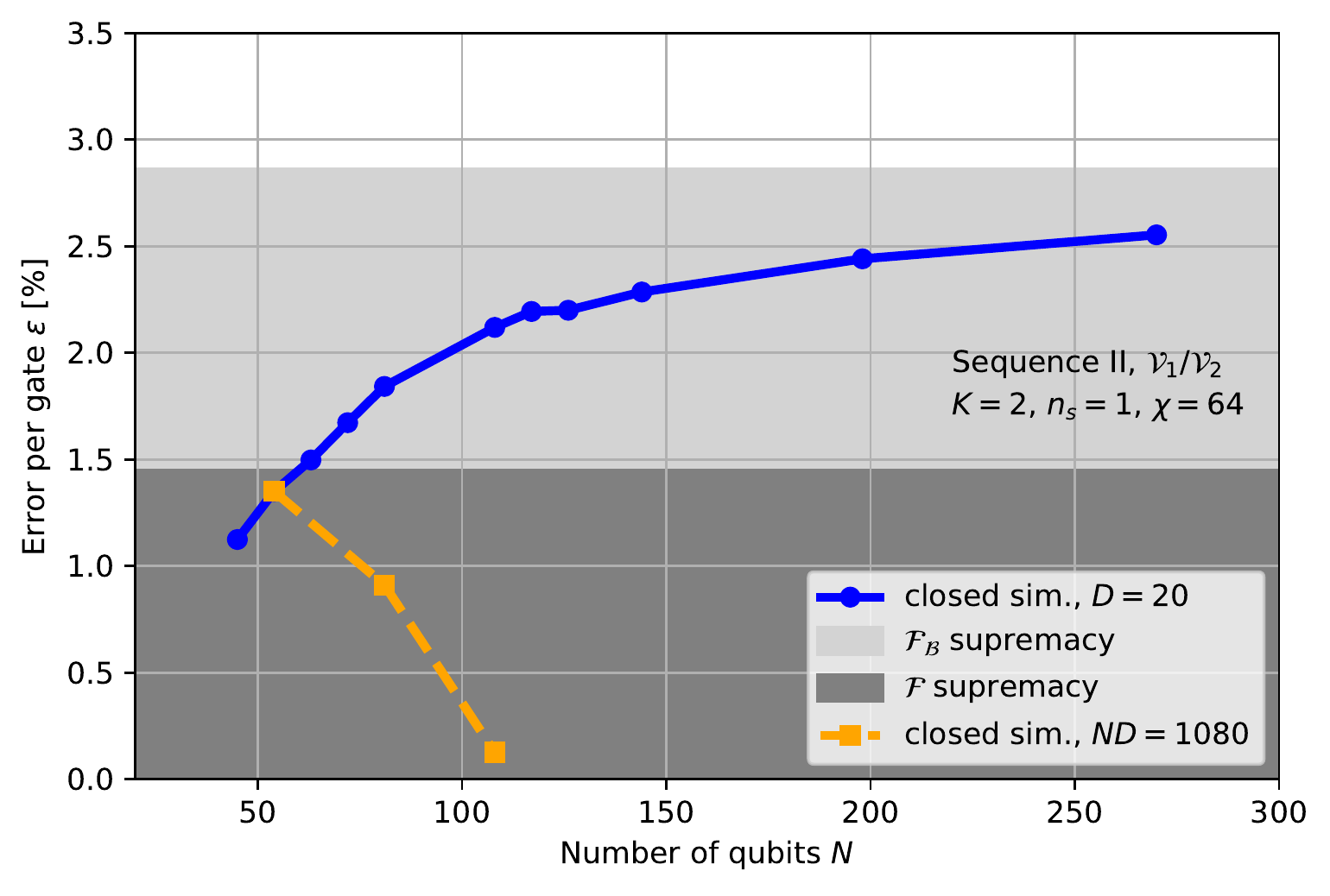}
	\caption{}
    \label{fig:panel_C}
    \end{subfigure}
    \begin{subfigure}{\columnwidth}
    \centering
    \includegraphics[width=\textwidth]{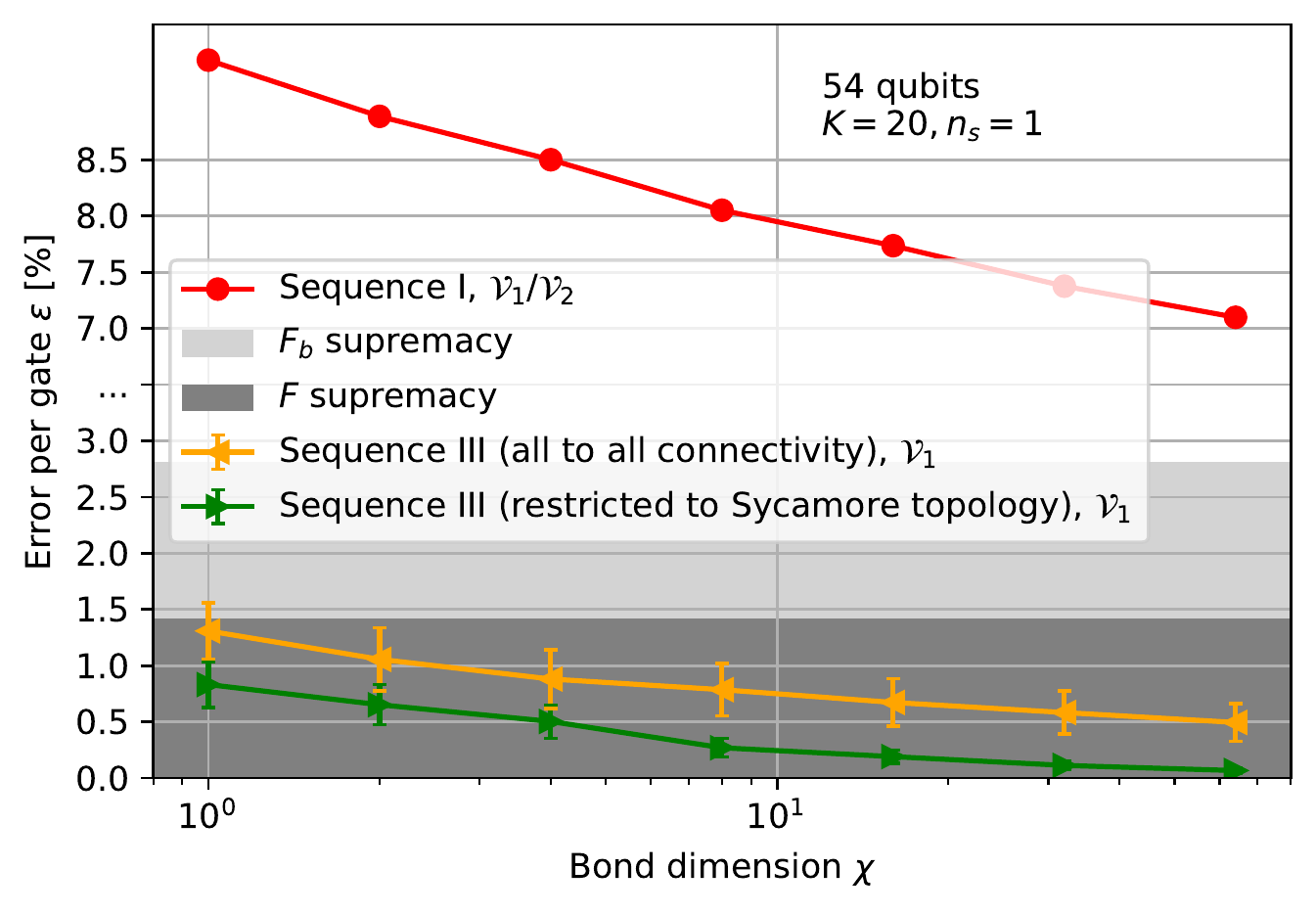}
	\caption{}
    \label{fig:seq_i_vs_seq_iii}
    \end{subfigure}
    \caption{\emph{Error rates achieved in our simulation}.
     (a) Error rate per gate $\epsilon$ as a function of bond dimension $\chi$ for the supremacy sequence I. In the gray regions the quality of the output is as good or better than the one produced in \cite{Arute2019}, see text.
     Light gray: $\epsilon\le 1.4\%$, dark gray: $\epsilon_{\cal B} \approx \epsilon/2 \le 1.4\%$.
     (b) Same as (a), for Sequence II.
     (c) Error rate per gate $\epsilon$ as a function of number of qubits $N$ for sequence II. $n_c$ is increased at fixed $n_b=5$ and $\chi=64$. Blue curve: fixed depth $D=20$. Orange curve: fixed number of two qubit gates i.e. $N D = 1080$. 
     (d) Comparison between the error rate of sequences I and III in open simulations. Each point in the sequence III curves is averaged over 10 graphs, with error bars corresponding to one standard deviation. Orange curve: QAOA circuit assuming perfect topology. Green curve: QAOA circuit using only the nearest neighbor connectivity of Sycamore chip.    }
\end{figure*}

\subsection{Simulating the supremacy sequence}

Fig~\ref{fig:strong_vs_chi_seqI} shows the error $\epsilon$ obtained in our simulations for the quantum supremacy sequence I with a system of $N = 54$ qubits. The $x$-axis is the ``bond dimension" $\chi$ that controls the level of compression of the quantum state, hence the accuracy. The computational cost of the simulation scales polynomially as $\propto \chi^2$ with a memory footprint $\propto \chi^2$. An exact simulation would correspond, at large depth, to an exponentially large \mbox{$\chi = 2^{N/2}$}, but we are here very far from this regime. 

The red curve shows the results in the ``open" mode. We find that the error rate for the largest 
$\chi=64$ studied is fairly high, around $8\%$, much higher than in state-of-the-art experiments. 
Going to ``closed" mode provides an important gain in error rate, typically by more than a factor two with our technique, enabling one to reach $\epsilon \approx 3\%$. By optimizing the closed mode (curve $D_2=3$, the details of the closed mode will be explained later), one reaches $\epsilon \approx 2.5\%$ at a computational cost that is still moderate.

To compare this error rate with the experimental one, we need to know the error rate associated with the experimentally measured cross-entropy benchmarking, i.e. $\epsilon_{\cal B} =  1 - {\cal F_B}^{1/N_\mathrm{2g}} \approx -(\log {\cal F_B})/N_{\rm 2g}$,
not $\epsilon$. 
The authors of Ref.~\cite{Arute2019} have argued that for their experiment ${\cal F_B} \approx {\cal F}$
for large enough depths. While this statement can be proven for certain classes of noise, it is not universal. For instance, it cannot hold at small depth since ${\cal F}(D=0) = 1$ while ${\cal F_B}(D=0) =2^N - 1$. Nor does it hold for systems consisting of disjoint pieces, for which fidelity composes multiplicatively while ${\cal F_B}$ composes additively when small \cite{Gao2021}.

{\it For our simulation technique}, a very different relation holds:
\begin{equation}
\label{eq:relation_between_F}
{\cal F_B} \approx \sqrt{{\cal F}}.
\end{equation} 
We shall provide strong evidence, both numerical and analytical, to support Eq.~(\ref{eq:relation_between_F}).
It follows that the $\epsilon = 2.5\%$ obtained in our simulations corresponds to $\epsilon_B = \epsilon/2 = 1.25\%
\le 1.4\%$ (light gray zone). Hence the bitstrings provided by these simulations have a higher cross-entropy benchmarking fidelity than those provided by Ref.~\cite{Arute2019} and in that sense, our technique can be considered as another breach in the claim of quantum supremacy. 
Note that since the relation between  ${\cal F}$ and
${\cal F_B}$ is highly non-universal, the fact that Eq.~(\ref{eq:relation_between_F}) holds in our simulation does not imply a similar relation in the experiments. 

\subsection{Scaling with the number $N$ of qubits}

A very important difference between the present work and previous attempts at bridging the quantum supremacy gap
is the scaling of the simulations with the number of qubits. Indeed, in our simulations, the exponential difficulty lies in increasing the fidelity, not the number of qubits.

In our present implementation, the computational cost of a simulation scales as $e^{\beta n_b} n_c D \chi^2$, where $n_b$ is the number of qubits in the first column and $n_c$ the number of columns
(see Fig.~\ref{fig:qubits}).  The memory required scales as  $e^{\beta n_b} n_c \chi^2$. The parameter $\beta \ge \log 2$ depends on the precise mode of calculation. Algorithms using Projected Entangled Pair States (PEPS) should provide a computational cost linear in both $n_b$ and $n_c$ \cite{Pang:2020,Lin:2021}. 

The scaling with $N$ is illustrated in Fig.~\ref{fig:panel_C}, where we show a calculation as a function of $N$ (by varying $n_c$) at fixed $\chi$ for sequence II. We perform simulations with more than $250$ qubits, while the error $\epsilon$ shows a limited increase before saturating (see the discussion in \cite{Zhou2020}). Such simulations would be totally out of the scope of standard simulation approaches. We emphasize again that the experiment corresponding to the blue curve in Fig.~\ref{fig:panel_C} would be exponentially difficult: one cannot increase $N$ at fixed $D$ experimentally (even assuming that so many qubits would be available) because the cross-entropy benchmarking would become too small to be measurable in a reasonable time. Working at fixed experimental measurement time, i.e. fixed ${\cal F_B}$ or equivalently keeping the product $ND$ constant corresponds to the orange curve in Fig.~\ref{fig:panel_C}.
We see here a first difference of behavior between our compression algorithms and actual experiments: our error rate $\epsilon$ actually drastically drops with $N$ in the orange curve, as our algorithm becomes essentially exact at small depth. This indicates that in order to beat compression algorithms, quantum hardware must improve in {\it fidelity} and/or {\it connectivity}: a mere increase of the number of qubit is not sufficient.

\subsection{Influence of the quantum circuit on the fidelity}

A second important difference between actual experiments and our compression algorithm appears upon considering different quantum circuits.
One of the chief results of \cite{Arute2019} is that the fidelity of the experiment  depends only on the number and type of gates applied and should to a large extent be agnostic to the type of circuit ran.
In practice, however, random circuits such as sequence I or II are experimentally easier than more structured circuits. This is due to several factors: $(i)$ these random circuits are optimally parallel without any idle time that could lead to further decoherence; $(ii)$ there is a compensation of errors due the random choice of gates; and $(iii)$ in the case of Ref.~\cite{Arute2019}, a pair-by-pair optimization of the fidelity of the two-qubit gates that could not have been performed had these gates corresponded to the prescription of an algorithm.  This increased difficulty---for experimental hardware---of running structured circuits is well known in the field of quantum benchmarking, see e.g \cite{Proctor2020}.

In our compression algorithm, we find, in sharp contrast with the above observations, that more structured quantum circuits are much easier to compress than random ones. While this result is not surprising on a qualitative level, the magnitude of the improvement in error rate that we observe is very high, with an error rate for open simulation dropping by a factor $100$ from $8\%$ for random circuits down to $0.07\%$ for QAOA circuits with the same $N=54$ and $N_{\rm 2g}=430$.

Fig.~\ref{fig:strong_vs_chi_seqII} shows a result for sequence II, which is only a slight modification of sequence I. We observe
that this slight modification of the sequence provides a twofold gain in $\epsilon$, bringing the open simulation almost down to the gray region and the closed simulations deep into the dark gray region. Fig.~\ref{fig:strong_vs_chi_seqII} also shows the result of the TEBD algorithm of Ref.~\onlinecite{Zhou2020} (green curve), showing that the DMRG algorithm presented in this article is a clear improvement over TEBD.

Fig.~\ref{fig:seq_i_vs_seq_iii} contrasts the results between sequence I and the QAOA sequence III. The results for the QAOA sequence correspond to the same number of qubits $N=54$ and same two-qubit gate count $N_{\rm 2g}=430$, i.e. such that experimentally one would expect a fidelity {\it lower} than the ${\cal F_B} =0.2\%$ observed for sequence I, for the reasons mentioned above.  In contrast, the results of the compression algorithm are drastically better for sequence III than
for sequence I. Two plots are presented in Fig.~\ref{fig:seq_i_vs_seq_iii}. 
In the orange curve, we have simulated the QAOA sequence supposing that the $N=54$ chip had perfect connectivity (a two-qubit gate can be applied between any pair of qubits). We find that
the error rate {\it in an open simulation} is reduced by a factor $14$ compared to sequence I, with $\epsilon = 0.5\%$ at $\chi=64$. The green curve corresponds to a circuit that respects the nearest-neighbor topology of the Sycamore chip. This topology puts additional constraints on the graphs that can be optimized with a given ``budget" $N_{\rm 2g}$ of two-qubit gates. It results in simpler graphs being simulated, and a further drop of the error rate down to $0.07\%$ for $\chi=64$ in the hardest {\it open } simulation mode. 
We note that two recent works \cite{Dupont2022,Sreedhar2022} considered a related problem (performance of a TEBD approach similar to \cite{Zhou2020} for a QAOA optimization) and arrived to conclusions that are, at least qualitatively, consistent with ours.
The simulation of QAOA circuits with up to 54 qubits was also tackled in a recent work~\cite{Medvidovi2021} using an alternative representation of the quantum state based on neural networks, the Restricted Boltzmann Machine \cite{Jnsson2018}, with similar conclusions.

While it is difficult to draw general conclusions from specific experiments, we conjecture that ``useful" quantum circuits, structured by nature, are generically {\it much} easier to simulate than random ones. It follows that in order for a quantum computer to show a genuine quantum advantage (i.e. quantum supremacy but for a useful task), far better fidelities will need to be demonstrated by the hardware.

\section{A Density-Matrix Renormalization Group algorithm for simulating quantum circuits}
\label{sec:dmrg}
\begin{figure}
	\includegraphics[width=\columnwidth]{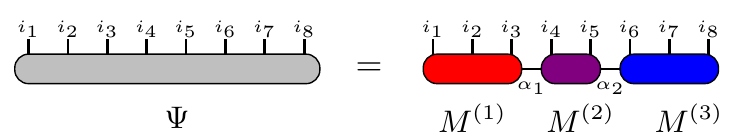}
	\caption{\emph{Decomposition of a 8-qubit state as a 3-tensor matrix product state (MPS).}}
	\label{fig:mps}
\end{figure}

We now describe the quantum state compression algorithm used in this article. This algorithm is inspired by the Density-Matrix Renormalization Group (DMRG) algorithm that has been highly instrumental for solving 1D quantum many-body problems \cite{White1992,Schollwoeck2011}. Our algorithm can be considered as a ``unitary" version of the original ``Hermitian" DMRG algorithm.

Our method combines a Schr\"odinger-type of simulation with compression steps where we approximate the quantum state with a Matrix Product State ansatz. This compression is performed every few layers of gates and requires one to find optimum contraction strategies for small tensor networks. The main new ingredient of this algorithm is the compression step. We emphasize that, although we introduce it in the context of Schr\"odinger-type of simulations, this step is in fact very general and could in principle be combined with other tensor-network simulations approaches. 

Since this article relies heavily on the tensor networks naturally associated to quantum circuits, the reader not familiar with these concepts may read the short introduction in Appendix~\ref{app:tensor_networks}.

\begin{figure*}
\includegraphics[width=0.8\textwidth]{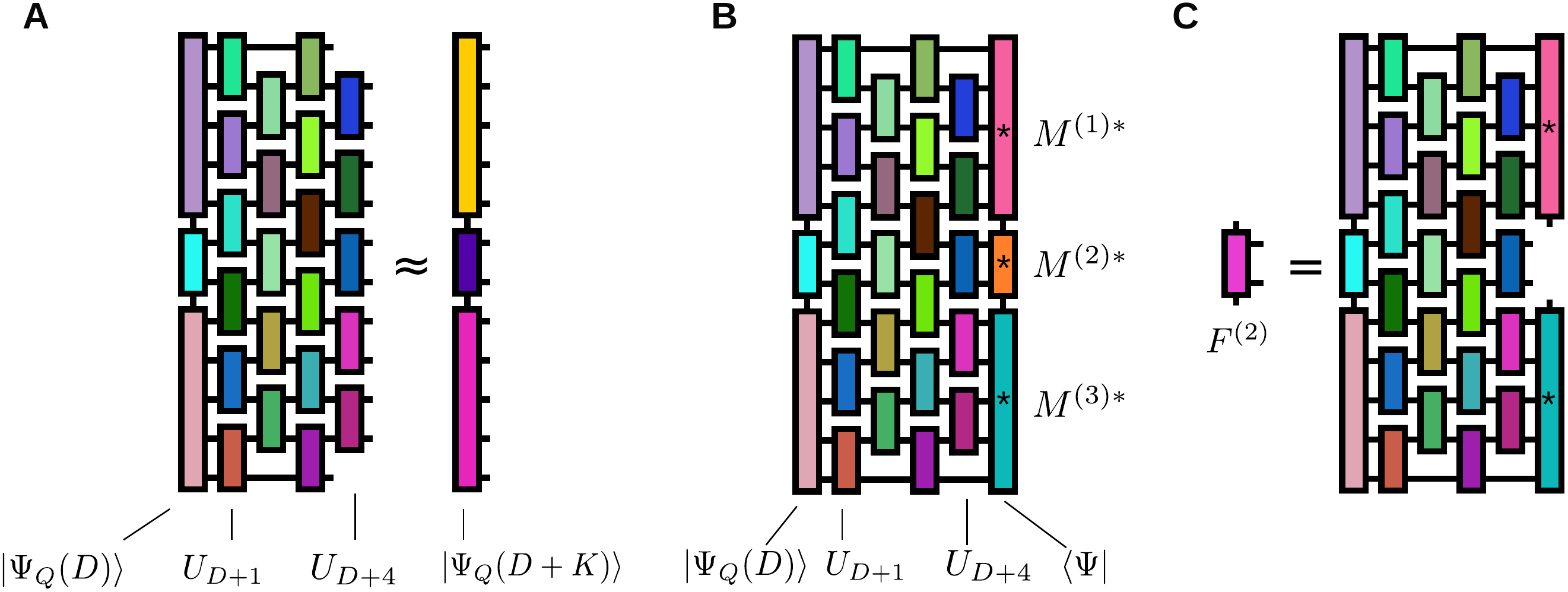}
\caption{ \emph{Compression step in the DMRG algorithm}.
	 (A) General schematic of the compression step: one adds $K$ layers of the quantum circuit, then approximates the resulting state with an MPS.
	 (B)  Tensor network representation of the scalar product to be optimized \eqref{eq:scalarproduct}.  The $*$ indicates the use of the complex conjugate of the tensor.
     (C) The central part of the calculation is the computation of the $F^{(\tau)}$ tensor.
\label{fig:flowchart}
}
\end{figure*}

\subsection{The Matrix Product State ansatz}

An arbitrary state $|\Psi\rangle$ with $N$ qubits
\begin{equation}
|\Psi\rangle = \sum_{i_1i_2i_3\dots i_N} \Psi_{i_1i_2i_3\dots i_N} 
|{i_1i_2i_3\dots i_{N}}\rangle,
\end{equation}
is described by a very large tensor $\Psi_{i_1i_2i_3\dots i_{N}}$. 
A matrix product state (MPS) factorizes and compresses this tensor by writing it as a product of $m$ tensors contracted in a chain-like geometry. See Fig.~\ref{fig:mps} for a schematic. 
Each tensor $\tau\in \{1, 2,\dots m\}$ contains the information on $r_\tau$ qubits:
\begin{eqnarray}
\Psi_{i_1i_2i_3\dots i_{N}} &= \sum_{\alpha_1,\dots,\alpha_{m-1}} 
M_{i_1i_2..i_{r_1},\alpha_1}^{(1)}
M_{\alpha_1,i_{r_1+1}\dots i_{r_1+r_2},\alpha_2}^{(2)}
\nonumber \\ 
&\cdots \ M_{\alpha_{m-1},i_{N-r_m+1}\dots i_{n_d-1}i_{N}}^{(m)}.
\end{eqnarray}

The  ``virtual'' indices $\alpha_\tau$ take at most $\chi$ values, where $\chi$ is known as the bond dimension. If $\chi$ is exponentially large, then a MPS can in fact describe any quantum state.
Here, however, we restrict ourselves to rather small values of $\chi$, in which case the MPS can only be an approximation of the entangled state that one aims at describing. There exists an important literature on many-body problems that can be successfully addressed by MPS variational ansatz~\cite{Schollwoeck2011}. The unentangled initial product state is naturally a MPS with $\chi=1$.

Note that in contrast to the approach of Ref.~\cite{Zhou2020}, grouping the qubits is not, in principle, necessary: one could use instead the conventional $\forall\tau,\  r_\tau = 1$ grouping and larger values of $\chi$ to obtain a variational ansatz as expressive as the one used with our non trivial grouping $r_\tau \ge 1$. 
We found however that, in practice, some groupings can be advantageous for some circuits, e.g. in the case where some qubits inside one group are highly entangled.
Since the first and last tensors only have $1$ virtual index instead of $2$, it is computationally advantageous to have more qubits in the corresponding groups.

\subsection{The main building block of the algorithm: the compression step}

The central part of the algorithm performs the following task: One starts from a MPS $| \Psi_Q(D) \rangle$, 
 supposedly a good approximation of the exact state $|\Psi_P(D) \rangle$. The problem is to find the best MPS
$|\Psi_Q(D+K)\rangle$ that approximates the state $U^{(D+K)}\cdots U^{(D+1)}| \Psi_Q(D) \rangle$ after one has applied $K$ layers of gates of the circuit, as illustrated on Fig.~\ref{fig:flowchart}A.
Namely, we want to determine
\begin{equation}
|\Psi_Q(D+K)\rangle \equiv 
\underset{|\Psi\rangle,\ \langle\Psi|\Psi\rangle=1}{\text{argmax}}
|\langle \Psi | U^{(D+K)}\cdots U^{(D+1)} | \Psi_Q(D) \rangle |^2.
\end{equation}

To perform this optimization, we optimize one given tensor $M^{(\tau)}$ of $|\Psi\rangle $ at a time while the remaining $m-1$ tensors are kept fixed. This optimization can be performed exactly using the simple formula 
Eq.~(\ref{eq:optimization}) derived below. It amounts to the contraction of a small tensor network.
To obtain the global optimum, we sweep over the choice of the tensor $\tau$ as in the single-site DMRG algorithm. 
Typically, a small number $n_s$ of sweeps is needed to obtained convergence towards  $|\Psi_Q(D+K)\rangle$.
Note that we have also tried variants of this algorithm analogous to original two-site DMRG algorithm (where two consecutive tensors are optimized simultaneously) but did not observe any significant improvement with respect to the simpler single-site version.

\subsubsection{Optimization of a single tensor}
Once we fix all the tensors $M^{(\tau')}$ of the MPS $|\Psi\rangle$ except for the tensor $M^{(\tau)}$, the scalar product to be optimized takes the form
\begin{equation}
\label{eq:scalarproduct}
	\langle \Psi | U^{(D+K)}\cdots U^{(D+1)} | \Psi_Q(D) \rangle =
\text{Tr} \ F^{(\tau)} M^{(\tau)*},
\end{equation}
where the trace means summation over all indices. Very importantly, this scalar product is a {\it linear} function of $M^{(\tau)}$. The tensor network for the left-hand side of Eq.~(\ref{eq:scalarproduct}) is shown in Fig.~\ref{fig:flowchart}B.
 It follows that $F^{(\tau)}$ is defined by the contraction of the tensor network shown in Fig.~\ref{fig:flowchart}C. In other words, $F^{(\tau)}$  simply corresponds to the tensor network for the full scalar product to which the 
$M^{(\tau)}$ tensor  has been removed. Note that in Fig.~\ref{fig:flowchart}C, the two tensors on the right (corresponding to $\langle\Psi|$) are complex conjugated.

Before doing the optimization, we need to enforce the fact that the MPS $|\Psi\rangle$ is a normalized state, i.e. $\langle\Psi|\Psi\rangle=1$. This is best done by performing a series of $QR$ factorizations on the tensors $M^{(\tau')}$ for $\tau'\ne \tau$ to bring the MPS in the so-called ``orthogonal form'', see \cite{Schollwoeck2011}. In this form, the norm of the MPS is simply given by
\begin{equation}
	\langle \Psi | \Psi \rangle =
\text{Tr}\  M^{(\tau)} M^{(\tau)*}.
\end{equation}
Introducing the Lagrange multiplier $\lambda$, the optimization over $M^{(\tau)}$ with the constraint $\langle\Psi|\Psi\rangle=1$ boils down to maximizing the function
$$
|\text{Tr} F^{(\tau)} M^{(\tau)*} - \lambda (1-\text{Tr} M^{(\tau)} M^{(\tau)*})|^2.
$$
i.e. we are maximizing a simple quadratic form. The optimum tensor $M_{\rm max}^{(\tau)}$ is easily found, and is related to the ``fitting'' approach used in the MPS literature \cite{PhysRevA.80.022334,Stoudenmire_2010}. It reads
\begin{equation}
\label{eq:optimization}
M_{\rm max}^{(\tau)} = \frac{1}{\sqrt{f_\tau}} F^{(\tau)},
\end{equation}
with $f_\tau = \text{Tr} \ F^{(\tau)} F^{(\tau)*}$. Eq.~(\ref{eq:optimization}) is the central equation around which
all this article is constructed. In addition, using Eqs~\eqref{eq:scalarproduct}, \eqref{eq:optimization}, one finds that the scalar $f_\tau$ is also the partial fidelity of the calculation 
\begin{equation}
f_\tau= |\langle \Psi | U^{(D+K)}\cdots U^{(D+1)} | \Psi_Q(D) \rangle|^2,
\end{equation}
which allows one to keep track of the progress of the optimization inside a sweep over $\tau$ or over different sweeps.

For each calculation of $F^{(\tau)}$, one obtains the local maximum of the partial fidelity with respect to $M^{(\tau)}$, hence $f_\tau$ can only increase as we sweep over the different tensors $\tau = 1 \dots m$.
We can then repeat the sweep over all tensors several times, yielding monotonically increasing fidelities $f^{(1)}_1, \dots, f^{(1)}_m, f^{(2)}_1, \dots, f^{(2)}_m, \dots, f^{(n_s)}_1, \dots, f^{(n_s)}_m$. 
The final value $f^{(n_s)}_m$ that we obtain after several sweeps over the different tensors is the partial fidelity $f_\delta$ that will enter our estimate of the fidelity $\tilde {\cal F}$, see Eq.~(\ref{eq:Ftilde-sup}), where $\delta$ indexes the number of compression steps.

\begin{figure}
	\includegraphics[width=7cm]{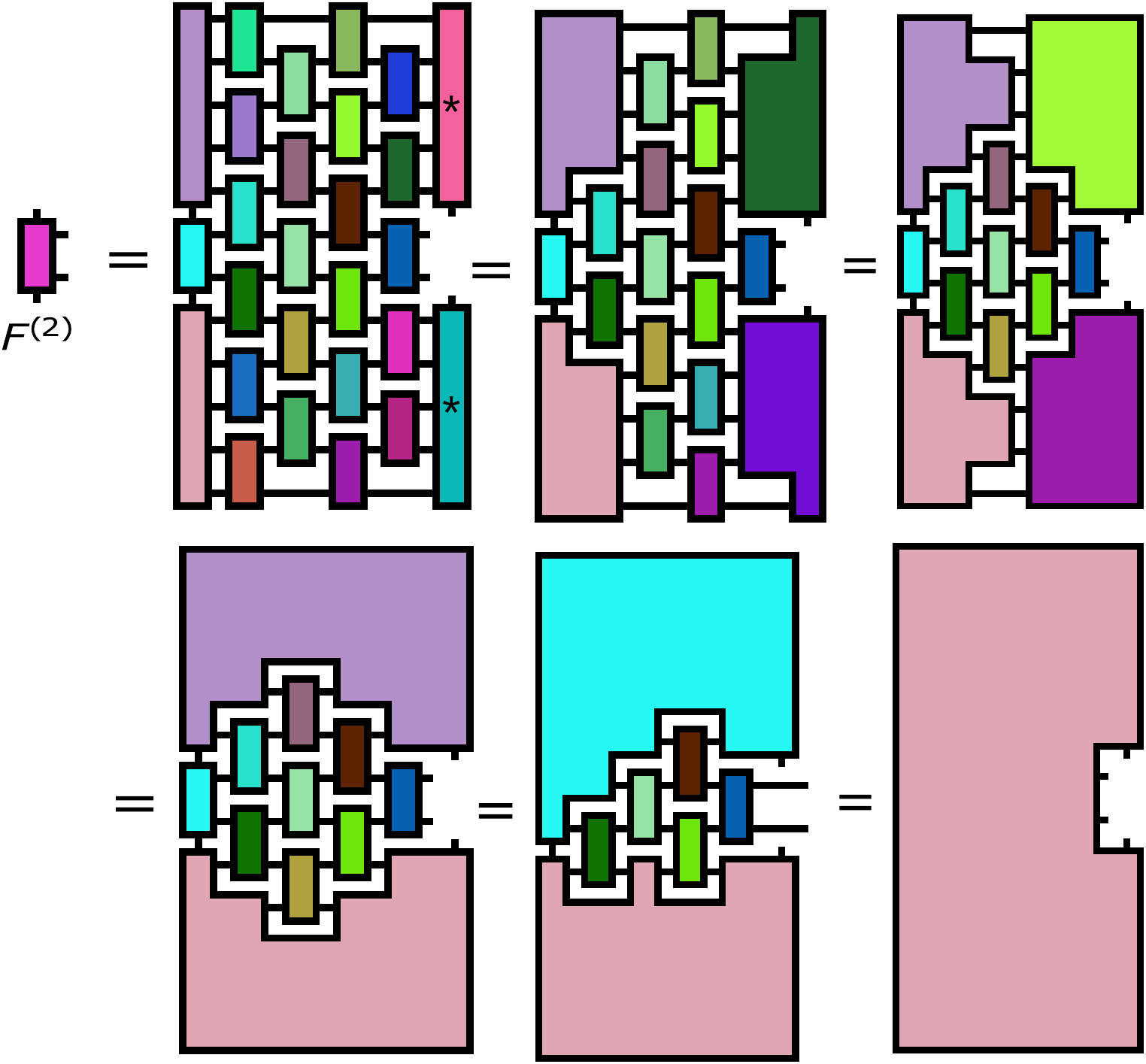}
	\caption{\textit{Contraction path to compute the $F^{(\tau)}$ tensor.}}
	\label{fig:contraction_order}
\end{figure}

\subsubsection{Contraction strategy for the tensor networks}

To complete the single tensor optimization, we need to perform the actual computation of the tensor $F^{(\tau)}$, i.e. we need a strategy for contracting the tensor network of Fig.~\ref{fig:flowchart}C.

The order in which the contractions are performed has a large impact on the final computation cost. In the case of a deep circuit with few qubits, a ``horizontal" contraction order (e.g. from left to right) will save computation cost. The horizontal contraction order corresponds to what is done in Schr\"odinger-like simulations. Its cost is prohibitive for a large number of qubits due to its exponential memory footprint $2^N$. Here, however, we consider only a shallow circuit of only a few $K$ layers at a time, so it is advantageous to perform the contraction in ``vertical" order (e.g. from top to bottom) since the exponential cost is with respect to $K$ instead of $N$.
This contraction algorithm is a direct adaptation of the well-known algorithm for calculating the scalar product between two MPS's \cite{Schollwoeck2011}. 

An example of contraction path for $K=4$ is shown in Fig.~\ref{fig:contraction_order}.
We first perform trivial contractions such as contracting one-qubit gates with nearby two-qubit gates.
Then, we contract the first top line of tensors and move down until we have reached the (missing) tensor $\tau$ that is being optimized. We repeat the same procedure from the bottom of the network upwards up to the missing $\tau$. Last, we merge the bottom part with the top part.  Through out the tensor network contraction, the largest tensors have $K$ physical indices (corresponding to the horizontal edges) and $2$ virtual indices (corresponding to the vertical edges). Each physical index represents $n_b$ qubits, and thus has dimension $2^{n_b}$; each virtual index has dimension $\chi$.
The typical maximum memory footprint for large $\chi$ thus scales as $\chi^2 2^{n_b K}$.  This is much smaller than
the $2^N$ scaling that one would be facing with a naive horizontal contraction path. 

Note that all the tensor network techniques discussed in Section \ref{sec:review} (heuristics for contraction paths, slicing...) could be used here to optimize and/or parallelize the calculation of $F^{(\tau)}$.

\subsection{Open versus closed simulation mode}

The algorithm can be used in two modes, open or closed, as discussed in section \ref{sec:review}. 
The open, ``Schr\"odinger-like'' mode provides the full quantum state after the $D$ layers of the circuit.
One simply adds $K$ layers at a time using the compression step until one has added all the $D$ layers.

In the closed simulation mode, we seek to calculate an amplitude
\begin{equation}
\Psi_x = \langle x | U_D U_{D-1}\cdots U_3 U_2 U_1 | 0 \rangle
\end{equation}
for a fixed output bitstring $x$. A closed simulation calculation corresponds to
the overlap of two MPS's, namely $U_{D/2+1}\cdots U_{D-1} U_{D} |x\rangle$ and $U_{D/2}\cdots U_1|0\rangle$,   which can be calculated with two separate open calculations whose circuit depths are halved compared to the open simulation mode.
Since calculations at small depths give much better fidelities with our technique (there is less entanglement at small depth)
the overall error rate $\epsilon$ is much lower. 

In practice, we first partition the circuit into three subcircuits with respectively $D_1$, $D_2$ and $D_3$ layers,
\begin{equation}
D = D_1 + D_2 + D_3
\end{equation} 
Then, we perform an open simulation with the first $D_1$ layers of the circuit (the forward part),
\begin{equation}
|\Psi_Q(D_1)\rangle \approx \prod_{d=1}^{D_1} U_d |0\rangle.
\end{equation}
Then, we perform a second open simulation starting from the $|x\rangle$ product state
with the last $D_3$ layers of the circuit (the backward part):
\begin{equation}
|\Psi_Q'(D_3)\rangle \approx \prod_{d=D}^{D-D_3+1} U_d^\dagger |x\rangle.
\end{equation}
Last, we add the remaining $D_2$ layers and compute the remaining scalar product without approximation, using a contraction strategy analogous to the one used for the calculation of the $F^{(\tau)}$:
\begin{equation}
\Psi_x \approx \langle \Psi_Q'(D_3) | \prod_{d=D_1+1}^{D_1+D_2} U_d | \Psi_Q(D_1) \rangle.
\end{equation}
The last calculation is performed exactly at the same cost as a compression step for $D_2=K$.
It may be advantageous to use $D_1 > D_3$
and/or the corresponding bond dimensions $\chi_1 > \chi_3$ if one wishes to calculate many different amplitudes $\Psi_x$. Indeed, the forward calculation needs to be done only once, while the backward and final calculations must be repeated for each bitstring $x$.
On the other hand, if one seeks the best possible fidelity, one should increase $D_2$ as much as possible in order to reduce the depth of the approximate parts of the calculation.

\section{Details on the numerical experiments}
\label{sec:details}

\subsection{Three quantum circuits}

\begin{figure*}
	\begin{center}
	\begin{subfigure}{0.3\textwidth}
		\includegraphics[width=\textwidth, angle=90, trim= 10 5 10 5]{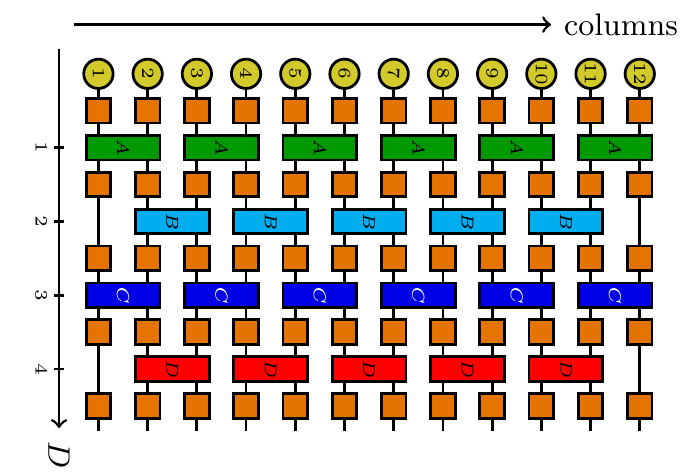}
		\caption{}
		\label{fig:circuit_seqI}
	\end{subfigure}
	\begin{subfigure}{0.3\textwidth}
		\includegraphics[width=\textwidth, angle=90, trim = 10 5 10 5]{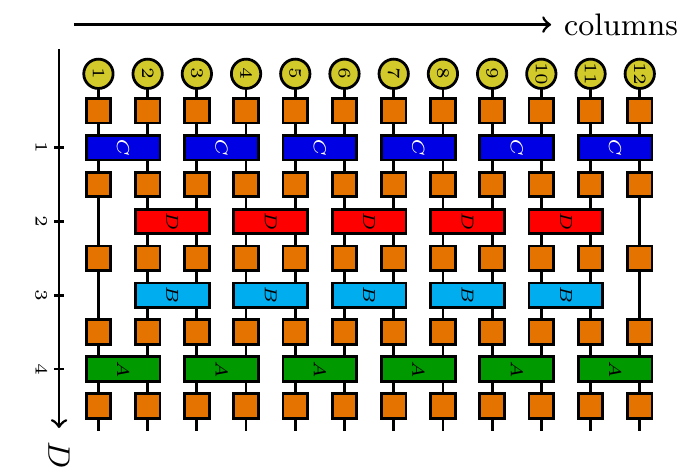}
		\caption{}	
		\label{fig:circuit_seqII}
	\end{subfigure}
	\begin{subfigure}{0.3\textwidth}
        \begin{center}
        \includegraphics[width=\textwidth]{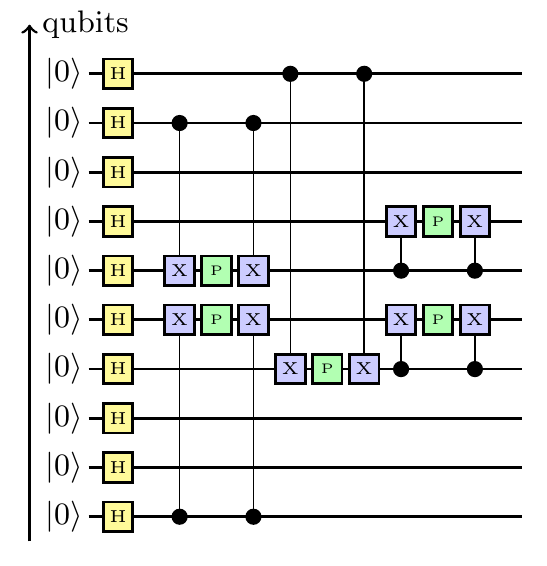}
        \caption{}
        \label{fig:circuit_seqIII}
        \end{center}
    \end{subfigure}
	\end{center}
	\caption{ \emph{The three sequences}.
		Random quantum circuits defined by: (a) Sequence I and (b) Sequence II. The circles represent columns of alternatively $5$ and $4$ qubits from the Sycamore chip, indexed from left to right. Orange squares correspond to applying a random one-qubit gate on each qubit (hence each column). Green, light blue, dark blue and red rectangles correspond to applying two-qubit gates between all qubits coupled by the associated coupler, respectively  A, B, C and D. (c) Beginning of a 10-qubit circuit implementing the QAOA algorithm for a MaxCut problem (where P gates are parametrized phase shift gates).}
	\label{fig:circuit}
\end{figure*}

In this article, we performed numerical experiments with three different quantum circuits, labeled sequence I, II and III as discussed in section \ref{sec:results}. A schematic of the three sequences is shown in Fig.~\ref{fig:circuit_seqI}, Fig.~\ref{fig:circuit_seqII} and Fig.~\ref{fig:circuit_seqIII} respectively.

Sequence I corresponds to the circuit of the quantum supremacy experiment, as shown on Fig.~\ref{fig:circuit_seqI}. It is designed to create a state as entangled as possible in as few steps as possible given the available nearest-neighbor connectivity. Each of the $D$ layers (where $D$ denotes the depth of the circuit, $D=20$ in the experiment) alternates between a one-qubit gate applied on all qubits (orange squares, drawn randomly from the $\sqrt{X}$, $\sqrt{Y}$ and $\sqrt{W}$ gates) and a two-qubit gate applied on a set of pairs of qubits, with four possible sets denoted by the letters A, B, C and D  as shown in Fig.~\ref{fig:qubits}  (A: green, B: light blue, C: dark blue and D: red rectangles). The two-qubit gate is the $\text{fsim}(\theta,\phi)$ gate defined in Ref.~\cite{Arute2019}. In the actual experiment, the values of $\theta$ and $\phi$ have been optimized for each pair of qubits to reach the best fidelity. Here we choose a constant value $\theta = 1$, $\phi=\pi/2$ that is close to the average experimental one, and deep in the difficult regime where the fsim gate has four different singular values. We also use the same sequence ABCD-CDAB-ABCD-... as \cite{Arute2019} in the supremacy regime.

The alternative sequence II is a variation on the quantum supremacy sequence where we have changed the order of the gates applied. Sequence II is rotated by 90 degrees compared to sequence I and reads CDBA-BACD-CDBA-...
This alternative sequence is just as useless but slightly ``less random" than the supremacy sequence, since we have not designed it to be optimally random. This slight modification of the ordering has a strong impact of the fidelity found in the simulations.

Finally, we benchmarked our DMRG algorithm on a useful task, sequence III. Sequence III is actually not a specific sequence of gates but a protocol for generating circuits implementing the Quantum Approximate Optimization Algorithm (QAOA)\cite{Farhi2014} for solving MaxCut problems of combinatorial optimization. We generated many such problems for the $N=54$ ``Sycamore" chip and selected instances where the associated QAOA circuit had a two-qubit gate count $N_{\rm 2g}\approx 430$ similar to the gate count of sequence I and II. Fig.~\ref{fig:circuit_seqIII} shows an example of such a circuit for a small number of qubits ($N=10$).

More specifically, we solve the MaxCut problem on Erdos-Renyi graphs $\mathcal{G}(N, \mathcal{P})$, a particular class of random graphs with $N$ vertices and a probability $\mathcal{P}$ for creating an edge between two vertices.
The QAOA ansatz circuit \cite{Farhi2014} is of the form $U = \prod_{k=1}^p U_B U_C$, with $U_B = \prod_{m=1 \dots N} e^{- i \beta X_m}$ and $U_C = \prod_{m,n \in E} e^{- i  \gamma Z_m Z_n} $, with $E$ the set of edges of the graph. Here, we are not interested in the result of the optimization itself. Hence,
we set the variational parameters $\beta$ and $\gamma$ to random values. For the same reason, we set the number of QAOA layers $p$ to 1.
The edge density $\mathcal{P}$ in the Erdos-Renyi graphs is adjusted so that the final two-qubit gate count is close to 430.
We consider two different cases: without and with compilation.
In the absence of compilation, the QAOA circuits do not necessarily comply with the grid connectivity of Sycamore (Fig.~\ref{fig:qubits}). To get a number of gates of about 430, we pick $\mathcal{P} = 32 \%$. 
In the second case, we compile the QAOA circuit to comply with the connectivity of Sycamore. The compilation uses SWAP insertion methods \cite{Hirata2009} to create a circuit that uses only the nearest neighbor two qubit gates available in Sycamore. As this procedure increases the depth of the circuit, we lower the edge probability $\mathcal{P}$ down to $5\%$ to keep the final two-qubit gate count to 430.

\subsection{Estimating the fidelity of a DMRG simulation}

A very interesting feature of the DMRG algorithm is that the fidelity 
${\cal F } = |\langle \Psi_P|\Psi_Q\rangle|^2$
of the calculation can be easily estimated, even though we do not necessarily have
access to the actual perfect state $|\Psi_P\rangle$. 
Inside a simulation, we estimate the fidelity with
\begin{equation}
\label{eq:Ftilde-sup}
\tilde{\cal F} = \prod_\delta f_\delta,
\end{equation}
where the partial fidelities $f_\delta$ are the final fidelities of compression step $\delta$
\begin{equation}
\label{eq:fa-sup}
f_\delta = |\langle \Psi_Q(\delta K + K) | U^{(D+K)}\cdots U^{(D+1)} | \Psi_Q(D =\delta K) \rangle|^2,
\end{equation}
at the end of the different optimization sweeps. $f_\delta$ is simply given by the norm of the 
last $F^{(\tau)}$ tensor calculated during the compression step.
It follows that our estimate of the error rate reads
\begin{equation}
\tilde{\epsilon}  = 1 - \tilde{\cal F}^{1/N_\mathrm{2g}}.
\end{equation}
It was shown in Ref.~\cite{Zhou2020} through a combination of analytical and numerical arguments, that to very good approximation, one has
\begin{equation}
\label{eq:FvsFtilde}
\tilde{\mathcal{F}} \approx \mathcal{F},
\end{equation} 
and the arguments and numerics given there remain valid for this article.
However, since we have used a different compression algorithm here, we have performed an additional extensive numerical study of the validity of the multiplicative law [Eq.~(\ref{eq:FvsFtilde})] for up to a maximum of $36$ qubits for which we can obtain the exact state $|\Psi_P\rangle$, hence the exact fidelity ${\cal F}$.
 The multiplicative law Eq.~(\ref{eq:Ftilde-sup}) has a very important role, as it allows us to perform estimations of the fidelities in regimes where the exact calculation is out of reach and only $\tilde{\cal F}$ can be obtained.
We have found that Eq.~(\ref{eq:FvsFtilde}) indeed holds in all regimes of interest.

 Fig.~\ref{fig:multi_panels_exact} shows the comparison between the exact fidelity ${\cal F}(D)$ (blue line) and our estimate $\tilde{\cal F}(D)$ (dashed blue line) for a large choice of values of $n_b$ and $n_c$ up to $N=36$ qubits (beyond that, we do not have access to the exact state $|\Psi_P\rangle$). We find a perfect match between the two curves in the relevant $1 \ge {\cal F} \ge 1/2^N$ regime.
For very large depth ${\cal F}$ saturates at an exponentially small value $1/2^N$  while $\tilde{\cal F}$ continues to decrease exponentially. The exponentially small asymptotic value  $1/2^N$ corresponds to the overlap between two independent Porter-Thomas chaotic vectors, see the derivation around Eq.~(\ref{eq:randomF}). It is essentially the lowest fidelity that can be reached in a random circuit.

Interestingly, Eq.~(\ref{eq:Ftilde-sup}) provides an estimate of the error $\epsilon_x$
even in the closed mode where we calculate a single amplitude $\Psi_x$. To check this assertion, we have computed the histogram of the error rates per gate in the closed mode for small systems where we could calculate
all the amplitudes $\Psi_x$ exactly. Typical results are shown in Figure \ref{fig:fidelity_exact_closed} for three different configurations. We find that our estimated fidelity closely matches the actual value with a precision of a few percent (typically less than $5\%$) for {\it all the points in the histogram}.  Note that in all regimes we have ${\cal \tilde{F}} \le {\cal {F}}$ so that ${\cal \tilde{F}}$ underestimates the actual fidelity.

\subsection{Different groupings of the qubits}

\begin{figure*}
	\centering
	\includegraphics[width=13cm]{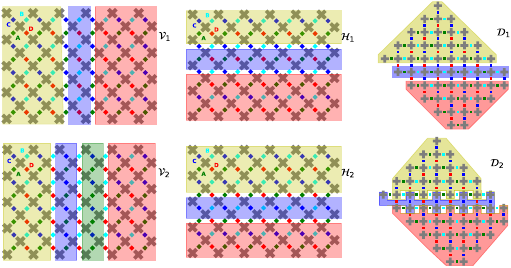}
	\caption{\emph{Different qubit groupings used in our simulation.} All qubits shaded with the same color (yellow, blue, green or red) belong to the same tensor $M^{(\tau)}$.}
	\label{fig:groupings}
\end{figure*}

Our DMRG algorithm gives us the freedom to define how we group the different qubits that
correspond to each tensor $M^{(\tau)}$, $1\le \tau\le m$. The different groupings that we have used 
are shown in Fig.~\ref{fig:groupings}, where each color corresponds to one tensor $M^{(\tau)}$.

Different groupings may have some advantage depending on the actual circuit ran. The vertical groupings $\mathcal{V}_1$ and $\mathcal{V}_2$ group the qubits by columns. For instance $\mathcal{V}_1$ contains three tensors with respectively $5$ ($23$), $2$ ($9$) and $5$ ($22$) columns (qubits). Layers B and D are ``trivial" for grouping $\mathcal{V}_1$, i.e. they
are internal to one tensor hence have a perfect fidelity $f_\delta =1$ for any value of the  bond dimension $\chi$. Likewise, layers A and C are trivial for grouping $\mathcal{V}_2$.
For the larger systems of Fig.~\ref{fig:panel_C} with more than $54$ qubits, we have used an extension of $\mathcal{V}_1$ and $\mathcal{V}_2$ by adding additional tensors of two columns to
obtain $N = 23 + 9(m-2) + 22$ qubits with $m$ tensors having respectively $23,9,9,\dots,9$ and $22$ qubits. We find that the vertical groupings are optimum for sequence II.

Another possibility is to group the qubits horizontally in rows as in $\mathcal{H}_1$ (for which A and D are trivial), or $\mathcal{H}_2$ (for which B and C are trivial). Last, we may group the qubits diagonally as in $\mathcal{D}_1$ and $\mathcal{D}_2$. In some calculations, we may try and alternate between two groupings e.g. $\mathcal{D}_1$ and $\mathcal{D}_2$ to optimize the number of trivial gates.

\subsection{Benchmark of the algorithm}

Let us now see how the algorithm performs in practice. All the simulations are carried out on a
Sycamore-like architecture with $n_c$ columns where each column has $n_b$ (odd) or $n_b-1$ (even) qubits, as shown in Fig.~\ref{fig:qubits}. The real Sycamore chip corresponds to $53$ qubits arranged in \mbox{$n_c=12$} columns with $n_b=5$ qubits in the first column. We also performed simulations on smaller systems where we could obtain the exact state (up to $N=35$ qubits) using Atos QLM's Schrödinger-style ``qat-linalg'' simulator.

\subsubsection{Convergence of the DMRG compression step}

\begin{figure*}
	\begin{center}
    \begin{subfigure}{\columnwidth}
        \includegraphics[width=\columnwidth]{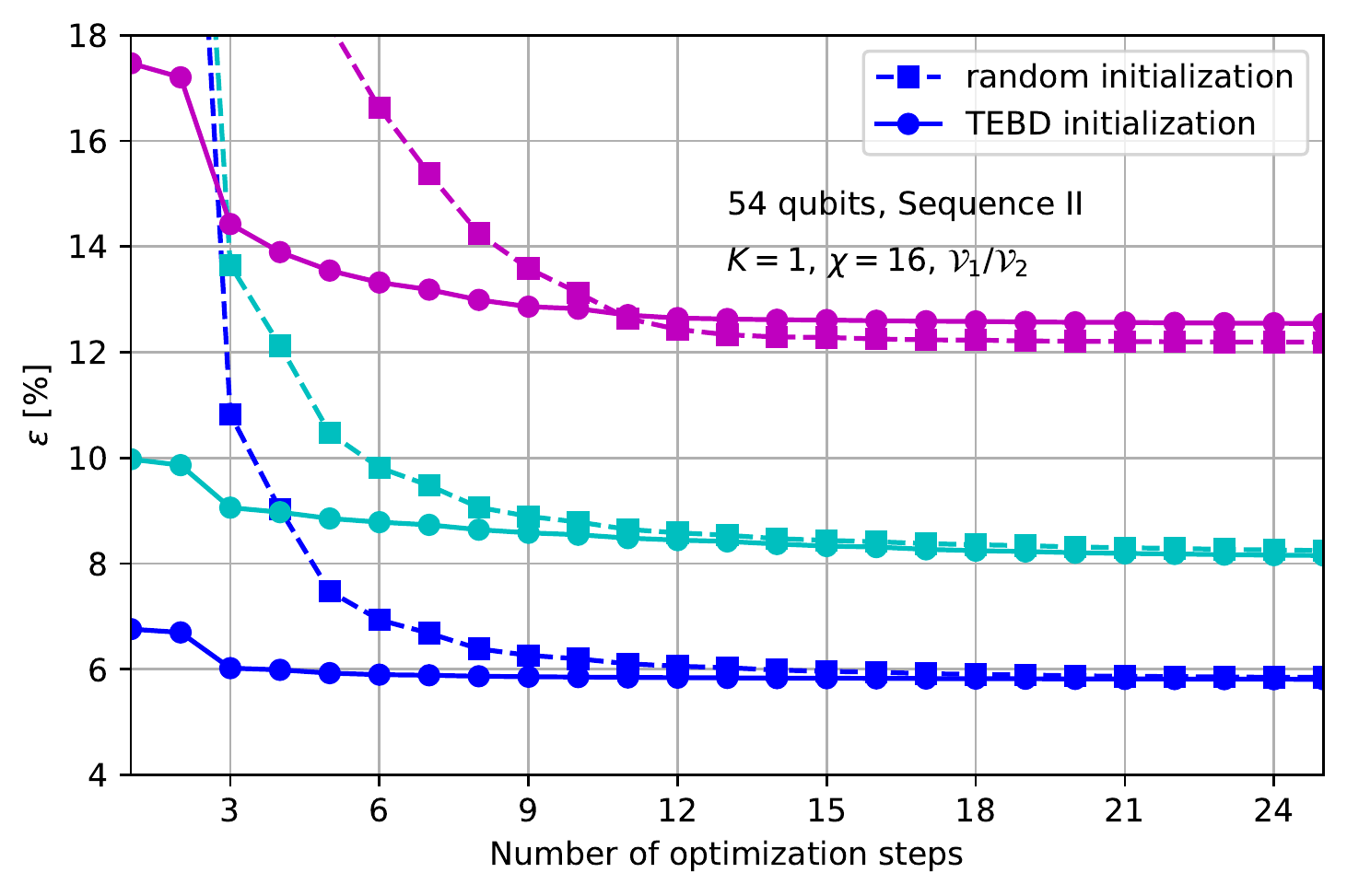}
        \caption{}
        \label{fig:optimization_tebd_vs_rnd}
    \end{subfigure}
     \begin{subfigure}{\columnwidth}
	\centering
	\includegraphics[width=\columnwidth]{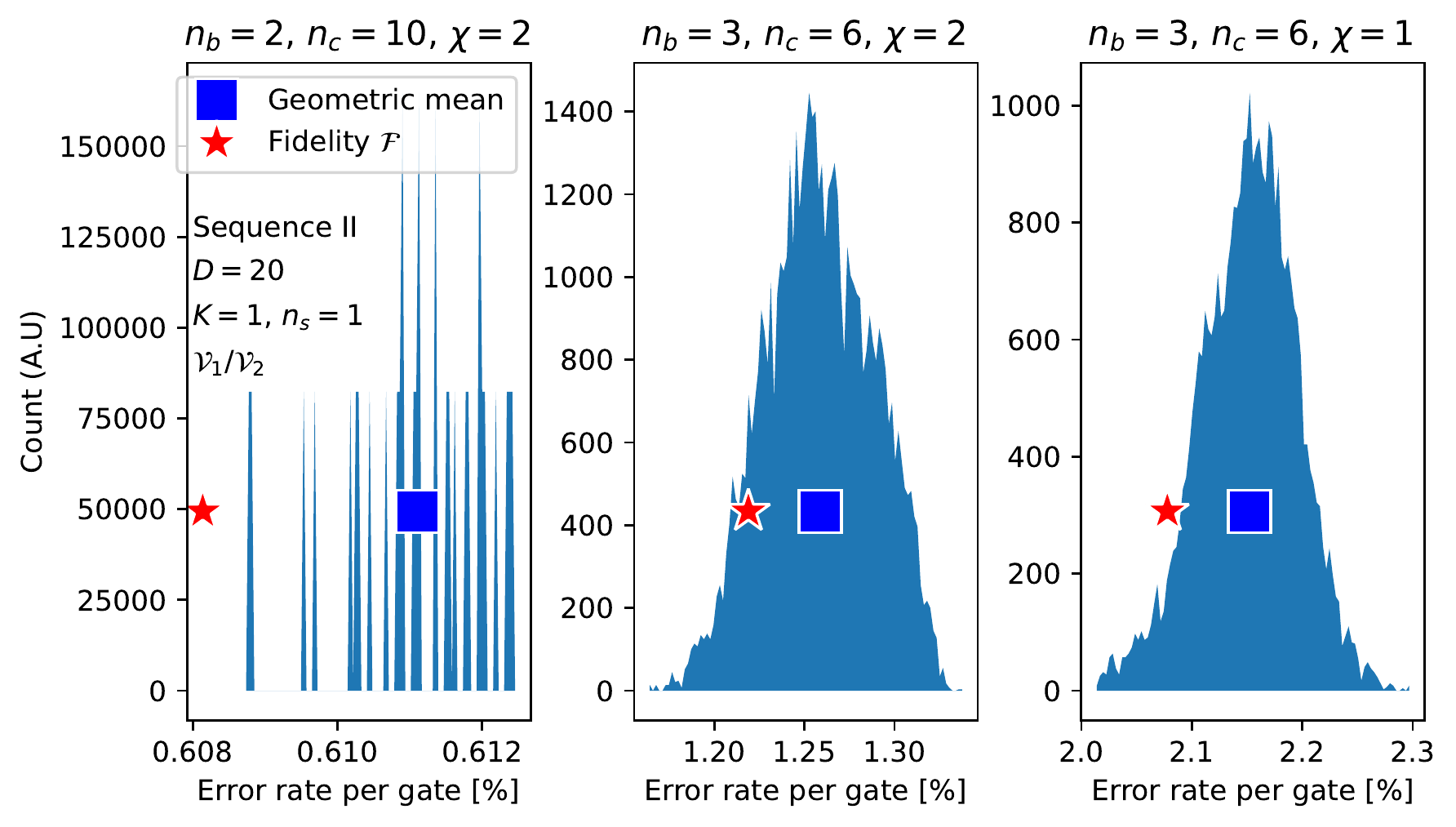}
	\caption{}
	\label{fig:fidelity_exact_closed}
    \end{subfigure}
	\begin{subfigure}{\columnwidth}
		\includegraphics[width=\textwidth]{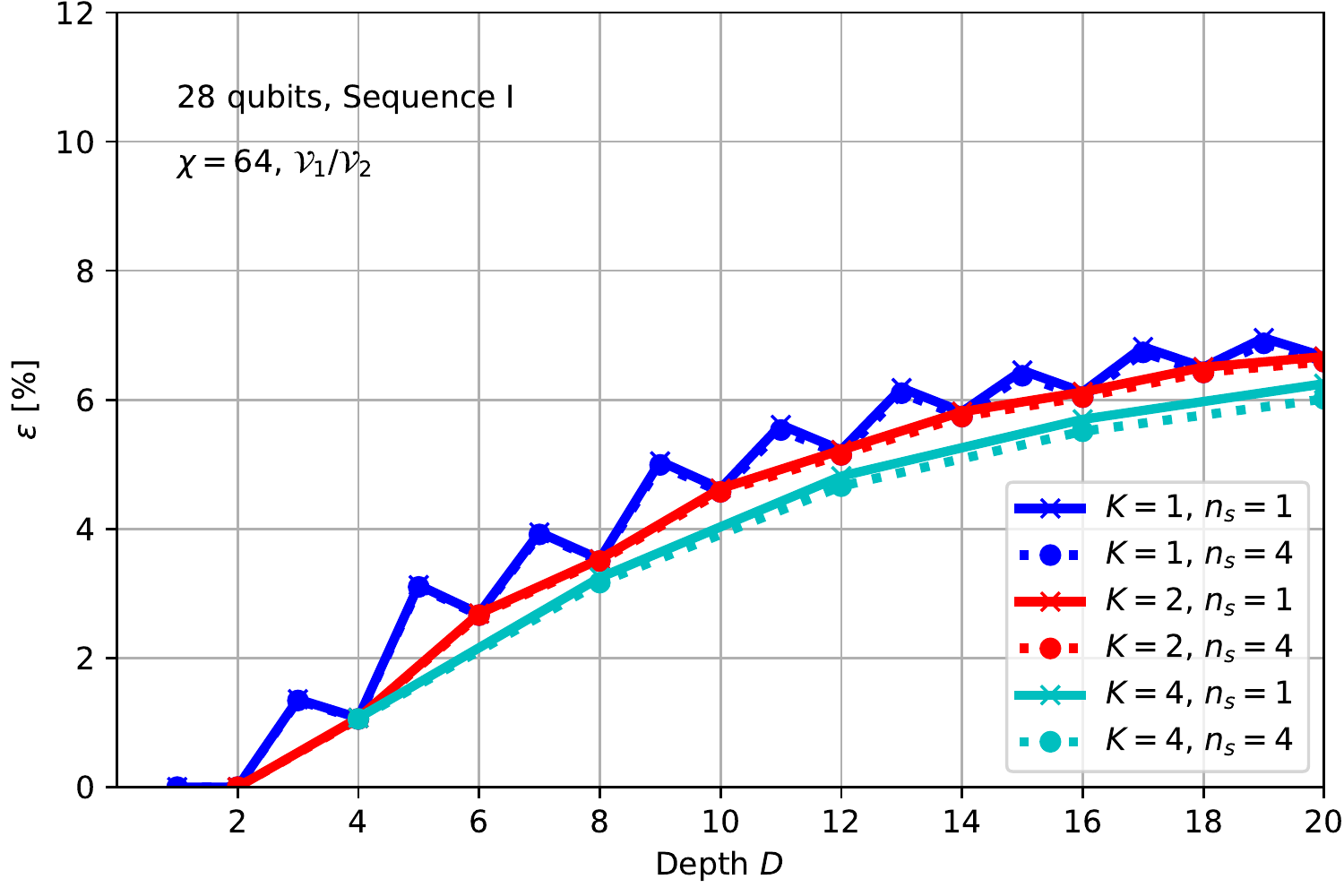}
		\caption{}
		\label{fig:effect_of_K_and_nsweeps_vertical}
	\end{subfigure}
	\begin{subfigure}{\columnwidth}
		\includegraphics[width=\textwidth]{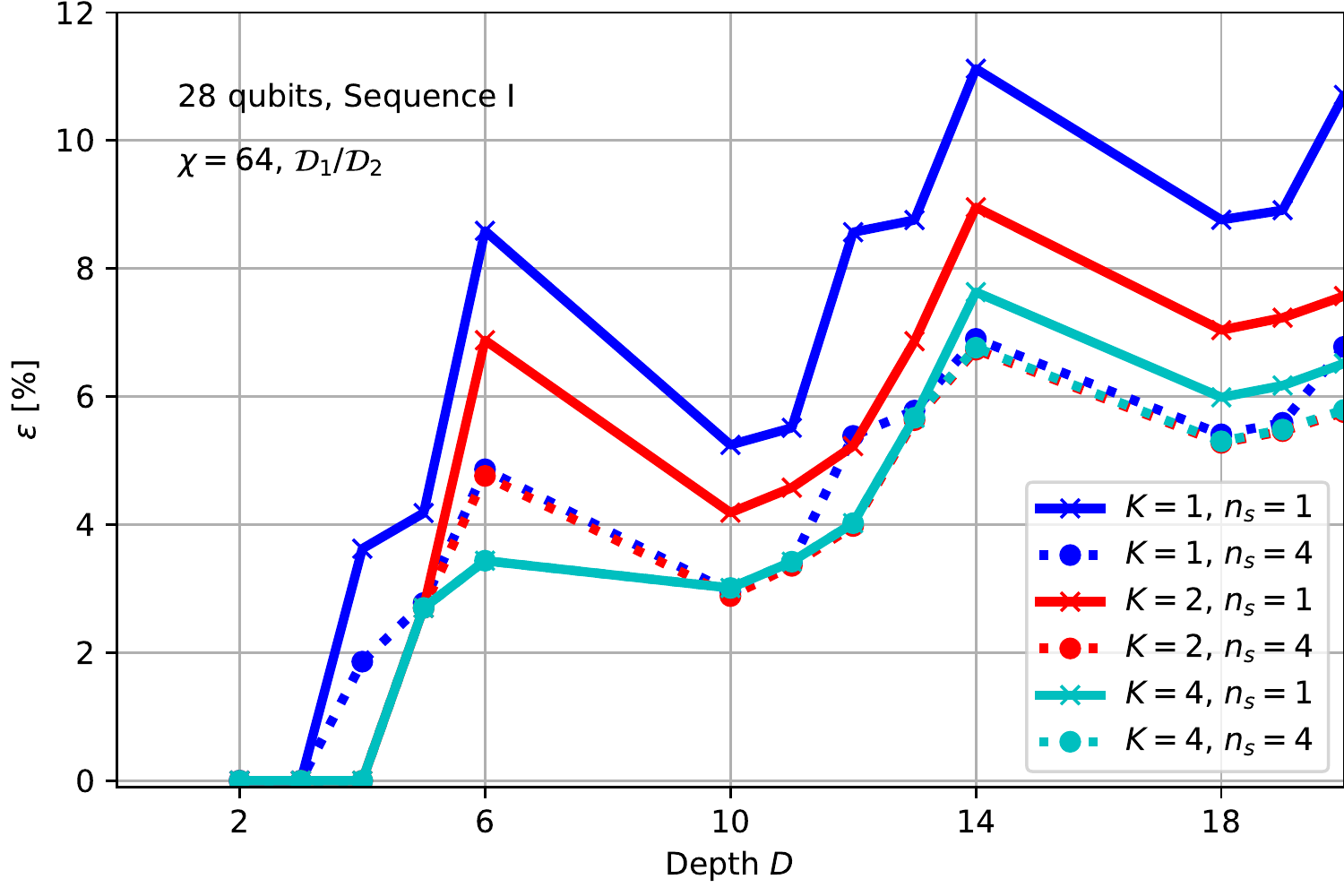}
		\caption{}
		\label{fig:effect_of_K_and_nsweeps_diagonal}
	\end{subfigure}
	\begin{subfigure}{\columnwidth}
        \includegraphics[width=\textwidth]{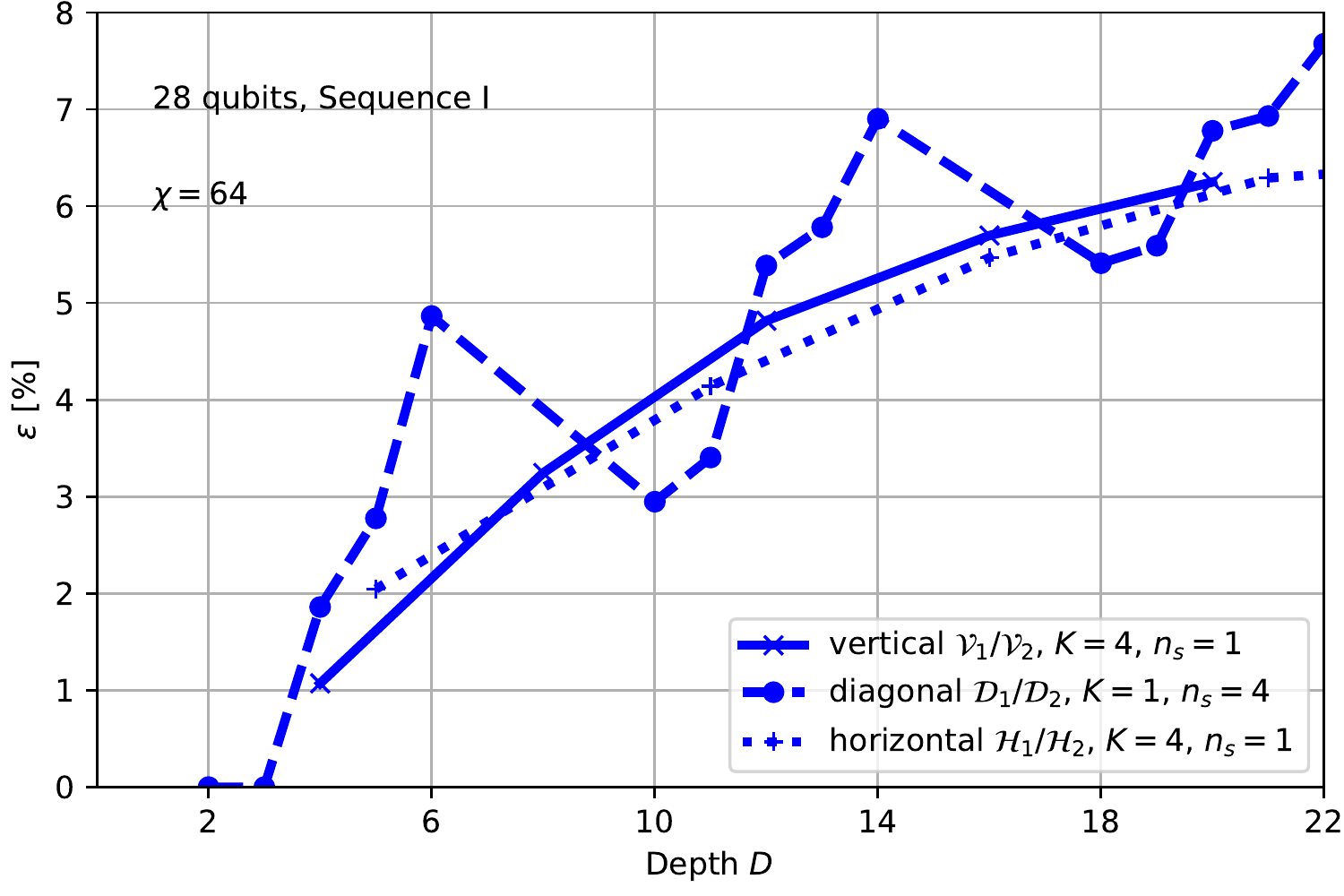}
        \caption{}
        \label{fig:strong_vs_depth_new}
    \end{subfigure}
	\begin{subfigure}{\columnwidth}
        \includegraphics[width=\textwidth]{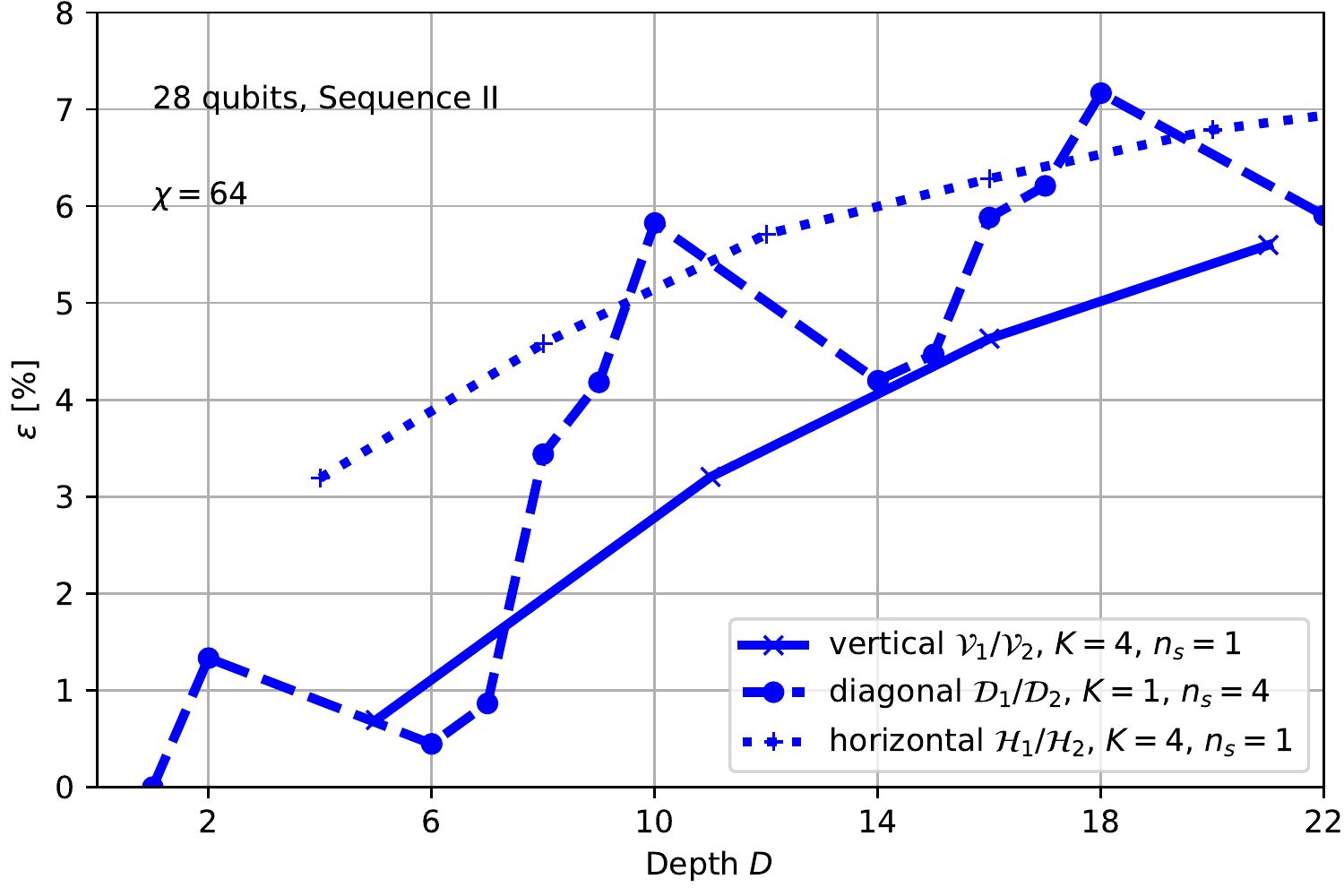}
        \caption{}
        \label{fig:strong_vs_depth_old}
    \end{subfigure}
   \end{center}
    \caption{\emph{Convergence of the DMRG algorithm.} 
        (a) Evolution of the error per gate versus the total number of local optimizations. Three steps correspond to a full sweep.  Different colors correspond to different compression steps for a random (squares and dashed lines) and for a TEBD (disks and solid lines) initialization. The different curves correspond to different compression steps in an open simulation (respectively the $4^{th}$ (blue), the $10^{th}$ (cyan) and the $12^{th}$ (magenta) step, with corresponding depths being $D=5$, $13$ and $16$.
        (b) Histogram of the error rates per gate $\epsilon_x$ of the amplitudes of individual bitstrings $x$ computed using closed simulation for various $n_b$, $n_c$ and $\chi$. Blue squares: geometric mean $\left(\prod_x \epsilon_x\right)^{1/2^N}$. Red stars: exact error per gate $1-\mathcal{F}^{2/(ND)}$.
        (c) and (d) Role of $K$ and number of sweeps $n_s$. Evolution of the error rate per gate $\epsilon$ versus depth $D$ for various numbers of layers $K$ and numbers of sweeps. Parameters:  $\chi=64$, $n_b = 4$ and $n_c = 8$. (c) Vertical ordering: groupings $\mathcal{V}_1$, $\mathcal{V}_2$. (d) Diagonal ordering: groupings $\mathcal{D}_1$, $\mathcal{D}_2$. 
        (e) and (f) Role of the choice of qubit groupings and of the sequence of gates applied. Evolution of the error rate per gate as a function of current depth. (e) Sequence I, used in Google supremacy experiment. (f) Sequence II, a variation on the supremacy sequence.
        }
\end{figure*}

Fig.~\ref{fig:optimization_tebd_vs_rnd} shows the convergence of the optimized MPS during the DMRG sweeps. We plot the error per gate $\epsilon = 1 - (f^{(k)}_\tau)^{1/N_\text{2g}^{(K)}}$ as a function of the number of optimization steps. Here  $N_\text{2g}^{(K)}$ is the number of two-qubit gates in the $K$ newly added layers
and $f^{(k)}_\tau$ is the fidelity obtained upon optimizing tensor $\tau$ in sweep $k$. Since the corresponding MPS contains $m=3$ tensors, $3$ optimization steps correspond to one full sweep.  Since each step provides a full optimization over one tensor, the error rates decreases monotonically, as expected. 

Fig.~\ref{fig:optimization_tebd_vs_rnd} shows two types of initialization of the MPS $|\Psi\rangle$
that is being optimized: either an arbitrary random MPS (squares and dashed lines) or an already partially optimized MPS obtained from the TEBD algorithm of Ref.~\cite{Zhou2020} (disks and solid lines). By construction, the DMRG error can only be lower than the TEBD error. Unsurprisingly, we find that the convergence is much faster when starting from TEBD (typically $1-3$ sweeps) than when starting from a random guess (typically $4-6$ sweeps). However, the final error found shows weak  dependence on the initial starting point
(in Fig.~\ref{fig:optimization_tebd_vs_rnd} we show one case (pink curves) where there is a visible difference between the TEBD initialization and the random ones, but it is seldom observed). We have also repeated the simulation with different random initializations and the error always converges to the same value. Since we optimize only one tensor at a time, we cannot dismiss the possibility to be trapped in a local minimum but these observations indicate that the DMRG algorithm gets at least very close to the global optimum MPS. Note that this is the global optimum MPS {\it for a given compression step}. In section \ref{sec:optimum}, we shall discuss how the algorithm manages to track the global optimum for the {\it full circuit} after several compression steps.

\subsubsection{Role of the number of layers per step $K$ and number of sweeps $n_s$}

In most of the numerical experiments shown in this article, we have used the vertical ordering with $K=2$, $n_s=1$ or the diagonal ordering with $K=1$, $n_s=4$. In the $\mathcal{V}_1$ grouping, ``even" layers of type B and D (see Fig.~\ref{fig:groupings}) can be absorbed trivially in the MPS without increasing the bond dimension. It follows that the fidelity for $K=1$ and $K=2$ is actually the same.
We have performed a few simulations in the $\mathcal{V}_1$ grouping with $K=4$. They present a small, typically $0.5\%$ gain with respect to $K=2$ in return for $16$-fold increase in computing time. Note that this gain reflects our ability to {\it find} the MPS closest to $|\Psi_P\rangle$, not the ability of the MPS to capture the exact state. Indeed, the $K=2$ and $K=4$ calculations share the same bond dimension, and hence the same maximum level of possible entanglement.

Fig.~\ref{fig:effect_of_K_and_nsweeps_vertical} shows the error rate $\epsilon$ as a function of the depth $D$ for different values of the number of layers $K$ added
inside the compression step, for the vertical grouping $\mathcal{V}_1$. 
The error rate for depth $D\le 2$ is zero as our MPS has a bond dimension large enough to accomodate the corresponding entanglement exactly. As one increases $D$ further, one starts to 
feel the approximation and the error rate increases.
As mentioned above, the error rates for $K=1$ and $K=2$ are identical with additional oscillations for the intermediate points for $K=1$. This increase in the error rate for intermediate points for $K=1$ corresponds to the fact that these depths do not benefit from an upcoming trivial layer, hence the average error rate is higher. For $K=4$, we observe a small gain at large depths, but since the corresponding computational time increases significantly, we have not used $K=4$ in practice. Similar calculations for the diagonal grouping $\mathcal{D}_1$ are shown in Fig.~\ref{fig:effect_of_K_and_nsweeps_diagonal}. The error rate shows oscillations due to the fact that, in this configuration, the $D$ gate is very costly. 
Overall, we find that for a large enough number of sweeps $n_s$, the $K$ dependence of the error rate is small. This is already a strong indication that the algorithm provides a MPS not far from optimum, even though we are carrying out multiple compression steps.
Increasing the number $K$ of layers provides a small gain of $<1\%$ in the error rate at $D=20$. 
However the computational cost increases exponentially with $K$. Calculations with 54 qubits and $K=2$ are beyond the scope of the present article for the diagonal grouping.

\subsubsection{Role of the qubit grouping}

Figs.~\ref{fig:strong_vs_depth_new} and~\ref{fig:strong_vs_depth_old} show the error rate versus depth for
three different groupings, vertical, diagonal and horizontal. We observe important variations of the error rate with the grouping as well as with the circuit (sequence I versus sequence II). Note that for this small system of
$28$ qubits ($n_b=4$, $n_c=8$), sequence II is only marginally easier than sequence I because the system is almost like a square (as opposed to a rectangle for the Sycamore case $n_b=5$, $n_c=12$). This difference between different groupings is in itself not surprising: different circuits tend to entangle some qubits more than others. Since the choice of the grouping amounts to choosing the position of ``entanglement bottleneck", there must be
an optimum grouping for each circuit.

\section{Does DMRG provide the optimal MPS?}
\label{sec:optimum}

\begin{figure}
	\centering
	\includegraphics[width=\columnwidth]{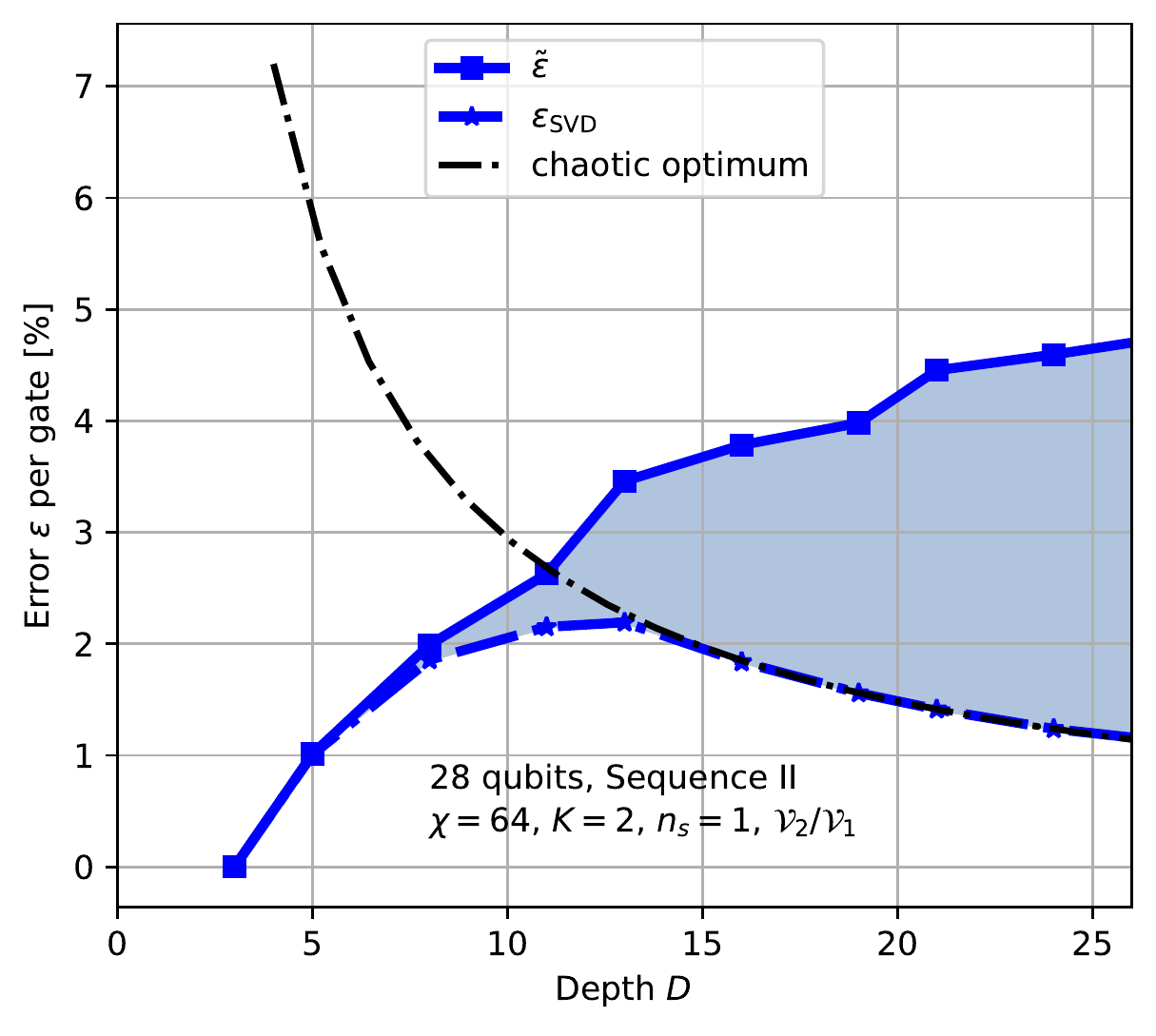}
	\caption{\emph{Comparison of the error per gate to the SVD error as a function of depth $D$.} Vertical qubit grouping $\mathcal{V}_2$. The chaotic optimum is Eq.~\eqref{eq:error_rate_asymptotics}}
	\label{fig:global_optimum}
\end{figure}

In this section, we discuss whether the MPS obtained by the DMRG algorithm
corresponds to the best possible MPS. Indeed, two possible causes may prevent the final MPS obtained to be optimal: the fact that the optimization is broken into different compression steps, and the fact that within a compression step the optimization is performed tensor by tensor, not globally. We analyze this problem in the context of the random circuit of sequence II.

To assess this point, Fig.~\ref{fig:global_optimum} shows three different errors $\epsilon$ versus depth $D$ curves:
\begin{itemize}
\item The blue squares (continuous line) show the error $\tilde{\varepsilon}$ obtained within the DMRG algorithm
\item The blue stars (dashed line) show the error $\varepsilon_\mathrm{SVD}$ corresponding to best possible approximation of the exact state with a MPS, as explained below. This reference curve can only be obtained for a small number of qubits and is not available in general.
\item The black dot-dashed line shows the error $\epsilon_{\rm opt}$,  the best possible approximation of a purely chaotic state with a MPS, as given by
\begin{equation}
\epsilon_{\rm opt}=\frac{1}{D}\left(\log2-\frac{\log4\chi}{2N}\right)\label{eq:error_rate_asymptotics}
\end{equation}
(see the analytical derivation below, appendix~\ref{app:chaotic_opt}). We refer to Eq.~(\ref{eq:error_rate_asymptotics}) as the ``chaotic optimum error". The fact that the chaotic optimum error decreases with $D$ stems from the simple fact that the best {\it fidelity} that one may obtain when approximating a chaotic state with a MPS is a finite number $\mathcal{F}_{\rm opt} ={4\chi}/{2^{N/2}}$ so that the error per gate must decrease.
\end{itemize}

\subsection{Best possible MPS calculation}

To obtain the blue stars (dashed line) ``best possible MPS" curve of 
Fig.~\ref{fig:global_optimum}, we performed an exact ``Schr\"odinger-like"
simulation of the small $28$-qubit system and obtained the exact state $|\Psi_P(D)\rangle$.
 
In a second step, we split the qubits into two groups A and B of equal size, so that $|\Psi_P(D)\rangle$ reads
\begin{equation}
|\Psi_P(D)\rangle=\sum_{\alpha \beta}\Psi_{\alpha \beta}(D) |\alpha\rangle_{A}|\beta \rangle_{B}
\end{equation}
where the states $|\alpha\rangle_{A}$ ( $|\beta\rangle_{B}$) form an orthonormal basis of A (B). We perform a singular value decomposition $\Psi=USV^\dagger$ of the $2^{N/2}\times2^{N/2}$  matrix, $\Psi_{ab}$, from which we get the Schmidt decomposition of $|\Psi\rangle$:
\begin{equation}
|\Psi\rangle=\sum_{\mu}S_{\mu}|\mu\rangle_{A}|\mu\rangle_{B}\label{eq:Psi_schmidt}
\end{equation}
with $|\mu\rangle_{A}=\sum_{\alpha}U_{\alpha \mu}|\alpha\rangle_{A}$ (and a similar
expression for $|\mu\rangle_{B}$). Sorting the singular values $S_\mu$
in decreasing order, we can obtain the best approximate MPS by truncating the above expression and keeping the $\chi$ largest singular values. 
 
 \subsection{Analytical calculation of the chaotic optimum Eq.~(\ref{eq:error_rate_asymptotics})}

The computation of the chaotic optimum Eq.~(\ref{eq:error_rate_asymptotics})
follows the same procedure as for the ``best possible MPS" discussed above
with a small modification: instead of starting with the exact state $|\Psi_P(D)\rangle$, we start with a fully chaotic state $|\Psi\rangle$ distributed according to the Porter-Thomas distribution, i.e. 
\begin{equation}
|\Psi\rangle = \sum_{x=0}^{\mathcal{N} - 1} \Psi_x |x\rangle
\end{equation}
where ${\cal N} = 2^N$ and the Porter-Thomas vector $\Psi_x$ corresponds to one column of a unitary matrix distributed according to the Haar (uniform) measure of $U({\cal N})$. 

The derivation of the ``chaotic optimum'' error formula follows from the
properties of the singular values of random Gaussian matrices. It is performed in Appendix~\ref{app:chaotic_opt}. 

\subsection{Numerical results}
We find in Fig.~\ref{fig:global_optimum} that the best possible error $\varepsilon_\mathrm{SVD}$ first increases as the system gets more and more entangled; reaches a maximum at $D\approx 12$; and then starts to decrease following closely the chaotic optimum. The maximum error at $D=12$ hence corresponds to the depth beyond which the state of the system is chaotic. 

In contrast, in our DMRG simulations the error $\tilde{\varepsilon}$ can, by construction, only increase. It eventually saturates to a finite value when the MPS becomes made of random tensors (see an in-depth discussion of this last point in \cite{Zhou2020}).
Hence, it must deviate from the best possible approximation  $\varepsilon_\mathrm{SVD}$ at some point.
 Fig.~\ref{fig:global_optimum} shows that this deviation appears around $D=12$ when the system becomes chaotic.
 At small depths, the DMRG results are indeed very close to the best possible approximation.
From Fig.~\ref{fig:global_optimum} data, we conjecture that the intersection between the DMRG error and the chaotic optimum can be used to estimate when the quantum state becomes chaotic. Before one reaches the chaotic regime, the DMRG algorithm is close to optimum. Conversely, Fig.~\ref{fig:global_optimum} can be seen as a strong indication that for non-chaotic states, DMRG will perform significantly better than for chaotic ones.

\section{Relation between fidelity and cross-entropy benchmarking}
\label{sec:FvsFB}

The relation between the actual fidelity ${\cal F}$ and the cross-entropy
benchmarking ${\cal F_B}$ is far from trivial. 
Ref.~\cite{Arute2019} has argued that {\it in their experiment} and {\it in the large-depth chaotic regime} both quantities are equivalent ${\cal F_B} \approx {\cal F}$, see also \cite{Zhou2020} for a discussion.

Here we show that {\it in our numerics} we have a very different relation
\begin{equation}
\label{eq:FvsFB}
{\cal F}_{\cal B}  \approx \sqrt{\cal F},
\end{equation}
which implies that ${\cal F}_{\cal B}  \gg  {\cal F}$ (since $\mathcal{F} \ll 1$).
We give two kinds of evidence for Eq.~(\ref{eq:FvsFB}):
\begin{itemize}
\item A large body of numerical caculations for systems up to $N=36$
qubits where we can simulate the exact state $|\Psi_P\rangle$, hence calculate both the left and right-hand side of Eq.~(\ref{eq:FvsFB})
\item An analytical calculation in the very large $D$ limit. In this limit, 
$|\Psi_P\rangle$ and $|\Psi_Q\rangle$ converge to two independent random chaotic states and one can calculate the two fidelities exactly and show that Eq.~(\ref{eq:FvsFB}) holds. It follows that the type of errors present in an experiment plays an important role to know which relation holds: assuming that one only makes precision errors experimentally (such as over rotations during a gate), one would retain a pure state at large depth and Eq.~(\ref{eq:FvsFB}) would hold.
\end{itemize}

These two bodies of evidence are presented in the rest of this section.

\subsection{Numerical evidence for $\mathcal{F}_\mathcal{B}\approx \sqrt{\mathcal{F}}$}

\begin{figure*}
    \centering
    \includegraphics[width=\textwidth]{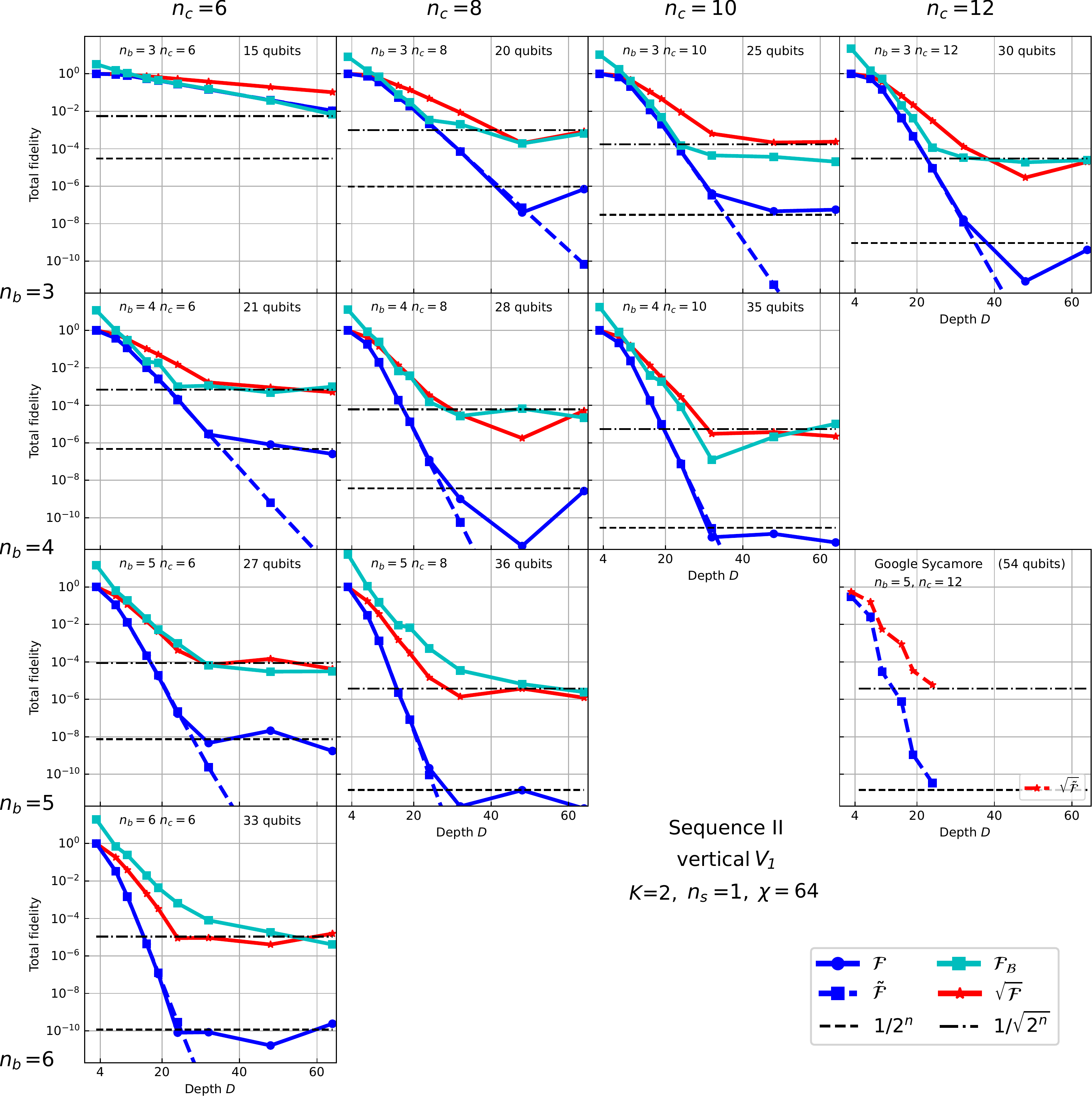}
    \caption{\emph{Comparison of various fidelities.} Fidelities as a function of total circuit depth $D$, at fixed bond dimension $\chi=64$. Exact fidelity $\mathcal{F}$ (Eq.~\eqref{eq:F-sup}), cross-entropy benchmarking fidelity $\mathcal{F}_\mathcal{B}$  (Eq.~\eqref{eq:FB-sup}), product fidelity $\tilde{\mathcal{F}}$  (Eq.~\eqref{eq:Ftilde-sup}). Open simulation mode. $K=2$ layers. }
    \label{fig:multi_panels_exact}
\end{figure*}

Fig.~\ref{fig:multi_panels_exact} shows  $\mathcal{F}$ (blue) and $\mathcal{F}_\mathcal{B}$ (green) for several values of $n_b$ and $n_c$ for which $N\le 36$, so that the exact state could be obtained using the large random access memory (RAM) of the Atos Quantum Learning Machine. Also shown
is $\sqrt{\cal F}$ (red), which is close to the green curve as well as the two exact asymptotic values $1/2^N$ for $\mathcal{F}$ and 
 $1/2^{N/2}$ for $\mathcal{F_B}$.
As one gets closer to the Sycamore chip regime ($n_b=5$, $n_c=12$), for which no exact statement can be made, we find that the relation ${\cal F}_{\cal B} \approx \sqrt{\cal F}$ becomes increasingly valid.

\subsection{Fidelity and cross-entropy benchmarking in the chaotic limit}

\begin{figure}
    \centering
    \includegraphics[width=\columnwidth]{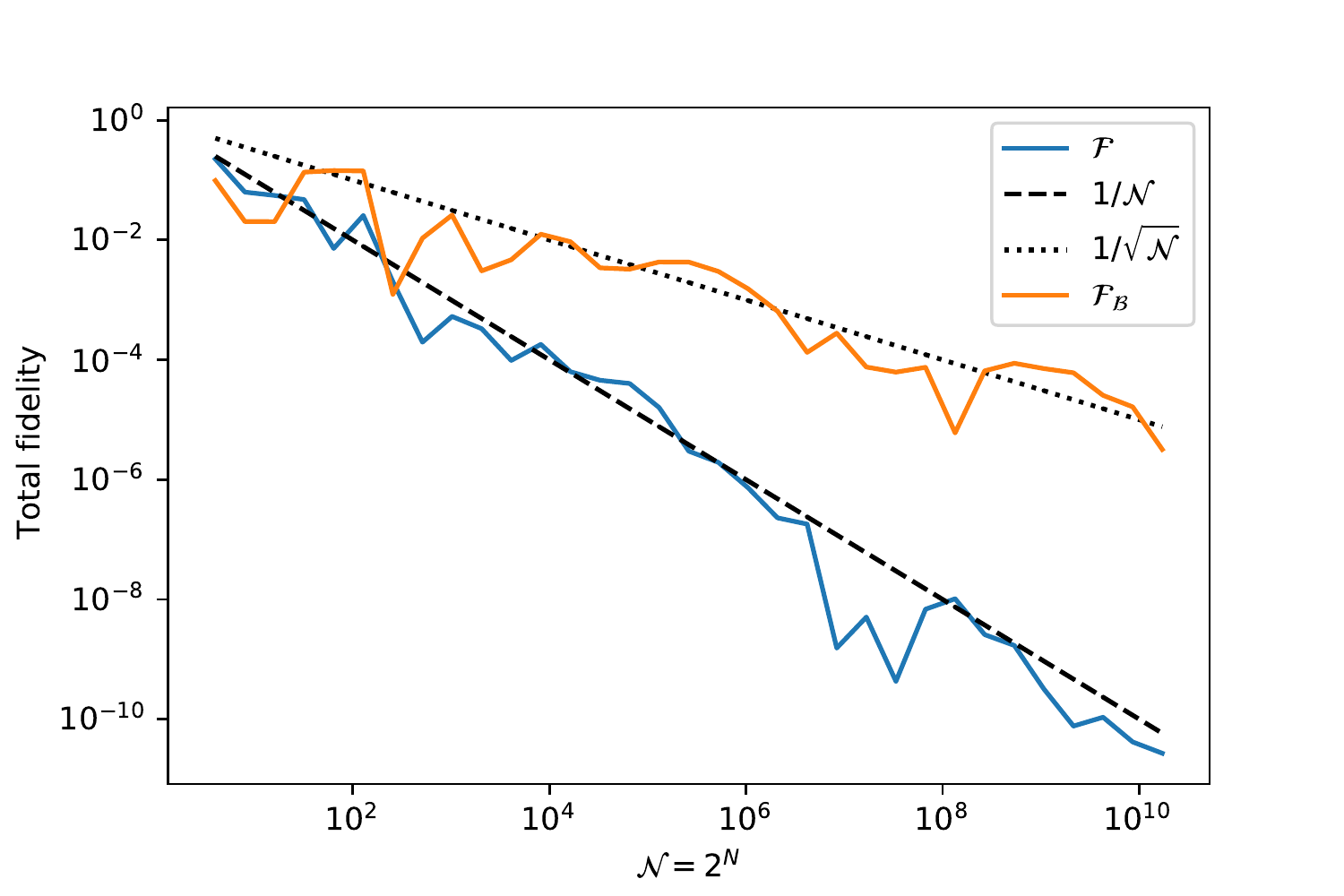}
    \caption{\emph{Various fidelities of random states.} Fidelity ${\cal F} = |\langle\Psi_P | \Psi_Q\rangle|^2$ and cross-entropy benchmarking ${\cal F}_{\cal B} = {\cal N} 
    \sum_x |\langle x|\Psi_P\rangle|^2 |\langle x|\Psi_Q\rangle|^2 - 1 $ for two independent random vectors $| \Psi_P\rangle$ and $| \Psi_Q\rangle$ distributed according to the Porter-Thomas distribution.}
    \label{fig:F_and_FB_chaotic}
\end{figure}

In this subsection, we calculate ${\cal F}$ and ${\cal F}_{\cal B}$ analytically in the $D\rightarrow\infty$ limit.
In this limit, the two states $|\Psi_P\rangle$
and $|\Psi_Q\rangle$ become essentially independent and distributed according to a Porter-Thomas distribution. 
Let us denote ${\cal N} =2^N$ and $U$ and $V$ two 
${\cal N} \times {\cal N}$ matrices distributed according to the Haar (uniform) measure of the $U({\cal N})$ group.
With these notations, the two states are, for very large depths, the first column of the matrices $U$ and $V$: $\langle x |\Psi_P\rangle \equiv U_{x1}$ and $\langle x |\Psi_Q\rangle \equiv V_{x1}$. We further denote $\langle X \rangle = \int X dU dV $ the average over these ensembles. We will make use of two integrals that can be found in \cite{Brouwer1996}: for any matrices $A,B, C$ and $D$, one has
\begin{equation}
\int dU \text{Tr} AUBU^\dagger = \frac{1}{\cal N} \text{Tr} A \ \text{Tr} B
\end{equation}
and
\begin{eqnarray}
\int dU \text{Tr} AUBU C U^\dagger D U^\dagger = \frac{1}{{\cal N}^2 -1} 
\bigg[ \text{Tr} A \ \text{Tr} BD \text{Tr} C  \nonumber \\
+ \left. \text{Tr} ADCB - \frac{1}{{\cal N}}\text{Tr} A \ \text{Tr} BDC  - \frac{1}{{\cal N}} \text{Tr} ADB \text{Tr} C \right]. \nonumber \\ 
\end{eqnarray}
Using the first of these two integrals, we obtain
\begin{equation}
\label{eq:randomF}
\langle {\cal F} \rangle = \int dU dV \sum_{x,x'} U_{x1} U_{1x'}^\dagger 
V_{x1} V_{1x'}^\dagger = \frac{1}{\cal N}
\end{equation}
and
\begin{equation}
\langle {\cal F}_{\cal B} \rangle ={\cal N} \int dU dV \sum_{x} U_{x1} U_{1x}^\dagger 
V_{x1} V_{1x}^\dagger  - 1 = 0.
\end{equation}
Since $\langle {\cal F}_{\cal B} \rangle = 0$ vanishes in average, we need to calculate its variance to estimate its typical value. We get
\begin{align}
\langle ({\cal F}_{\cal B})^2 \rangle & =  \nonumber \\
 & \!\!\!\!\!\!{\cal N}^2 \!\!\int\! dU dV \sum_{xx'}\! U_{x1}U_{x'1} U_{1x}^\dagger U_{1x'}^\dagger
V_{x1} V_{x'1} V_{1x}^\dagger V_{1x'}^\dagger \!-\! 1.
\end{align}
After some straightforward algebra, we get
\begin{equation}
\langle ({\cal F}_{\cal B})^2 \rangle = \frac{{\cal N} - 1}{({\cal N} +1)^2}.
\end{equation}
It follows that at very large depths, ${\cal F}_{\cal B} \approx 1/\sqrt{\cal N} \approx \sqrt{\cal F}$. Fig.~\ref{fig:F_and_FB_chaotic} shows a numerical calculation of ${\cal F}$ and ${\cal F}_{\cal B}$ for two independent Porter-Thomas vectors together with the analytical expressions derived above.

\section{Conclusions}
\label{sec:conclusions}

We have presented an algorithm to efficiently simulate quantum circuits with finite fidelity. This algorithm extends Ref.~\cite{Zhou2020}, where some of us adapted the time-evolving bond-decimation (TEBD) technique from many-body physics to the context of quantum circuits. Here we have introduced a generalization of the density-matrix renormalization group (DMRG) algorithm to quantum circuits. This new algorithm also has a simulation cost that scales polynomially with the number of qubits $N$ and the depth of the circuit $D$. In addition, it is more general and more efficient than the previous TEBD-like algorithm. From the simulation point of view, we emphasize that the main building block of our DMRG algorithm, the ``compression step", is completely general and could be used in other contexts. In particular, it may be combined with other tensor network techniques such as slicing or contraction heuristics
as in Refs.~\cite{Zhang2021simulating,Gray2021hyperoptimized}. These generalisations have not been attempted yet. 

We have benchmarked our algorithm on the supremacy sequence designed by Google and found that we can produce amplitudes (hence bitstrings) of quality as good as in the quantum supremacy experiment \cite{Arute2019}.
More importantly, for QAOA sequences, representative of useful applications of quantum computers, we obtain much better fidelities than the supremacy threshold set by Google. Our results provide strong evidence that quantum advantage (the ability for a quantum computer to perform a useful task better than a classical computer) might be much harder than reaching quantum supremacy (the ability for a quantum computer to perform a given, not necessarily useful, task that no classical computer can perform), despite what the words seem to indicate. In particular, since our algorithm scales polynomially with the number of qubits, an improvement in the experimental fidelity is needed in order for the experiments to reach better results than the DMRG algorithm for useful tasks.

Our work emphasizes the need for benchmarks of quantum computers that test the actual usefulness of quantum processors, rather than their ability to perform a relatively contrived task, in order to incentivize hardware and software efforts towards concrete applications. Such benchmarks should strive to be application-centric, hardware-agnostic and scalable. Some of us recently proposed a protocol fulfilling these criteria~\cite{Martiel2021}.

\section*{Acknowledgments} 

We thank Giuseppe Carleo, Lei Wang, and Pan Zhang for helpful feedback about prior work in simulating quantum circuits and for pointing out several errors in the literature discussion.
XW acknowledges funding from the French ANR QCONTROL as wel as the Plan France 2030 ANR-22-PETQ-0007.  TL, XW and TA acknowledge funding from ANR QPEG. 
XW acknowledges funding from the European Union’s Horizon 2020 research and innovation programme under grant agreement No. 862683 (UltraFastNano). EMS acknowledges helpful discussions with Soonwon Choi on the topic of cross-entropy benchmarking and fidelity.
The Flatiron Institute is a division of the Simons Foundation. The computations were performed on the Atos Quantum Learning Machine.

\appendix
\renewcommand\thefigure{S\arabic{figure}}
\setcounter{figure}{0}

\begin{figure*}
	\begin{subfigure}{5cm}
		\includegraphics[width=\textwidth]{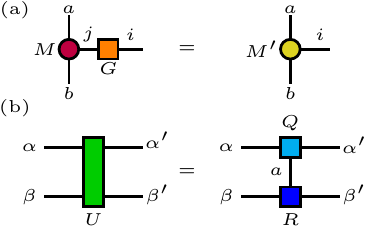}
	\end{subfigure}
	\hspace{0.5cm}
	\begin{subfigure}{8.5cm}
		\includegraphics[width=\textwidth]{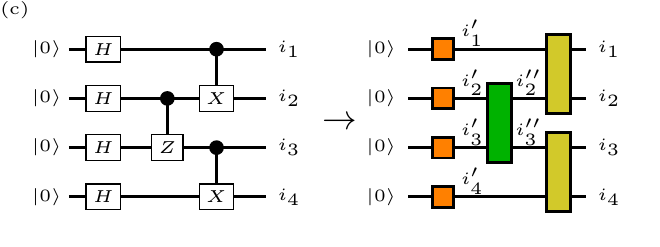}
	\end{subfigure}
	\caption{\textit{Tensor networks and quantum circuits.}  (a) Contraction of a three-index tensor $M$ with a one-qubit gate $G$. (b) QR decomposition of a two-qubit gate $U$. (c) From a quantum circuit to its tensor network representation.}
	\label{fig:app-circuit_and_tn}
\end{figure*}	
\begin{figure}
	\begin{subfigure}{7.6cm}
		\includegraphics[width=\textwidth]{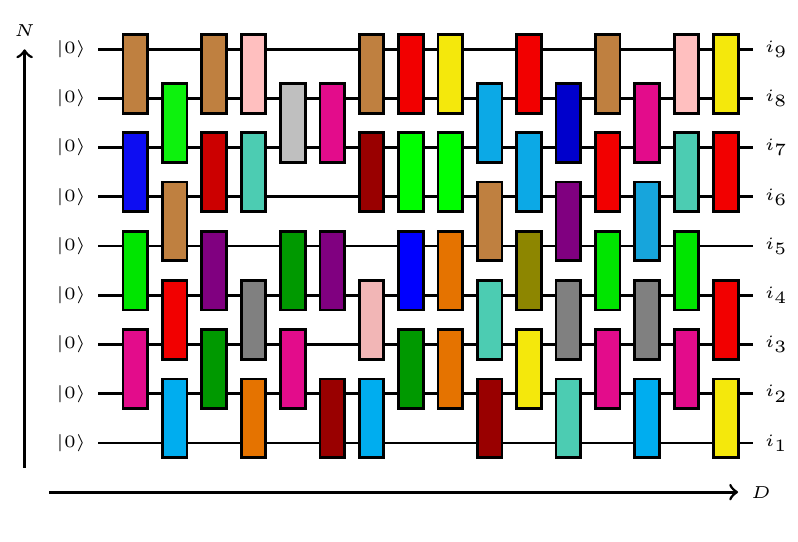}
		\caption{}
		\label{fig:app-circuit_plain}
	\end{subfigure}
	\begin{subfigure}{7.6cm}
		\includegraphics[width=\textwidth]{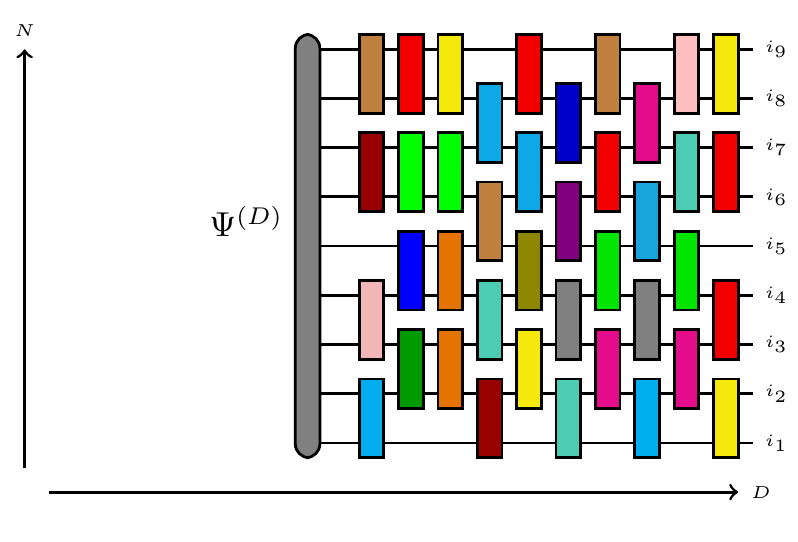}
		\caption{}
		\label{fig:app-circuit_schrodinger}
	\end{subfigure}
	\begin{subfigure}{7.6cm}
		\includegraphics[width=\textwidth]{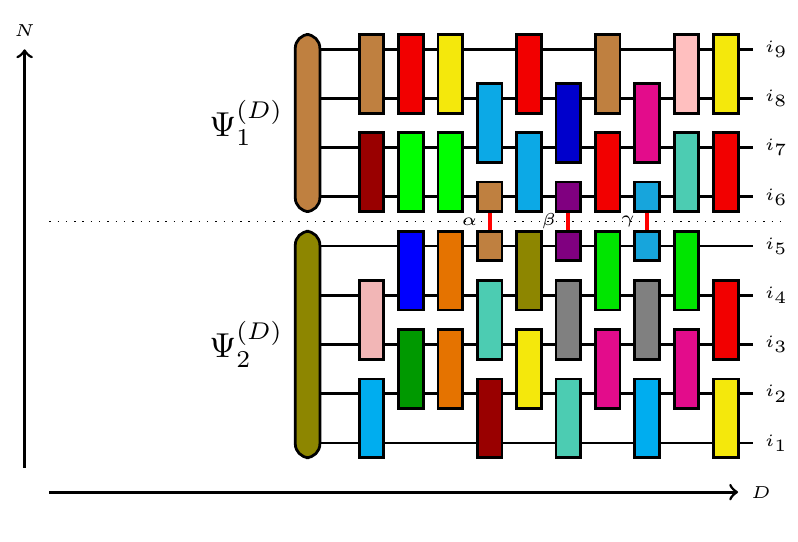}
		\caption{}
		\label{fig:app-circuit_schrodinger_feynman}
	\end{subfigure}
	\caption{\label{fig:sim} \emph{Schematic of two contraction strategies for simulating  a quantum circuit.} (a) Quantum circuit to contract. (b) Schr\"odinger-like simulation. (c) Schr\"odinger Feynman-like simulation.}
\end{figure}	

\section{A short introduction to tensor networks for simulating quantum circuits}
\label{app:tensor_networks}
In this appendix, we briefly review the main  aspects of tensor networks
in the context of quantum circuit simulation.

\subsection{Basic definitions and actions: contracting and splitting}
A tensor is simply an array of complex numbers that generalizes the concepts of vectors (1D array) and matrices (2D arrays) to an arbitray number of indices. A tensor $V_i$ with one index $i$ (that takes a finite number of values) is simply a vector; a tensor with two indices $A_{ij}$ is a matrix; a tensor $M_{ijk}$ ($P_{ijkl}$) is a 3D (4D) array of numbers. A tensor is represented graphically by a box (here a rectangle or a circle)
with as many legs (outgoing lines) as there are indices, see examples in Fig.~\ref{fig:app-circuit_and_tn}.

There are two basic operations that one can do with tensors: contracting and splitting.
Contractions of two tensors is the generalization of matrix-matrix multiplication. 
An example is shown in Fig.~\ref{fig:app-circuit_and_tn}b for the contraction of 
$M_{abj}$ with $G_{ij}$. The resulting tensor $M'_{abi}$ is simply given by
\begin{equation}
 M'_{a b i} = \sum_j G_{ij} M_{a b j}.
\end{equation}
i.e. one performs a summation over the index $j$ that links the two tensors.

The second operation, splitting of e.g. a tensor $U_{\alpha\beta\alpha'\beta'}$, is illustrated in Fig.~\ref{fig:app-circuit_and_tn}c. It consists of three steps. First,
one constructs two meta-indices $i$ and $j$ that group several indices together. For instance, one may choose $i = \alpha + N_\alpha \alpha'$ and $j = \beta + N_\beta \beta'$
where $N_\alpha$ and $N_\beta$ are the number of different values that the indices $\alpha$ and $\beta$ take, respectively. This allows us to define a one-to-one mapping between the tensor $U_{\alpha\beta\alpha'\beta'}$ and a {\it matrix} $\hat U$ defined as
\begin{equation}
\label{eq:mapping}
\hat U_{ij} \equiv U_{\alpha(i)\beta(j)\alpha'(i)\beta'(j)}.
\end{equation} 
Second, we may use any result known from linear algebra on the matrix $\hat U$,
for instance a QR decomposition, a SVD decomposition or any other decomposition.
Let us suppose we use a QR decomposition and write $\hat U = \hat Q \hat R$.
Third, we use the mapping Eq.~(\ref{eq:mapping}) backward to go back to the original indices and obtain
\begin{equation}
U_{\alpha\beta\alpha'\beta'}= \sum_{a} Q_{\alpha \alpha' a} R_{a \beta \beta'}, 
\end{equation}
i.e. we have split the tensor $U_{\alpha\beta\alpha'\beta'}$ in terms of the contraction
of two tensors $Q_{\alpha \alpha' a}$ and $R_{a \beta \beta'}$.

\subsection{Tensor networks for quantum circuits}
There is a natural correspondence between the usual representation for quantum circuits and tensor networks. The left-hand side of Fig.~\ref{fig:app-circuit_and_tn}c shows a small quantum circuit for four qubits that uses the standard Hadamard gate $H$, the control-NOT gate $C^X$ and the control-Z gate $C^Z$. The system wave-function $\Psi_{i_1i_2i_3i_4}$ is a tensor whose explicit form is given by
\begin{eqnarray}
\Psi_{i_1i_2i_3i_4} &= \sum_{i_1'i_2'i_3'i_4'i_2''i_3''} 
C^X_{i_1i_2i_1'i_2''} C^X_{i_3i_4i_3''i_4'} C^Z_{i_2''i_3''i_2'i_3'}  \nonumber \\
& H_{i_1'0} H_{i_2'0} H_{i_3'0} H_{i_4'0} \label{eq:psi_qn}
\end{eqnarray}

i.e. it corresponds to the contraction of the tensor network shown on the right-hand side of
 Fig.~\ref{fig:app-circuit_and_tn}c. The problem of computing the wave-function is reduced to the problem of performing the summation over the internal indices, i.e. the contraction of the tensor network \cite{Markov2008}. Finding the best order to perform the contraction is in general a difficult (NP hard) problem for which there nevertheless exists good heuristics.

\subsection{Schr\"odinger versus Schr\"odinger-Feynman-like simulations}
There are several possible different strategies to contract the tensor network associated with a quantum circuit. Fig.~\ref{fig:sim} shows two examples
for the Schr\"odinger and the Schr\"odinger-Feynman approaches discussed in 
section \ref{sec:review}. In the Schr\"odinger approach,  the contraction of the network is performed from left to right, as shown in Fig.~\ref{fig:app-circuit_schrodinger}.
In the Schr\"odinger-Feynman approach, shown in Fig.~\ref{fig:app-circuit_schrodinger_feynman}, one divides the qubits into two groups and splits the two-qubit gates that connect the two groups.
For a given value $\alpha, \beta, \gamma$ of the indices cut by the dotted line, 
one may propagate the two sub states $\Psi_1^{(\alpha, \beta, \gamma)}$ and $\Psi_2^{(\alpha, \beta, \gamma)}$ independently, see \eqref{eq:schrofeynman1} and \eqref{eq:schrofeynman2} from the main text. 
Thus instead of one complex simulation of the whole circuit, we perform $\chi_p$ easier simulations, where $\chi_p$ is the number of values taken by the extra bond indices.

\section{Derivation of the chaotic optimum error (Eq.~(\ref{eq:error_rate_asymptotics}))}
\label{app:chaotic_opt}

In this appendix, we prove that for a chaotic state $|\Psi\rangle$ distributed according to the Porter-Thomas distribution, the best possible fidelity that one may obtain by approximating it with a $m=2$ MPS is
\begin{equation}
\label{eq:chaotic_opt}
\mathcal{F}_{\rm opt} =\frac{4\chi}{2^{N/2}}.
\end{equation}

To establish this result, the wavefunction $\Psi_x$ is considered as a matrix $\Psi_{\alpha\beta}$ where index $\alpha$ labels half of the qubits and $\beta$ labels the other half. Performing a singular value decomposition of the $2^{N/2} \times 2^{N/2}$ matrix $\Psi_{\alpha\beta}$ to obtain its singular values $S_\mu$, the fidelity for a bond dimension $\chi$ is given by the 
 largest $\chi$ singular values
\begin{equation}
\mathcal{F}_{\rm opt} = \sum_{\mu=1}^\chi S_\mu^2.
\end{equation}
 The proof contains two parts:
\begin{itemize}
\item establish that the matrix $\Psi_{\alpha\beta}$ is, in the large ${\cal N} =2^N$ limit, a complex Gaussian random matrix;
\item use known results from random matrix theory to obtain the distribution of singular values from which one can obtain Eq.~(\ref{eq:chaotic_opt}) after a little algebra.

\end{itemize}
We perform these tasks below.

\subsection{Construction of a Porter-Thomas state from random Gaussian variables}
We want to establish that a Porter-Thomas state can be constructed from random Gaussian variables.
We recall that the sum $S$ of the squares of $k$ random normal variables
$X_{i}\sim\mathcal{N}\left(0,\sigma^{2}\right)$ follows a (generalized)
$\chi^{2}$ distribution with mean and variance
\begin{align}
\mathbb{E}(S) & =k/\sigma^{2},\label{eq:mean_chi_squared-2}\\
\mathrm{Var}(S) & =\frac{2k}{\sigma^{4}}.\label{eq:variance_chi_squared-2}
\end{align}
Its probability density function is
\begin{equation}
P(s)=\frac{1}{2^{k/2}\Gamma(k/2)\sigma^{2}}\left(\frac{s}{\sigma^{2}}\right)^{k/2-1}e^{-s/(2\sigma^{2})}.\label{eq:chi_squared_pdf-1}
\end{equation}

Let us construct a wavefunction $\Psi$ with $2^{N}$ complex amplitudes
\begin{equation}
\Psi_{x}=\psi_{x}'+i\psi_{x}''
\end{equation}
and choose the real and imaginary components to be normally distributed:
\begin{align}
\psi_{x}' & \sim\mathcal{N}\left(0,1/(2\cdot2^{N})\right),\\
\psi_{x}'' & \sim\mathcal{N}\left(0,1/(2\cdot2^{N})\right).
\end{align}

Let us first consider the probability
\begin{equation}
p_{x}=\left|\Psi_{x}\right|^{2}=\left(\psi_{x}'\right)^{2}+\left(\psi_{x}''\right)^{2}.
\end{equation}
This random variable is a sum of normal variables. Applying formula
(\ref{eq:chi_squared_pdf-1}) with $k=2$ and $\sigma^{2}=1/(2\cdot2^{N})$,
we find
\begin{equation}
P(p_{x})=2^{N}e^{-2^{N}p_{x}},
\end{equation}
i.e the Porter-Thomas distribution, as expected.

Let us now check that $\Psi$ is normalized in the large $N$ limit. Let us consider its norm
\begin{equation}
\text{\ensuremath{\left\Vert \Psi\right\Vert }}^{2}=\sum_{x}\left|\Psi_{x}\right|^{2}=\sum_{x}\left(\psi_{x}'\right)^{2}+\left(\psi_{x}''\right)^{2}.
\end{equation}
This random variable is also a sum of normal variables. We can apply
formulae (\ref{eq:mean_chi_squared-2}-\ref{eq:variance_chi_squared-2})
with $k=2\cdot2^{N}$ and $\sigma^{2}=1/(2\cdot2^{N})$, we find
\begin{align}
\mathbb{E}(\text{\ensuremath{\left\Vert \Psi\right\Vert }}^{2}) & =1,\label{eq:mean_chi_squared-1}\\
\mathrm{Var}(\text{\ensuremath{\left\Vert \Psi\right\Vert }}^{2}) & =\frac{1}{2^{N}}.\label{eq:variance_chi_squared-1}
\end{align}

We have constructed a wavefunction whose norm is 1 on average, with
a deviation to the average that vanishes exponentially fast as the
number of qubits $n$ increases.

As a result, the matrix $\Psi_{\alpha \beta}$ is a random complex Gaussian matrix, up to exponentially small corrections. Let us note that the Gaussian probability distribution that
we have used,
\begin{equation}
P(\Psi) \propto \exp\left(-2^N \sum_x |\Psi_x|^2\right)
\end{equation} 
obviously respects the Haar invariance 
\begin{equation}
P(\Psi)\prod_x d\Psi_x'\Psi_x'' = P(\bar \Psi)\prod_x d\bar\Psi_x'\bar\Psi_x''
\end{equation}
for any unitary rotation $\bar\Psi = U\Psi$ with $UU^\dagger=1$ and for all $N$. 
It is only the constraint $\Vert \Psi\Vert  = 1$ which is enforced only in average with an exponentially small variance.

\subsection{Scaling law for the singular values of a Porter-Thomas state}

\begin{figure}
	\begin{centering}
		\includegraphics[width=1.0\columnwidth]{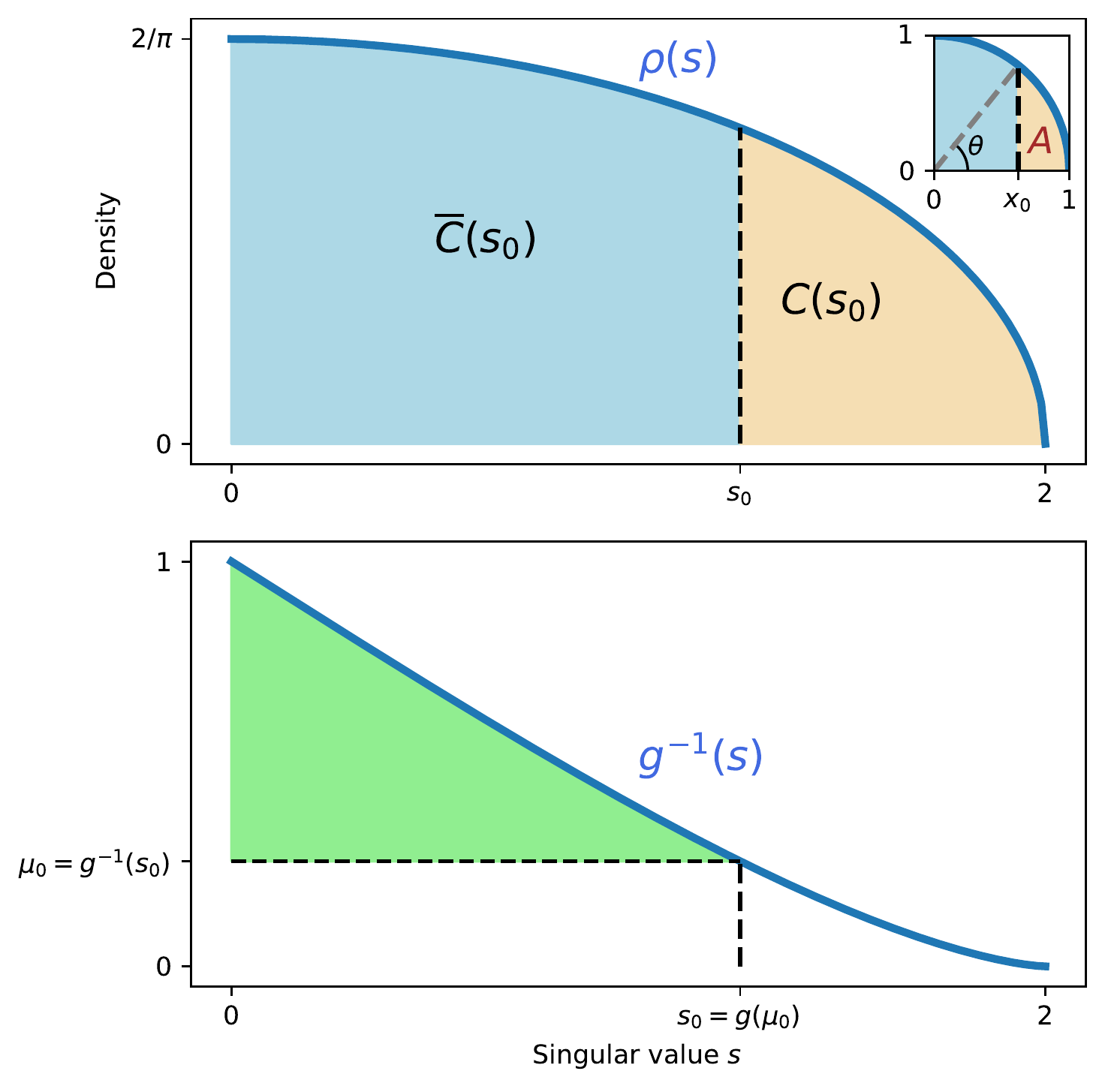}
		\par\end{centering}
	\caption{Quadrant law (upper panel) and dispersion of the singular values (lower panel).\label{fig:Definition-of-areas}}
\end{figure}
\begin{figure}
	\centering
	\includegraphics[width=\columnwidth]{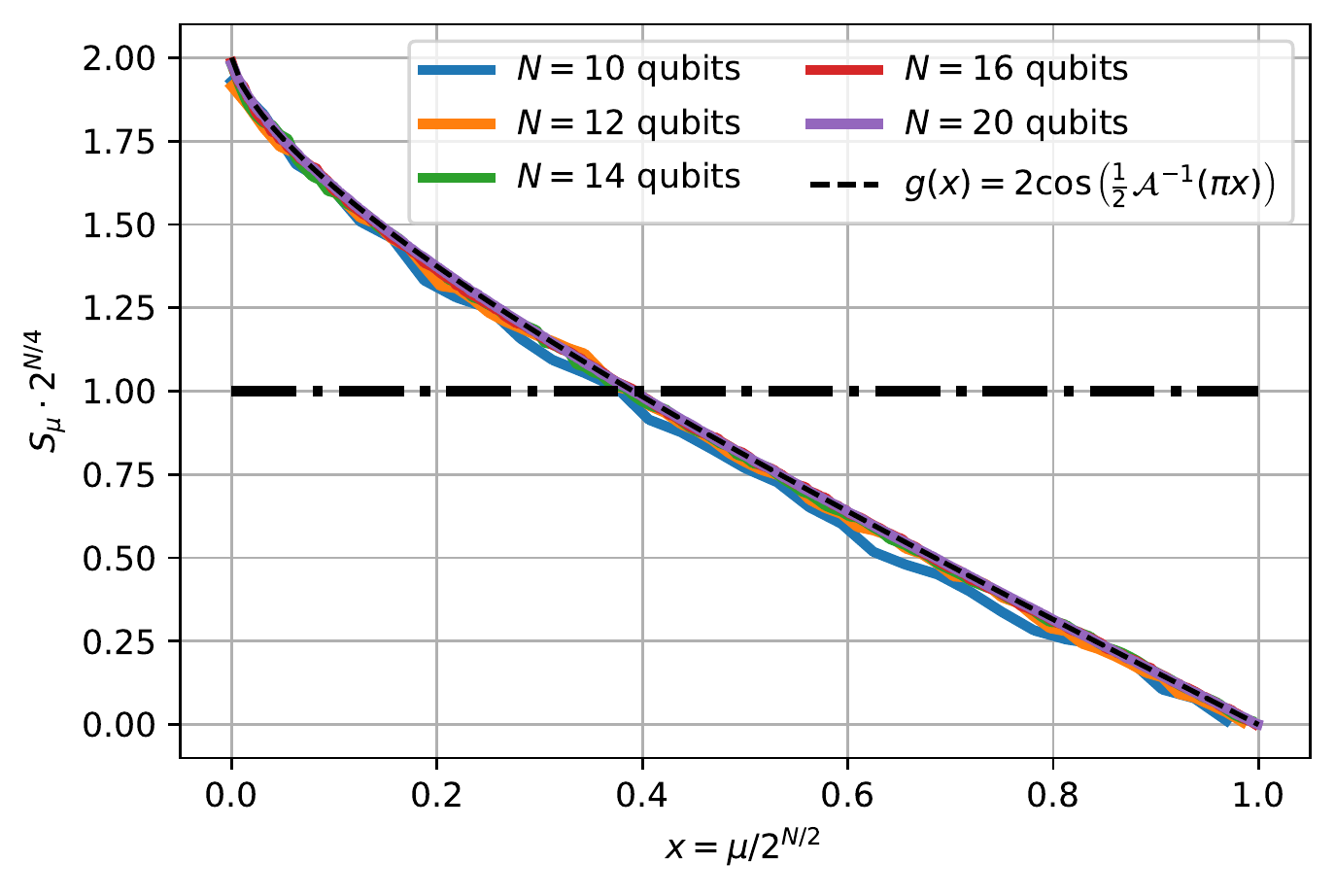}
	\caption{Scaling behavior of the Schmidt coefficients (singular values) $S_\mu$ of a Porter-Thomas wavevector for various numbers of qubits $N$. Black dashed line: scaling function $g(x)$. Black dash-dotted line: constant singular values case ($S_\mu = 1/2^{N/4}$).}
	\label{fig:figure_S5}
\end{figure}



In this subsection, our goal is to understand how the Schmidt coefficients $S_\mu$ of the Porter-Thomas state constructed in the previous subsection  decrease as a function of the index $\mu$: a fast decrease will be synonymous of a high MPS quality.

For this purpose, we make use of known results from random matrix theory.
More specifically, we use the fact that, in the large-$N$ limit, the 
average density $\rho(S) \equiv 1/2^{N/2} \sum_{\mu=1}^{2^{N/2}} \delta(S - S_\mu)$ of the singular values $S_\mu$ of a random
complex $2^{N/2} \times 2^{N/2}$ Gaussian matrix (with matrix elements $\Psi_{\alpha, \beta} \sim \mathcal{N}(0, \sigma^2) + j \mathcal{N}(0, \sigma^2)$, $\sigma^{2}=1/(2\cdot2^{N}$), as discussed in the previous subsection) follows
a quadrant law (see e.g \cite{Shen2001}):
\begin{equation}
\lim_{N \rightarrow \infty} 1/ 2^{N/2} \rho(s/2^{N/2})=\frac{1}{\pi} \sqrt{4 - s^2},~s \in [0,2].
\end{equation}

In the large-$N$ limit, the number $\mu$ of singular values above a given threshold $S_0$ is
given by
\begin{equation}
\mu(S_0) = 2^{N/2} \int_{S_0}^{2 \times 2^{-N/4}} \rho(S) dS.
\end{equation}
Inverting the above function $\mu(S_0)$ provides the sought-after scaling of the
singular values, $S_\mu$. Introducing the rescaled singular value $s = 2^{N/4} S$, we get
\begin{equation}
\mu(s_0) = 2^{N/2} C(s_0)
\end{equation}
with
\begin{equation}
C(s_0) = \frac{1}{\pi}\int_{s_0}^{2} ds \sqrt{4-s^{2}}.
\end{equation}
It follows that $S_\mu$ follows a scaling law
\begin{equation}
S_{\mu}=2^{-N/4} g\left(\mu/2^{N/2}\right),\label{eq:svd_collapse}
\end{equation}
where the function $g(x)= C^{-1}(x)$ is the inverse of $C(s_0)$, 
as illustrated in the lower panel of Fig.~\ref{fig:Definition-of-areas}.
The function $C(s_{0})$ corresponds to the area of a portion of a (distorted) circle and can be computed using a simple geometrical argument. Introducing the angle $\theta$ (see inset of Fig.~\ref{fig:Definition-of-areas}), one obtains
\begin{equation}
C(s_{0})=\frac{1}{\pi}\mathcal{A}\left(2\mathrm{arccos}\left(\frac{s_{0}}{2}\right)\right)\label{eq:C_expr2}
\end{equation}
with $\mathcal{A}(\theta)=\theta-\sin\left(\theta\right)$.
We therefore obtain
\begin{equation}
g(x)=2\cos\left(\frac{1}{2}\mathcal{A}^{-1}\left(\pi x\right)\right).
\end{equation}
In particular, one has $g(0)=2$ and $g(1)=0$.

This scaling law can be used to derive
the behavior of the error rate in the chaotic limit, Eq. (\ref{eq:error_rate_asymptotics}).
Truncating the original wave vector $|\Psi\rangle$ (Eq. (\ref{eq:Psi_schmidt}))
to its first $\chi$ eigenvalues ($|\tilde{\Psi}\rangle$) yields
the fidelity 
$\mathcal{F}(\chi)=|\langle\tilde{\Psi}|\Psi\rangle|^{2}=\sum_{\mu=0}^{\chi-1}S_{\mu}^{2}$.
Thus, using Eq.~(\ref{eq:svd_collapse}), 
\begin{equation}
\mathcal{F}(\chi)=\int_{0}^{\chi/2^{N/2}}g^{2}\left(x\right)dx,\label{eq:F_chi_rel_int}
\end{equation}
and in the $\chi\ll2^{N/2}$ limit we obtain the advertised chaotic limit:
\begin{equation}
\mathcal{F}(\chi)  =\frac{4\chi}{2^{N/2}}+O\left(\frac{\chi}{2^{N/2}}\right)^{2}\label{eq:F_chi_rel}
\end{equation}
and the associated value for the error rate $\epsilon = 1 - f = 1 - \mathcal{F}^{2/(ND)}$.
Interestingly, $\epsilon$ plateaus at $\log(2)/D$ for large $N$.

We note that if all the singular values were equal (\mbox{$S_\mu = 1/2^{N/4}$}), then we would have $g(x)=1$ and hence \mbox{$\mathcal{F} = \chi / 2^{N/2}$} (following Eq.~\eqref{eq:F_chi_rel_int}), namely one fourth of the leading term of Eq.~\eqref{eq:F_chi_rel}. In other terms, the fidelity in the chaotic limit is only $4$ times larger than in the worst possible situation where all the singular values are of equal importance.

We check in Fig. \ref{fig:figure_S5} that this scaling law of the singular values can indeed be observed. We perform a SVD decomposition of the random vector obtained by the procedure described in the previous subsection, and plot the corresponding function $S_\mu$ properly rescaled. The result is almost indistinguishable from the analytical function $g(x)$ calculated above.



\section{Metropolis sampling for closed simulations}
\label{sec:metropolis}

To produce the same output as a quantum computer within the framework of closed simulations, one must be able to sample from the distribution $Q(x) = |\Psi_x|^2$. This means not only computing amplitudes for a given bitstring but producing bitstrings distributed following $Q(x)$. A simple general algorithm for this task has been proposed in \cite{Bravyi2021} (see also the discussion in section \ref{sec:review}). The algorithm of \cite{Bravyi2021} requires the calculations of $\propto N_{\rm 2g}$ amplitudes per bitstring. In the case of random circuits such as sequence I and II, this is far from optimum. A possible strategy is to use the Metropolis algorithm to construct a Markov chain of bitstrings $x_t$: one picks $x_{t+1}$ at random
and accepts the proposed value with the probability 
$p_\mathrm{acc}= \min(|\Psi_{x_{t+1}}/\Psi_{x_{t}}|^2,1)$ (acceptance ratio).
If the proposed move is refused then $x_{t+1}\equiv x_{t}$. In the random circuit of quantum supremacy, one quickly reaches a Porter-Thomas distribution, i.e. the amplitudes $\Psi_x$ are themselves distributed according to an exponential law 
$P(|\Psi_x|^2=p)= 2^N\exp (-2^N p)$.
It follows that the average acceptance ratio is fairly large:
\begin{equation}
\langle p_\mathrm{acc} \rangle  = \int_0^\infty dx \int_0^\infty dy \  \min( x/y,1)e^{-x} e^{-y} \approx 70\%
\end{equation}
i.e. in average $1/0.7\approx 1.5$ bitstrings must be calculated in order to get a new accepted bitstring.

If one wanted to pretend that the bitstrings came from an actual quantum computer, one would want to be almost certain that a given bitstring could not appear twice consecutively. In that case, one would want to keep only one bitstring every $L$ Metropolis update, hence lowering the probability of repetition to about $ (1-0.7)^L$. For $L=10$, this would give a very low repetition probability of $6\cdot 10^{-4} \%$.
However, this precaution can probably be skipped as the cross-entropy benchmarking fidelity can easily be spoofed: simply ignoring the repeated values would create a bias in the distribution that would probably be very hard if not impossible to detect. Hence one or two amplitudes per bitstring should be enough in practice.

\bibliography{bibliography}

\begin{thebibliography}{40}%
\makeatletter
\providecommand \@ifxundefined [1]{%
 \@ifx{#1\undefined}
}%
\providecommand \@ifnum [1]{%
 \ifnum #1\expandafter \@firstoftwo
 \else \expandafter \@secondoftwo
 \fi
}%
\providecommand \@ifx [1]{%
 \ifx #1\expandafter \@firstoftwo
 \else \expandafter \@secondoftwo
 \fi
}%
\providecommand \natexlab [1]{#1}%
\providecommand \enquote  [1]{``#1''}%
\providecommand \bibnamefont  [1]{#1}%
\providecommand \bibfnamefont [1]{#1}%
\providecommand \citenamefont [1]{#1}%
\providecommand \href@noop [0]{\@secondoftwo}%
\providecommand \href [0]{\begingroup \@sanitize@url \@href}%
\providecommand \@href[1]{\@@startlink{#1}\@@href}%
\providecommand \@@href[1]{\endgroup#1\@@endlink}%
\providecommand \@sanitize@url [0]{\catcode `\\12\catcode `\$12\catcode
  `\&12\catcode `\#12\catcode `\^12\catcode `\_12\catcode `\%12\relax}%
\providecommand \@@startlink[1]{}%
\providecommand \@@endlink[0]{}%
\providecommand \url  [0]{\begingroup\@sanitize@url \@url }%
\providecommand \@url [1]{\endgroup\@href {#1}{\urlprefix }}%
\providecommand \urlprefix  [0]{URL }%
\providecommand \Eprint [0]{\href }%
\providecommand \doibase [0]{http://dx.doi.org/}%
\providecommand \selectlanguage [0]{\@gobble}%
\providecommand \bibinfo  [0]{\@secondoftwo}%
\providecommand \bibfield  [0]{\@secondoftwo}%
\providecommand \translation [1]{[#1]}%
\providecommand \BibitemOpen [0]{}%
\providecommand \bibitemStop [0]{}%
\providecommand \bibitemNoStop [0]{.\EOS\space}%
\providecommand \EOS [0]{\spacefactor3000\relax}%
\providecommand \BibitemShut  [1]{\csname bibitem#1\endcsname}%
\let\auto@bib@innerbib\@empty
\bibitem [{\citenamefont {Arute}\ \emph {et~al.}(2019)\citenamefont {Arute},
  \citenamefont {Arya}, \citenamefont {Babbush}, \citenamefont {Bacon},
  \citenamefont {Bardin}, \citenamefont {Barends}, \citenamefont {Biswas},
  \citenamefont {Boixo}, \citenamefont {Brandao}, \citenamefont {Buell},\ and\
  \citenamefont {et~al.}}]{Arute2019}%
  \BibitemOpen
  \bibfield  {author} {\bibinfo {author} {\bibfnamefont {Frank}\ \bibnamefont
  {Arute}}, \bibinfo {author} {\bibfnamefont {Kunal}\ \bibnamefont {Arya}},
  \bibinfo {author} {\bibfnamefont {Ryan}\ \bibnamefont {Babbush}}, \bibinfo
  {author} {\bibfnamefont {Dave}\ \bibnamefont {Bacon}}, \bibinfo {author}
  {\bibfnamefont {Joseph~C.}\ \bibnamefont {Bardin}}, \bibinfo {author}
  {\bibfnamefont {Rami}\ \bibnamefont {Barends}}, \bibinfo {author}
  {\bibfnamefont {Rupak}\ \bibnamefont {Biswas}}, \bibinfo {author}
  {\bibfnamefont {Sergio}\ \bibnamefont {Boixo}}, \bibinfo {author}
  {\bibfnamefont {Fernando G. S.~L.}\ \bibnamefont {Brandao}}, \bibinfo
  {author} {\bibfnamefont {David~A.}\ \bibnamefont {Buell}}, \ and\ \bibinfo
  {author} {\bibnamefont {et~al.}},\ }\bibfield  {title} {\enquote {\bibinfo
  {title} {Quantum supremacy using a programmable superconducting processor},}\
  }\href {\doibase 10.1038/s41586-019-1666-5} {\bibfield  {journal} {\bibinfo
  {journal} {Nature}\ }\textbf {\bibinfo {volume} {574}},\ \bibinfo {pages}
  {505–510} (\bibinfo {year} {2019})}\BibitemShut {NoStop}%
\bibitem [{\citenamefont {Bauer}\ \emph {et~al.}(2020)\citenamefont {Bauer},
  \citenamefont {Bravyi}, \citenamefont {Motta},\ and\ \citenamefont {{Kin-Lic
  Chan}}}]{Bauer2020}%
  \BibitemOpen
  \bibfield  {author} {\bibinfo {author} {\bibfnamefont {Bela}\ \bibnamefont
  {Bauer}}, \bibinfo {author} {\bibfnamefont {Sergey}\ \bibnamefont {Bravyi}},
  \bibinfo {author} {\bibfnamefont {Mario}\ \bibnamefont {Motta}}, \ and\
  \bibinfo {author} {\bibfnamefont {Garnet}\ \bibnamefont {{Kin-Lic Chan}}},\
  }\bibfield  {title} {\enquote {\bibinfo {title} {{Quantum Algorithms for
  Quantum Chemistry and Quantum Materials Science}},}\ }\href {\doibase
  10.1021/acs.chemrev.9b00829} {\bibfield  {journal} {\bibinfo  {journal}
  {Chemical Reviews}\ }\textbf {\bibinfo {volume} {120}},\ \bibinfo {pages}
  {12685--12717} (\bibinfo {year} {2020})},\ \Eprint
  {http://arxiv.org/abs/2001.03685} {arXiv:2001.03685} \BibitemShut {NoStop}%
\bibitem [{\citenamefont {Reiher}\ \emph {et~al.}(2017)\citenamefont {Reiher},
  \citenamefont {Wiebe}, \citenamefont {Svore}, \citenamefont {Wecker},\ and\
  \citenamefont {Troyer}}]{Reiher2016}%
  \BibitemOpen
  \bibfield  {author} {\bibinfo {author} {\bibfnamefont {Markus}\ \bibnamefont
  {Reiher}}, \bibinfo {author} {\bibfnamefont {Nathan}\ \bibnamefont {Wiebe}},
  \bibinfo {author} {\bibfnamefont {Krysta~M}\ \bibnamefont {Svore}}, \bibinfo
  {author} {\bibfnamefont {Dave}\ \bibnamefont {Wecker}}, \ and\ \bibinfo
  {author} {\bibfnamefont {Matthias}\ \bibnamefont {Troyer}},\ }\bibfield
  {title} {\enquote {\bibinfo {title} {{Elucidating reaction mechanisms on
  quantum computers}},}\ }\href {\doibase 10.1073/pnas.1619152114} {\bibfield
  {journal} {\bibinfo  {journal} {Proceedings of the National Academy of
  Sciences}\ }\textbf {\bibinfo {volume} {114}},\ \bibinfo {pages} {7555--7560}
  (\bibinfo {year} {2017})},\ \Eprint {http://arxiv.org/abs/1605.03590}
  {arXiv:1605.03590} \BibitemShut {NoStop}%
\bibitem [{\citenamefont {Gray}\ and\ \citenamefont
  {Kourtis}(2021)}]{Gray2021hyperoptimized}%
  \BibitemOpen
  \bibfield  {author} {\bibinfo {author} {\bibfnamefont {Johnnie}\ \bibnamefont
  {Gray}}\ and\ \bibinfo {author} {\bibfnamefont {Stefanos}\ \bibnamefont
  {Kourtis}},\ }\bibfield  {title} {\enquote {\bibinfo {title} {Hyper-optimized
  tensor network contraction},}\ }\href {\doibase 10.22331/q-2021-03-15-410}
  {\bibfield  {journal} {\bibinfo  {journal} {{Quantum}}\ }\textbf {\bibinfo
  {volume} {5}},\ \bibinfo {pages} {410} (\bibinfo {year} {2021})}\BibitemShut
  {NoStop}%
\bibitem [{\citenamefont {Pan}\ and\ \citenamefont
  {Zhang}(2021)}]{Zhang2021simulating}%
  \BibitemOpen
  \bibfield  {author} {\bibinfo {author} {\bibfnamefont {Feng}\ \bibnamefont
  {Pan}}\ and\ \bibinfo {author} {\bibfnamefont {Pan}\ \bibnamefont {Zhang}},\
  }\href@noop {} {\enquote {\bibinfo {title} {Simulating the sycamore quantum
  supremacy circuits},}\ } (\bibinfo {year} {2021}),\ \Eprint
  {http://arxiv.org/abs/2103.03074} {arXiv:2103.03074 [quant-ph]} \BibitemShut
  {NoStop}%
\bibitem [{\citenamefont {Pan}\ \emph {et~al.}(2022)\citenamefont {Pan},
  \citenamefont {Chen},\ and\ \citenamefont {Zhang}}]{Pan2021solving}%
  \BibitemOpen
  \bibfield  {author} {\bibinfo {author} {\bibfnamefont {Feng}\ \bibnamefont
  {Pan}}, \bibinfo {author} {\bibfnamefont {Keyang}\ \bibnamefont {Chen}}, \
  and\ \bibinfo {author} {\bibfnamefont {Pan}\ \bibnamefont {Zhang}},\
  }\bibfield  {title} {\enquote {\bibinfo {title} {Solving the sampling problem
  of the sycamore quantum circuits},}\ }\href {\doibase
  10.1103/PhysRevLett.129.090502} {\bibfield  {journal} {\bibinfo  {journal}
  {Phys. Rev. Lett.}\ }\textbf {\bibinfo {volume} {129}},\ \bibinfo {pages}
  {090502} (\bibinfo {year} {2022})}\BibitemShut {NoStop}%
\bibitem [{\citenamefont {Zhou}\ \emph {et~al.}(2020)\citenamefont {Zhou},
  \citenamefont {Stoudenmire},\ and\ \citenamefont {Waintal}}]{Zhou2020}%
  \BibitemOpen
  \bibfield  {author} {\bibinfo {author} {\bibfnamefont {Yiqing}\ \bibnamefont
  {Zhou}}, \bibinfo {author} {\bibfnamefont {E.~Miles}\ \bibnamefont
  {Stoudenmire}}, \ and\ \bibinfo {author} {\bibfnamefont {Xavier}\
  \bibnamefont {Waintal}},\ }\bibfield  {title} {\enquote {\bibinfo {title}
  {What limits the simulation of quantum computers?}}\ }\href {\doibase
  10.1103/PhysRevX.10.041038} {\bibfield  {journal} {\bibinfo  {journal} {Phys.
  Rev. X}\ }\textbf {\bibinfo {volume} {10}},\ \bibinfo {pages} {041038}
  (\bibinfo {year} {2020})}\BibitemShut {NoStop}%
\bibitem [{\citenamefont {Daley}\ \emph {et~al.}(2004)\citenamefont {Daley},
  \citenamefont {Kollath}, \citenamefont {Schollwöck},\ and\ \citenamefont
  {Vidal}}]{Daley_2004}%
  \BibitemOpen
  \bibfield  {author} {\bibinfo {author} {\bibfnamefont {A~J}\ \bibnamefont
  {Daley}}, \bibinfo {author} {\bibfnamefont {C}~\bibnamefont {Kollath}},
  \bibinfo {author} {\bibfnamefont {U}~\bibnamefont {Schollwöck}}, \ and\
  \bibinfo {author} {\bibfnamefont {G}~\bibnamefont {Vidal}},\ }\bibfield
  {title} {\enquote {\bibinfo {title} {Time-dependent density-matrix
  renormalization-group using adaptive effective hilbert spaces},}\ }\href
  {\doibase 10.1088/1742-5468/2004/04/p04005} {\bibfield  {journal} {\bibinfo
  {journal} {Journal of Statistical Mechanics: Theory and Experiment}\ }\textbf
  {\bibinfo {volume} {2004}},\ \bibinfo {pages} {P04005} (\bibinfo {year}
  {2004})}\BibitemShut {NoStop}%
\bibitem [{\citenamefont {Vidal}(2004)}]{Vidal2004efficient}%
  \BibitemOpen
  \bibfield  {author} {\bibinfo {author} {\bibfnamefont {Guifr\'e}\
  \bibnamefont {Vidal}},\ }\bibfield  {title} {\enquote {\bibinfo {title}
  {Efficient simulation of one-dimensional quantum many-body systems},}\ }\href
  {\doibase 10.1103/PhysRevLett.93.040502} {\bibfield  {journal} {\bibinfo
  {journal} {Phys. Rev. Lett.}\ }\textbf {\bibinfo {volume} {93}},\ \bibinfo
  {pages} {040502} (\bibinfo {year} {2004})}\BibitemShut {NoStop}%
\bibitem [{\citenamefont {White}\ and\ \citenamefont
  {Feiguin}(2004)}]{White2004realtime}%
  \BibitemOpen
  \bibfield  {author} {\bibinfo {author} {\bibfnamefont {Steven~R.}\
  \bibnamefont {White}}\ and\ \bibinfo {author} {\bibfnamefont {Adrian~E.}\
  \bibnamefont {Feiguin}},\ }\bibfield  {title} {\enquote {\bibinfo {title}
  {Real-time evolution using the density matrix renormalization group},}\
  }\href {\doibase 10.1103/PhysRevLett.93.076401} {\bibfield  {journal}
  {\bibinfo  {journal} {Phys. Rev. Lett.}\ }\textbf {\bibinfo {volume} {93}},\
  \bibinfo {pages} {076401} (\bibinfo {year} {2004})}\BibitemShut {NoStop}%
\bibitem [{\citenamefont {White}(1993)}]{White:1993}%
  \BibitemOpen
  \bibfield  {author} {\bibinfo {author} {\bibfnamefont {Steven~R.}\
  \bibnamefont {White}},\ }\bibfield  {title} {\enquote {\bibinfo {title}
  {Density-matrix algorithms for quantum renormalization groups},}\ }\href
  {\doibase 10.1103/PhysRevB.48.10345} {\bibfield  {journal} {\bibinfo
  {journal} {Phys. Rev. B}\ }\textbf {\bibinfo {volume} {48}},\ \bibinfo
  {pages} {10345--10356} (\bibinfo {year} {1993})}\BibitemShut {NoStop}%
\bibitem [{\citenamefont {Schollwöck}(2011)}]{Schollwoeck2011}%
  \BibitemOpen
  \bibfield  {author} {\bibinfo {author} {\bibfnamefont {U.}~\bibnamefont
  {Schollwöck}},\ }\bibfield  {title} {\enquote {\bibinfo {title} {The
  density-matrix renormalization group in the age of matrix product states},}\
  }\href {\doibase https://doi.org/10.1016/j.aop.2010.09.012} {\bibfield
  {journal} {\bibinfo  {journal} {Annals of Physics}\ }\textbf {\bibinfo
  {volume} {326}},\ \bibinfo {pages} {96--192} (\bibinfo {year}
  {2011})}\BibitemShut {NoStop}%
\bibitem [{\citenamefont {Edward}\ \emph {et~al.}(2014)\citenamefont {Edward},
  \citenamefont {Jeffrey},\ and\ \citenamefont {Sam}}]{Farhi2014}%
  \BibitemOpen
  \bibfield  {author} {\bibinfo {author} {\bibfnamefont {Farhi}\ \bibnamefont
  {Edward}}, \bibinfo {author} {\bibfnamefont {Goldstone}\ \bibnamefont
  {Jeffrey}}, \ and\ \bibinfo {author} {\bibfnamefont {Gutmann}\ \bibnamefont
  {Sam}},\ }\bibfield  {title} {\enquote {\bibinfo {title} {{A Quantum
  Approximate Optimization Algorithm}},}\ }\href
  {https://arxiv.org/abs/1411.4028} {\  (\bibinfo {year} {2014})},\ \Eprint
  {http://arxiv.org/abs/1411.4028} {arXiv:1411.4028} \BibitemShut {NoStop}%
\bibitem [{\citenamefont {Wu}\ \emph {et~al.}(2021)\citenamefont {Wu},
  \citenamefont {Bao}, \citenamefont {Cao}, \citenamefont {Chen}, \citenamefont
  {Chen}, \citenamefont {Chen}, \citenamefont {Chung}, \citenamefont {Deng},
  \citenamefont {Du}, \citenamefont {Fan}, \citenamefont {Gong}, \citenamefont
  {Guo}, \citenamefont {Guo}, \citenamefont {Guo}, \citenamefont {Han},
  \citenamefont {Hong}, \citenamefont {Huang}, \citenamefont {Huo},
  \citenamefont {Li}, \citenamefont {Li}, \citenamefont {Li}, \citenamefont
  {Li}, \citenamefont {Liang}, \citenamefont {Lin}, \citenamefont {Lin},
  \citenamefont {Qian}, \citenamefont {Qiao}, \citenamefont {Rong},
  \citenamefont {Su}, \citenamefont {Sun}, \citenamefont {Wang}, \citenamefont
  {Wang}, \citenamefont {Wu}, \citenamefont {Xu}, \citenamefont {Yan},
  \citenamefont {Yang}, \citenamefont {Yang}, \citenamefont {Ye}, \citenamefont
  {Yin}, \citenamefont {Ying}, \citenamefont {Yu}, \citenamefont {Zha},
  \citenamefont {Zhang}, \citenamefont {Zhang}, \citenamefont {Zhang},
  \citenamefont {Zhang}, \citenamefont {Zhao}, \citenamefont {Zhao},
  \citenamefont {Zhou}, \citenamefont {Zhu}, \citenamefont {Lu}, \citenamefont
  {Peng}, \citenamefont {Zhu},\ and\ \citenamefont {Pan}}]{Wu2021}%
  \BibitemOpen
  \bibfield  {author} {\bibinfo {author} {\bibfnamefont {Yulin}\ \bibnamefont
  {Wu}}, \bibinfo {author} {\bibfnamefont {Wan-su}\ \bibnamefont {Bao}},
  \bibinfo {author} {\bibfnamefont {Sirui}\ \bibnamefont {Cao}}, \bibinfo
  {author} {\bibfnamefont {Fusheng}\ \bibnamefont {Chen}}, \bibinfo {author}
  {\bibfnamefont {Ming-Cheng}\ \bibnamefont {Chen}}, \bibinfo {author}
  {\bibfnamefont {Xiawei}\ \bibnamefont {Chen}}, \bibinfo {author}
  {\bibfnamefont {Tung-Hsun}\ \bibnamefont {Chung}}, \bibinfo {author}
  {\bibfnamefont {Hui}\ \bibnamefont {Deng}}, \bibinfo {author} {\bibfnamefont
  {Yajie}\ \bibnamefont {Du}}, \bibinfo {author} {\bibfnamefont {Daojin}\
  \bibnamefont {Fan}}, \bibinfo {author} {\bibfnamefont {Ming}\ \bibnamefont
  {Gong}}, \bibinfo {author} {\bibfnamefont {Cheng}\ \bibnamefont {Guo}},
  \bibinfo {author} {\bibfnamefont {Chu}\ \bibnamefont {Guo}}, \bibinfo
  {author} {\bibfnamefont {Shaojun}\ \bibnamefont {Guo}}, \bibinfo {author}
  {\bibfnamefont {Lianchen}\ \bibnamefont {Han}}, \bibinfo {author}
  {\bibfnamefont {Linyin}\ \bibnamefont {Hong}}, \bibinfo {author}
  {\bibfnamefont {He-liang}\ \bibnamefont {Huang}}, \bibinfo {author}
  {\bibfnamefont {Yong-heng}\ \bibnamefont {Huo}}, \bibinfo {author}
  {\bibfnamefont {Liping}\ \bibnamefont {Li}}, \bibinfo {author} {\bibfnamefont
  {Na}~\bibnamefont {Li}}, \bibinfo {author} {\bibfnamefont {Shaowei}\
  \bibnamefont {Li}}, \bibinfo {author} {\bibfnamefont {Yuan}\ \bibnamefont
  {Li}}, \bibinfo {author} {\bibfnamefont {Futian}\ \bibnamefont {Liang}},
  \bibinfo {author} {\bibfnamefont {Chun}\ \bibnamefont {Lin}}, \bibinfo
  {author} {\bibfnamefont {Jin}\ \bibnamefont {Lin}}, \bibinfo {author}
  {\bibfnamefont {Haoran}\ \bibnamefont {Qian}}, \bibinfo {author}
  {\bibfnamefont {Dan}\ \bibnamefont {Qiao}}, \bibinfo {author} {\bibfnamefont
  {Hao}\ \bibnamefont {Rong}}, \bibinfo {author} {\bibfnamefont {Hong}\
  \bibnamefont {Su}}, \bibinfo {author} {\bibfnamefont {Lihua}\ \bibnamefont
  {Sun}}, \bibinfo {author} {\bibfnamefont {Liangyuan}\ \bibnamefont {Wang}},
  \bibinfo {author} {\bibfnamefont {Shiyu}\ \bibnamefont {Wang}}, \bibinfo
  {author} {\bibfnamefont {Dachao}\ \bibnamefont {Wu}}, \bibinfo {author}
  {\bibfnamefont {Yu}~\bibnamefont {Xu}}, \bibinfo {author} {\bibfnamefont
  {Kai}\ \bibnamefont {Yan}}, \bibinfo {author} {\bibfnamefont {Weifeng}\
  \bibnamefont {Yang}}, \bibinfo {author} {\bibfnamefont {Yang}\ \bibnamefont
  {Yang}}, \bibinfo {author} {\bibfnamefont {Yangsen}\ \bibnamefont {Ye}},
  \bibinfo {author} {\bibfnamefont {Jianghan}\ \bibnamefont {Yin}}, \bibinfo
  {author} {\bibfnamefont {Chong}\ \bibnamefont {Ying}}, \bibinfo {author}
  {\bibfnamefont {Jiale}\ \bibnamefont {Yu}}, \bibinfo {author} {\bibfnamefont
  {Chen}\ \bibnamefont {Zha}}, \bibinfo {author} {\bibfnamefont {Cha}\
  \bibnamefont {Zhang}}, \bibinfo {author} {\bibfnamefont {Haibin}\
  \bibnamefont {Zhang}}, \bibinfo {author} {\bibfnamefont {Kaili}\ \bibnamefont
  {Zhang}}, \bibinfo {author} {\bibfnamefont {Yiming}\ \bibnamefont {Zhang}},
  \bibinfo {author} {\bibfnamefont {Han}\ \bibnamefont {Zhao}}, \bibinfo
  {author} {\bibfnamefont {Youwei}\ \bibnamefont {Zhao}}, \bibinfo {author}
  {\bibfnamefont {Liang}\ \bibnamefont {Zhou}}, \bibinfo {author}
  {\bibfnamefont {Qingling}\ \bibnamefont {Zhu}}, \bibinfo {author}
  {\bibfnamefont {Chao-yang}\ \bibnamefont {Lu}}, \bibinfo {author}
  {\bibfnamefont {Cheng-zhi}\ \bibnamefont {Peng}}, \bibinfo {author}
  {\bibfnamefont {Xiaobo}\ \bibnamefont {Zhu}}, \ and\ \bibinfo {author}
  {\bibfnamefont {Jian-wei}\ \bibnamefont {Pan}},\ }\bibfield  {title}
  {\enquote {\bibinfo {title} {{Strong Quantum Computational Advantage Using a
  Superconducting Quantum Processor}},}\ }\href {\doibase
  10.1103/PhysRevLett.127.180501} {\bibfield  {journal} {\bibinfo  {journal}
  {Physical Review Letters}\ }\textbf {\bibinfo {volume} {127}},\ \bibinfo
  {pages} {180501} (\bibinfo {year} {2021})},\ \Eprint
  {http://arxiv.org/abs/2106.14734} {arXiv:2106.14734} \BibitemShut {NoStop}%
\bibitem [{\citenamefont {Zhong}\ \emph {et~al.}(2020)\citenamefont {Zhong},
  \citenamefont {Wang}, \citenamefont {Deng}, \citenamefont {Chen},
  \citenamefont {Peng}, \citenamefont {Luo}, \citenamefont {Qin}, \citenamefont
  {Wu}, \citenamefont {Ding}, \citenamefont {Hu}, \citenamefont {Hu},
  \citenamefont {Yang}, \citenamefont {Zhang}, \citenamefont {Li},
  \citenamefont {Li}, \citenamefont {Jiang}, \citenamefont {Gan}, \citenamefont
  {Yang}, \citenamefont {You}, \citenamefont {Wang}, \citenamefont {Li},
  \citenamefont {Liu}, \citenamefont {Lu},\ and\ \citenamefont
  {Pan}}]{Zhong2020}%
  \BibitemOpen
  \bibfield  {author} {\bibinfo {author} {\bibfnamefont {Han-Sen}\ \bibnamefont
  {Zhong}}, \bibinfo {author} {\bibfnamefont {Hui}\ \bibnamefont {Wang}},
  \bibinfo {author} {\bibfnamefont {Yu-Hao}\ \bibnamefont {Deng}}, \bibinfo
  {author} {\bibfnamefont {Ming-Cheng}\ \bibnamefont {Chen}}, \bibinfo {author}
  {\bibfnamefont {Li-Chao}\ \bibnamefont {Peng}}, \bibinfo {author}
  {\bibfnamefont {Yi-Han}\ \bibnamefont {Luo}}, \bibinfo {author}
  {\bibfnamefont {Jian}\ \bibnamefont {Qin}}, \bibinfo {author} {\bibfnamefont
  {Dian}\ \bibnamefont {Wu}}, \bibinfo {author} {\bibfnamefont {Xing}\
  \bibnamefont {Ding}}, \bibinfo {author} {\bibfnamefont {Yi}~\bibnamefont
  {Hu}}, \bibinfo {author} {\bibfnamefont {Peng}\ \bibnamefont {Hu}}, \bibinfo
  {author} {\bibfnamefont {Xiao-Yan}\ \bibnamefont {Yang}}, \bibinfo {author}
  {\bibfnamefont {Wei-Jun}\ \bibnamefont {Zhang}}, \bibinfo {author}
  {\bibfnamefont {Hao}\ \bibnamefont {Li}}, \bibinfo {author} {\bibfnamefont
  {Yuxuan}\ \bibnamefont {Li}}, \bibinfo {author} {\bibfnamefont {Xiao}\
  \bibnamefont {Jiang}}, \bibinfo {author} {\bibfnamefont {Lin}\ \bibnamefont
  {Gan}}, \bibinfo {author} {\bibfnamefont {Guangwen}\ \bibnamefont {Yang}},
  \bibinfo {author} {\bibfnamefont {Lixing}\ \bibnamefont {You}}, \bibinfo
  {author} {\bibfnamefont {Zhen}\ \bibnamefont {Wang}}, \bibinfo {author}
  {\bibfnamefont {Li}~\bibnamefont {Li}}, \bibinfo {author} {\bibfnamefont
  {Nai-Le}\ \bibnamefont {Liu}}, \bibinfo {author} {\bibfnamefont {Chao-Yang}\
  \bibnamefont {Lu}}, \ and\ \bibinfo {author} {\bibfnamefont {Jian-Wei}\
  \bibnamefont {Pan}},\ }\bibfield  {title} {\enquote {\bibinfo {title}
  {Quantum computational advantage using photons},}\ }\href {\doibase
  10.1126/science.abe8770} {\bibfield  {journal} {\bibinfo  {journal}
  {Science}\ }\textbf {\bibinfo {volume} {370}},\ \bibinfo {pages} {1460--1463}
  (\bibinfo {year} {2020})},\ \Eprint
  {http://arxiv.org/abs/https://science.sciencemag.org/content/370/6523/1460.full.pdf}
  {https://science.sciencemag.org/content/370/6523/1460.full.pdf} \BibitemShut
  {NoStop}%
\bibitem [{\citenamefont {Popova}\ and\ \citenamefont
  {Rubtsov}(2021)}]{Popova2021}%
  \BibitemOpen
  \bibfield  {author} {\bibinfo {author} {\bibfnamefont {A.~S.}\ \bibnamefont
  {Popova}}\ and\ \bibinfo {author} {\bibfnamefont {A.~N.}\ \bibnamefont
  {Rubtsov}},\ }\href {\doibase 10.48550/ARXIV.2106.01445} {\enquote {\bibinfo
  {title} {Cracking the quantum advantage threshold for gaussian boson
  sampling},}\ } (\bibinfo {year} {2021})\BibitemShut {NoStop}%
\bibitem [{\citenamefont {Villalonga}\ \emph {et~al.}(2021)\citenamefont
  {Villalonga}, \citenamefont {Niu}, \citenamefont {Li}, \citenamefont {Neven},
  \citenamefont {Platt}, \citenamefont {Smelyanskiy},\ and\ \citenamefont
  {Boixo}}]{Villalonga2021}%
  \BibitemOpen
  \bibfield  {author} {\bibinfo {author} {\bibfnamefont {Benjamin}\
  \bibnamefont {Villalonga}}, \bibinfo {author} {\bibfnamefont
  {Murphy~Yuezhen}\ \bibnamefont {Niu}}, \bibinfo {author} {\bibfnamefont
  {Li}~\bibnamefont {Li}}, \bibinfo {author} {\bibfnamefont {Hartmut}\
  \bibnamefont {Neven}}, \bibinfo {author} {\bibfnamefont {John~C.}\
  \bibnamefont {Platt}}, \bibinfo {author} {\bibfnamefont {Vadim~N.}\
  \bibnamefont {Smelyanskiy}}, \ and\ \bibinfo {author} {\bibfnamefont
  {Sergio}\ \bibnamefont {Boixo}},\ }\href {\doibase 10.48550/ARXIV.2109.11525}
  {\enquote {\bibinfo {title} {Efficient approximation of experimental gaussian
  boson sampling},}\ } (\bibinfo {year} {2021})\BibitemShut {NoStop}%
\bibitem [{\citenamefont {Oh}\ \emph {et~al.}(2022)\citenamefont {Oh},
  \citenamefont {Lim}, \citenamefont {Fefferman},\ and\ \citenamefont
  {Jiang}}]{Oh2022}%
  \BibitemOpen
  \bibfield  {author} {\bibinfo {author} {\bibfnamefont {Changhun}\
  \bibnamefont {Oh}}, \bibinfo {author} {\bibfnamefont {Youngrong}\
  \bibnamefont {Lim}}, \bibinfo {author} {\bibfnamefont {Bill}\ \bibnamefont
  {Fefferman}}, \ and\ \bibinfo {author} {\bibfnamefont {Liang}\ \bibnamefont
  {Jiang}},\ }\bibfield  {title} {\enquote {\bibinfo {title} {Classical
  simulation of boson sampling based on graph structure},}\ }\href {\doibase
  10.1103/PhysRevLett.128.190501} {\bibfield  {journal} {\bibinfo  {journal}
  {Phys. Rev. Lett.}\ }\textbf {\bibinfo {volume} {128}},\ \bibinfo {pages}
  {190501} (\bibinfo {year} {2022})}\BibitemShut {NoStop}%
\bibitem [{\citenamefont {Yong}\ \emph {et~al.}()\citenamefont {Yong},
  \citenamefont {Liu}, \citenamefont {Xin}, \citenamefont {Liu}, \citenamefont
  {Fang}, \citenamefont {Li}, \citenamefont {Fu}, \citenamefont {Yang},
  \citenamefont {Song}, \citenamefont {Zhao}, \citenamefont {Wang},
  \citenamefont {Peng}, \citenamefont {Chen}, \citenamefont {Guo},
  \citenamefont {Huang}, \citenamefont {Wu},\ and\ \citenamefont
  {Chen}}]{Yong2021closingsupremacygap}%
  \BibitemOpen
  \bibfield  {author} {\bibinfo {author} {\bibnamefont {Yong}}, \bibinfo
  {author} {\bibnamefont {Liu}}, \bibinfo {author} {\bibnamefont {Xin}},
  \bibinfo {author} {\bibnamefont {Liu}}, \bibinfo {author} {\bibnamefont
  {Fang}}, \bibinfo {author} {\bibnamefont {Li}}, \bibinfo {author}
  {\bibfnamefont {Haohuan}\ \bibnamefont {Fu}}, \bibinfo {author}
  {\bibfnamefont {Yuling}\ \bibnamefont {Yang}}, \bibinfo {author}
  {\bibfnamefont {Jiawei}\ \bibnamefont {Song}}, \bibinfo {author}
  {\bibfnamefont {Pengpeng}\ \bibnamefont {Zhao}}, \bibinfo {author}
  {\bibfnamefont {Zhen}\ \bibnamefont {Wang}}, \bibinfo {author} {\bibfnamefont
  {Dajia}\ \bibnamefont {Peng}}, \bibinfo {author} {\bibfnamefont {Huarong}\
  \bibnamefont {Chen}}, \bibinfo {author} {\bibfnamefont {Chu}\ \bibnamefont
  {Guo}}, \bibinfo {author} {\bibfnamefont {Heliang}\ \bibnamefont {Huang}},
  \bibinfo {author} {\bibfnamefont {Wenzhao}\ \bibnamefont {Wu}}, \ and\
  \bibinfo {author} {\bibfnamefont {Dexun}\ \bibnamefont {Chen}},\ }\bibfield
  {title} {\enquote {\bibinfo {title} {Closing the "quantum supremacy" gap:
  Achieving real-time simulation of a random quantum circuit using a new sunway
  supercomputer},}\ }\href {\doibase 10.1145/3458817.3487399} {\
  10.1145/3458817.3487399},\ \Eprint {http://arxiv.org/abs/2110.14502v2}
  {2110.14502v2} \BibitemShut {NoStop}%
\bibitem [{\citenamefont {Arute}\ \emph {et~al.}(2020)\citenamefont {Arute},
  \citenamefont {Arya}, \citenamefont {Babbush}, \citenamefont {Bacon},
  \citenamefont {Bardin}, \citenamefont {Barends}, \citenamefont {Boixo},
  \citenamefont {Broughton}, \citenamefont {Buckley}, \citenamefont {Buell},
  \citenamefont {Burkett}, \citenamefont {Bushnell}, \citenamefont {Chen},
  \citenamefont {Chen}, \citenamefont {Chiaro}, \citenamefont {Collins},
  \citenamefont {Courtney}, \citenamefont {Demura}, \citenamefont {Dunsworth},
  \citenamefont {Farhi}, \citenamefont {Fowler}, \citenamefont {Foxen},
  \citenamefont {Gidney}, \citenamefont {Giustina}, \citenamefont {Graff},
  \citenamefont {Habegger}, \citenamefont {Harrigan}, \citenamefont {Ho},
  \citenamefont {Hong}, \citenamefont {Huang}, \citenamefont {Huggins},
  \citenamefont {Ioffe}, \citenamefont {Isakov}, \citenamefont {Jeffrey},
  \citenamefont {Jiang}, \citenamefont {Jones}, \citenamefont {Kafri},
  \citenamefont {Kechedzhi}, \citenamefont {Kelly}, \citenamefont {Kim},
  \citenamefont {Klimov}, \citenamefont {Korotkov}, \citenamefont {Kostritsa},
  \citenamefont {Landhuis}, \citenamefont {Laptev}, \citenamefont {Lindmark},
  \citenamefont {Lucero}, \citenamefont {Martin}, \citenamefont {Martinis},
  \citenamefont {McClean}, \citenamefont {McEwen}, \citenamefont {Megrant},
  \citenamefont {Mi}, \citenamefont {Mohseni}, \citenamefont {Mruczkiewicz},
  \citenamefont {Mutus}, \citenamefont {Naaman}, \citenamefont {Neeley},
  \citenamefont {Neill}, \citenamefont {Neven}, \citenamefont {Niu},
  \citenamefont {O'Brien}, \citenamefont {Ostby}, \citenamefont {Petukhov},
  \citenamefont {Putterman}, \citenamefont {Quintana}, \citenamefont {Roushan},
  \citenamefont {Rubin}, \citenamefont {Sank}, \citenamefont {Satzinger},
  \citenamefont {Smelyanskiy}, \citenamefont {Strain}, \citenamefont {Sung},
  \citenamefont {Szalay}, \citenamefont {Takeshita}, \citenamefont
  {Vainsencher}, \citenamefont {White}, \citenamefont {Wiebe}, \citenamefont
  {Yao}, \citenamefont {Yeh},\ and\ \citenamefont {Zalcman}}]{Arute2020}%
  \BibitemOpen
  \bibfield  {author} {\bibinfo {author} {\bibfnamefont {Frank}\ \bibnamefont
  {Arute}}, \bibinfo {author} {\bibfnamefont {Kunal}\ \bibnamefont {Arya}},
  \bibinfo {author} {\bibfnamefont {Ryan}\ \bibnamefont {Babbush}}, \bibinfo
  {author} {\bibfnamefont {Dave}\ \bibnamefont {Bacon}}, \bibinfo {author}
  {\bibfnamefont {Joseph~C.}\ \bibnamefont {Bardin}}, \bibinfo {author}
  {\bibfnamefont {Rami}\ \bibnamefont {Barends}}, \bibinfo {author}
  {\bibfnamefont {Sergio}\ \bibnamefont {Boixo}}, \bibinfo {author}
  {\bibfnamefont {Michael}\ \bibnamefont {Broughton}}, \bibinfo {author}
  {\bibfnamefont {Bob~B.}\ \bibnamefont {Buckley}}, \bibinfo {author}
  {\bibfnamefont {David~A.}\ \bibnamefont {Buell}}, \bibinfo {author}
  {\bibfnamefont {Brian}\ \bibnamefont {Burkett}}, \bibinfo {author}
  {\bibfnamefont {Nicholas}\ \bibnamefont {Bushnell}}, \bibinfo {author}
  {\bibfnamefont {Yu}~\bibnamefont {Chen}}, \bibinfo {author} {\bibfnamefont
  {Zijun}\ \bibnamefont {Chen}}, \bibinfo {author} {\bibfnamefont {Benjamin}\
  \bibnamefont {Chiaro}}, \bibinfo {author} {\bibfnamefont {Roberto}\
  \bibnamefont {Collins}}, \bibinfo {author} {\bibfnamefont {William}\
  \bibnamefont {Courtney}}, \bibinfo {author} {\bibfnamefont {Sean}\
  \bibnamefont {Demura}}, \bibinfo {author} {\bibfnamefont {Andrew}\
  \bibnamefont {Dunsworth}}, \bibinfo {author} {\bibfnamefont {Edward}\
  \bibnamefont {Farhi}}, \bibinfo {author} {\bibfnamefont {Austin}\
  \bibnamefont {Fowler}}, \bibinfo {author} {\bibfnamefont {Brooks}\
  \bibnamefont {Foxen}}, \bibinfo {author} {\bibfnamefont {Craig}\ \bibnamefont
  {Gidney}}, \bibinfo {author} {\bibfnamefont {Marissa}\ \bibnamefont
  {Giustina}}, \bibinfo {author} {\bibfnamefont {Rob}\ \bibnamefont {Graff}},
  \bibinfo {author} {\bibfnamefont {Steve}\ \bibnamefont {Habegger}}, \bibinfo
  {author} {\bibfnamefont {Matthew~P.}\ \bibnamefont {Harrigan}}, \bibinfo
  {author} {\bibfnamefont {Alan}\ \bibnamefont {Ho}}, \bibinfo {author}
  {\bibfnamefont {Sabrina}\ \bibnamefont {Hong}}, \bibinfo {author}
  {\bibfnamefont {Trent}\ \bibnamefont {Huang}}, \bibinfo {author}
  {\bibfnamefont {William~J.}\ \bibnamefont {Huggins}}, \bibinfo {author}
  {\bibfnamefont {Lev}\ \bibnamefont {Ioffe}}, \bibinfo {author} {\bibfnamefont
  {Sergei~V.}\ \bibnamefont {Isakov}}, \bibinfo {author} {\bibfnamefont {Evan}\
  \bibnamefont {Jeffrey}}, \bibinfo {author} {\bibfnamefont {Zhang}\
  \bibnamefont {Jiang}}, \bibinfo {author} {\bibfnamefont {Cody}\ \bibnamefont
  {Jones}}, \bibinfo {author} {\bibfnamefont {Dvir}\ \bibnamefont {Kafri}},
  \bibinfo {author} {\bibfnamefont {Kostyantyn}\ \bibnamefont {Kechedzhi}},
  \bibinfo {author} {\bibfnamefont {Julian}\ \bibnamefont {Kelly}}, \bibinfo
  {author} {\bibfnamefont {Seon}\ \bibnamefont {Kim}}, \bibinfo {author}
  {\bibfnamefont {Paul~V.}\ \bibnamefont {Klimov}}, \bibinfo {author}
  {\bibfnamefont {Alexander}\ \bibnamefont {Korotkov}}, \bibinfo {author}
  {\bibfnamefont {Fedor}\ \bibnamefont {Kostritsa}}, \bibinfo {author}
  {\bibfnamefont {David}\ \bibnamefont {Landhuis}}, \bibinfo {author}
  {\bibfnamefont {Pavel}\ \bibnamefont {Laptev}}, \bibinfo {author}
  {\bibfnamefont {Mike}\ \bibnamefont {Lindmark}}, \bibinfo {author}
  {\bibfnamefont {Erik}\ \bibnamefont {Lucero}}, \bibinfo {author}
  {\bibfnamefont {Orion}\ \bibnamefont {Martin}}, \bibinfo {author}
  {\bibfnamefont {John~M.}\ \bibnamefont {Martinis}}, \bibinfo {author}
  {\bibfnamefont {Jarrod~R.}\ \bibnamefont {McClean}}, \bibinfo {author}
  {\bibfnamefont {Matt}\ \bibnamefont {McEwen}}, \bibinfo {author}
  {\bibfnamefont {Anthony}\ \bibnamefont {Megrant}}, \bibinfo {author}
  {\bibfnamefont {Xiao}\ \bibnamefont {Mi}}, \bibinfo {author} {\bibfnamefont
  {Masoud}\ \bibnamefont {Mohseni}}, \bibinfo {author} {\bibfnamefont
  {Wojciech}\ \bibnamefont {Mruczkiewicz}}, \bibinfo {author} {\bibfnamefont
  {Josh}\ \bibnamefont {Mutus}}, \bibinfo {author} {\bibfnamefont {Ofer}\
  \bibnamefont {Naaman}}, \bibinfo {author} {\bibfnamefont {Matthew}\
  \bibnamefont {Neeley}}, \bibinfo {author} {\bibfnamefont {Charles}\
  \bibnamefont {Neill}}, \bibinfo {author} {\bibfnamefont {Hartmut}\
  \bibnamefont {Neven}}, \bibinfo {author} {\bibfnamefont {Murphy~Yuezhen}\
  \bibnamefont {Niu}}, \bibinfo {author} {\bibfnamefont {Thomas~E.}\
  \bibnamefont {O'Brien}}, \bibinfo {author} {\bibfnamefont {Eric}\
  \bibnamefont {Ostby}}, \bibinfo {author} {\bibfnamefont {Andre}\ \bibnamefont
  {Petukhov}}, \bibinfo {author} {\bibfnamefont {Harald}\ \bibnamefont
  {Putterman}}, \bibinfo {author} {\bibfnamefont {Chris}\ \bibnamefont
  {Quintana}}, \bibinfo {author} {\bibfnamefont {Pedram}\ \bibnamefont
  {Roushan}}, \bibinfo {author} {\bibfnamefont {Nicholas~C.}\ \bibnamefont
  {Rubin}}, \bibinfo {author} {\bibfnamefont {Daniel}\ \bibnamefont {Sank}},
  \bibinfo {author} {\bibfnamefont {Kevin~J.}\ \bibnamefont {Satzinger}},
  \bibinfo {author} {\bibfnamefont {Vadim}\ \bibnamefont {Smelyanskiy}},
  \bibinfo {author} {\bibfnamefont {Doug}\ \bibnamefont {Strain}}, \bibinfo
  {author} {\bibfnamefont {Kevin~J.}\ \bibnamefont {Sung}}, \bibinfo {author}
  {\bibfnamefont {Marco}\ \bibnamefont {Szalay}}, \bibinfo {author}
  {\bibfnamefont {Tyler~Y.}\ \bibnamefont {Takeshita}}, \bibinfo {author}
  {\bibfnamefont {Amit}\ \bibnamefont {Vainsencher}}, \bibinfo {author}
  {\bibfnamefont {Theodore}\ \bibnamefont {White}}, \bibinfo {author}
  {\bibfnamefont {Nathan}\ \bibnamefont {Wiebe}}, \bibinfo {author}
  {\bibfnamefont {Z.~Jamie}\ \bibnamefont {Yao}}, \bibinfo {author}
  {\bibfnamefont {Ping}\ \bibnamefont {Yeh}}, \ and\ \bibinfo {author}
  {\bibfnamefont {Adam}\ \bibnamefont {Zalcman}},\ }\bibfield  {title}
  {\enquote {\bibinfo {title} {{Hartree-Fock on a superconducting qubit quantum
  computer}},}\ }\href {http://arxiv.org/abs/2004.04174} {\  (\bibinfo {year}
  {2020})},\ \Eprint {http://arxiv.org/abs/2004.04174} {arXiv:2004.04174}
  \BibitemShut {NoStop}%
\bibitem [{\citenamefont {Pednault}\ \emph {et~al.}(2019)\citenamefont
  {Pednault}, \citenamefont {Gunnels}, \citenamefont {Nannicini}, \citenamefont
  {Horesh},\ and\ \citenamefont {Wisnieff}}]{Pednault2019}%
  \BibitemOpen
  \bibfield  {author} {\bibinfo {author} {\bibfnamefont {Edwin}\ \bibnamefont
  {Pednault}}, \bibinfo {author} {\bibfnamefont {John~A.}\ \bibnamefont
  {Gunnels}}, \bibinfo {author} {\bibfnamefont {Giacomo}\ \bibnamefont
  {Nannicini}}, \bibinfo {author} {\bibfnamefont {Lior}\ \bibnamefont
  {Horesh}}, \ and\ \bibinfo {author} {\bibfnamefont {Robert}\ \bibnamefont
  {Wisnieff}},\ }\bibfield  {title} {\enquote {\bibinfo {title} {Leveraging
  secondary storage to simulate deep 54-qubit sycamore circuits},}\ }\href@noop
  {} {\  (\bibinfo {year} {2019})},\ \Eprint {http://arxiv.org/abs/1910.09534}
  {arXiv:1910.09534} \BibitemShut {NoStop}%
\bibitem [{\citenamefont {Bravyi}\ \emph {et~al.}(2021)\citenamefont {Bravyi},
  \citenamefont {Gosset},\ and\ \citenamefont {Liu}}]{Bravyi2021}%
  \BibitemOpen
  \bibfield  {author} {\bibinfo {author} {\bibfnamefont {Sergey}\ \bibnamefont
  {Bravyi}}, \bibinfo {author} {\bibfnamefont {David}\ \bibnamefont {Gosset}},
  \ and\ \bibinfo {author} {\bibfnamefont {Yinchen}\ \bibnamefont {Liu}},\
  }\bibfield  {title} {\enquote {\bibinfo {title} {How to simulate quantum
  measurement without computing marginals},}\ }\href@noop {} {\  (\bibinfo
  {year} {2021})},\ \Eprint {http://arxiv.org/abs/2112.08499v2} {2112.08499v2}
  \BibitemShut {NoStop}%
\bibitem [{\citenamefont {Huang}\ \emph {et~al.}(2020)\citenamefont {Huang},
  \citenamefont {Zhang}, \citenamefont {Newman}, \citenamefont {Cai},
  \citenamefont {Gao}, \citenamefont {Tian}, \citenamefont {Wu}, \citenamefont
  {Xu}, \citenamefont {Yu}, \citenamefont {Yuan}, \citenamefont {Szegedy},
  \citenamefont {Shi},\ and\ \citenamefont {Chen}}]{Huang2020}%
  \BibitemOpen
  \bibfield  {author} {\bibinfo {author} {\bibfnamefont {Cupjin}\ \bibnamefont
  {Huang}}, \bibinfo {author} {\bibfnamefont {Fang}\ \bibnamefont {Zhang}},
  \bibinfo {author} {\bibfnamefont {Michael}\ \bibnamefont {Newman}}, \bibinfo
  {author} {\bibfnamefont {Junjie}\ \bibnamefont {Cai}}, \bibinfo {author}
  {\bibfnamefont {Xun}\ \bibnamefont {Gao}}, \bibinfo {author} {\bibfnamefont
  {Zhengxiong}\ \bibnamefont {Tian}}, \bibinfo {author} {\bibfnamefont
  {Junyin}\ \bibnamefont {Wu}}, \bibinfo {author} {\bibfnamefont {Haihong}\
  \bibnamefont {Xu}}, \bibinfo {author} {\bibfnamefont {Huanjun}\ \bibnamefont
  {Yu}}, \bibinfo {author} {\bibfnamefont {Bo}~\bibnamefont {Yuan}}, \bibinfo
  {author} {\bibfnamefont {Mario}\ \bibnamefont {Szegedy}}, \bibinfo {author}
  {\bibfnamefont {Yaoyun}\ \bibnamefont {Shi}}, \ and\ \bibinfo {author}
  {\bibfnamefont {Jianxin}\ \bibnamefont {Chen}},\ }\bibfield  {title}
  {\enquote {\bibinfo {title} {Classical simulation of quantum supremacy
  circuits},}\ }\href@noop {} {\  (\bibinfo {year} {2020})},\ \Eprint
  {http://arxiv.org/abs/2005.06787v1} {2005.06787v1} \BibitemShut {NoStop}%
\bibitem [{\citenamefont {Harrigan}\ \emph {et~al.}(2021)\citenamefont
  {Harrigan}, \citenamefont {Sung}, \citenamefont {Neeley}, \citenamefont
  {Satzinger}, \citenamefont {Arute}, \citenamefont {Arya}, \citenamefont
  {Atalaya}, \citenamefont {Bardin}, \citenamefont {Barends}, \citenamefont
  {Boixo}, \citenamefont {Broughton}, \citenamefont {Buckley}, \citenamefont
  {Buell}, \citenamefont {Burkett}, \citenamefont {Bushnell}, \citenamefont
  {Chen}, \citenamefont {Chen}, \citenamefont {{Ben Chiaro}}, \citenamefont
  {Collins}, \citenamefont {Courtney}, \citenamefont {Demura}, \citenamefont
  {Dunsworth}, \citenamefont {Eppens}, \citenamefont {Fowler}, \citenamefont
  {Foxen}, \citenamefont {Gidney}, \citenamefont {Giustina}, \citenamefont
  {Graff}, \citenamefont {Habegger}, \citenamefont {Ho}, \citenamefont {Hong},
  \citenamefont {Huang}, \citenamefont {Ioffe}, \citenamefont {Isakov},
  \citenamefont {Jeffrey}, \citenamefont {Jiang}, \citenamefont {Jones},
  \citenamefont {Kafri}, \citenamefont {Kechedzhi}, \citenamefont {Kelly},
  \citenamefont {Kim}, \citenamefont {Klimov}, \citenamefont {Korotkov},
  \citenamefont {Kostritsa}, \citenamefont {Landhuis}, \citenamefont {Laptev},
  \citenamefont {Lindmark}, \citenamefont {Leib}, \citenamefont {Martin},
  \citenamefont {Martinis}, \citenamefont {McClean}, \citenamefont {McEwen},
  \citenamefont {Megrant}, \citenamefont {Mi}, \citenamefont {Mohseni},
  \citenamefont {Mruczkiewicz}, \citenamefont {Mutus}, \citenamefont {Naaman},
  \citenamefont {Neill}, \citenamefont {Neukart}, \citenamefont {Niu},
  \citenamefont {O'Brien}, \citenamefont {O'Gorman}, \citenamefont {Ostby},
  \citenamefont {Petukhov}, \citenamefont {Putterman}, \citenamefont
  {Quintana}, \citenamefont {Roushan}, \citenamefont {Rubin}, \citenamefont
  {Sank}, \citenamefont {Skolik}, \citenamefont {Smelyanskiy}, \citenamefont
  {Strain}, \citenamefont {Streif}, \citenamefont {Szalay}, \citenamefont
  {Vainsencher}, \citenamefont {White}, \citenamefont {Yao}, \citenamefont
  {Yeh}, \citenamefont {Zalcman}, \citenamefont {Zhou}, \citenamefont {Neven},
  \citenamefont {Bacon}, \citenamefont {Lucero}, \citenamefont {Farhi},\ and\
  \citenamefont {Babbush}}]{Harrigan2021}%
  \BibitemOpen
  \bibfield  {author} {\bibinfo {author} {\bibfnamefont {Matthew~P.}\
  \bibnamefont {Harrigan}}, \bibinfo {author} {\bibfnamefont {Kevin~J.}\
  \bibnamefont {Sung}}, \bibinfo {author} {\bibfnamefont {Matthew}\
  \bibnamefont {Neeley}}, \bibinfo {author} {\bibfnamefont {Kevin~J.}\
  \bibnamefont {Satzinger}}, \bibinfo {author} {\bibfnamefont {Frank}\
  \bibnamefont {Arute}}, \bibinfo {author} {\bibfnamefont {Kunal}\ \bibnamefont
  {Arya}}, \bibinfo {author} {\bibfnamefont {Juan}\ \bibnamefont {Atalaya}},
  \bibinfo {author} {\bibfnamefont {Joseph~C.}\ \bibnamefont {Bardin}},
  \bibinfo {author} {\bibfnamefont {Rami}\ \bibnamefont {Barends}}, \bibinfo
  {author} {\bibfnamefont {Sergio}\ \bibnamefont {Boixo}}, \bibinfo {author}
  {\bibfnamefont {Michael}\ \bibnamefont {Broughton}}, \bibinfo {author}
  {\bibfnamefont {Bob~B.}\ \bibnamefont {Buckley}}, \bibinfo {author}
  {\bibfnamefont {David~A.}\ \bibnamefont {Buell}}, \bibinfo {author}
  {\bibfnamefont {Brian}\ \bibnamefont {Burkett}}, \bibinfo {author}
  {\bibfnamefont {Nicholas}\ \bibnamefont {Bushnell}}, \bibinfo {author}
  {\bibfnamefont {Yu}~\bibnamefont {Chen}}, \bibinfo {author} {\bibfnamefont
  {Zijun}\ \bibnamefont {Chen}}, \bibinfo {author} {\bibnamefont {{Ben
  Chiaro}}}, \bibinfo {author} {\bibfnamefont {Roberto}\ \bibnamefont
  {Collins}}, \bibinfo {author} {\bibfnamefont {William}\ \bibnamefont
  {Courtney}}, \bibinfo {author} {\bibfnamefont {Sean}\ \bibnamefont {Demura}},
  \bibinfo {author} {\bibfnamefont {Andrew}\ \bibnamefont {Dunsworth}},
  \bibinfo {author} {\bibfnamefont {Daniel}\ \bibnamefont {Eppens}}, \bibinfo
  {author} {\bibfnamefont {Austin}\ \bibnamefont {Fowler}}, \bibinfo {author}
  {\bibfnamefont {Brooks}\ \bibnamefont {Foxen}}, \bibinfo {author}
  {\bibfnamefont {Craig}\ \bibnamefont {Gidney}}, \bibinfo {author}
  {\bibfnamefont {Marissa}\ \bibnamefont {Giustina}}, \bibinfo {author}
  {\bibfnamefont {Rob}\ \bibnamefont {Graff}}, \bibinfo {author} {\bibfnamefont
  {Steve}\ \bibnamefont {Habegger}}, \bibinfo {author} {\bibfnamefont {Alan}\
  \bibnamefont {Ho}}, \bibinfo {author} {\bibfnamefont {Sabrina}\ \bibnamefont
  {Hong}}, \bibinfo {author} {\bibfnamefont {Trent}\ \bibnamefont {Huang}},
  \bibinfo {author} {\bibfnamefont {L.~B.}\ \bibnamefont {Ioffe}}, \bibinfo
  {author} {\bibfnamefont {Sergei~V.}\ \bibnamefont {Isakov}}, \bibinfo
  {author} {\bibfnamefont {Evan}\ \bibnamefont {Jeffrey}}, \bibinfo {author}
  {\bibfnamefont {Zhang}\ \bibnamefont {Jiang}}, \bibinfo {author}
  {\bibfnamefont {Cody}\ \bibnamefont {Jones}}, \bibinfo {author}
  {\bibfnamefont {Dvir}\ \bibnamefont {Kafri}}, \bibinfo {author}
  {\bibfnamefont {Kostyantyn}\ \bibnamefont {Kechedzhi}}, \bibinfo {author}
  {\bibfnamefont {Julian}\ \bibnamefont {Kelly}}, \bibinfo {author}
  {\bibfnamefont {Seon}\ \bibnamefont {Kim}}, \bibinfo {author} {\bibfnamefont
  {Paul~V.}\ \bibnamefont {Klimov}}, \bibinfo {author} {\bibfnamefont
  {Alexander~N.}\ \bibnamefont {Korotkov}}, \bibinfo {author} {\bibfnamefont
  {Fedor}\ \bibnamefont {Kostritsa}}, \bibinfo {author} {\bibfnamefont {David}\
  \bibnamefont {Landhuis}}, \bibinfo {author} {\bibfnamefont {Pavel}\
  \bibnamefont {Laptev}}, \bibinfo {author} {\bibfnamefont {Mike}\ \bibnamefont
  {Lindmark}}, \bibinfo {author} {\bibfnamefont {Martin}\ \bibnamefont {Leib}},
  \bibinfo {author} {\bibfnamefont {Orion}\ \bibnamefont {Martin}}, \bibinfo
  {author} {\bibfnamefont {John~M.}\ \bibnamefont {Martinis}}, \bibinfo
  {author} {\bibfnamefont {Jarrod~R.}\ \bibnamefont {McClean}}, \bibinfo
  {author} {\bibfnamefont {Matt}\ \bibnamefont {McEwen}}, \bibinfo {author}
  {\bibfnamefont {Anthony}\ \bibnamefont {Megrant}}, \bibinfo {author}
  {\bibfnamefont {Xiao}\ \bibnamefont {Mi}}, \bibinfo {author} {\bibfnamefont
  {Masoud}\ \bibnamefont {Mohseni}}, \bibinfo {author} {\bibfnamefont
  {Wojciech}\ \bibnamefont {Mruczkiewicz}}, \bibinfo {author} {\bibfnamefont
  {Josh}\ \bibnamefont {Mutus}}, \bibinfo {author} {\bibfnamefont {Ofer}\
  \bibnamefont {Naaman}}, \bibinfo {author} {\bibfnamefont {Charles}\
  \bibnamefont {Neill}}, \bibinfo {author} {\bibfnamefont {Florian}\
  \bibnamefont {Neukart}}, \bibinfo {author} {\bibfnamefont {Murphy~Yuezhen}\
  \bibnamefont {Niu}}, \bibinfo {author} {\bibfnamefont {Thomas~E.}\
  \bibnamefont {O'Brien}}, \bibinfo {author} {\bibfnamefont {Bryan}\
  \bibnamefont {O'Gorman}}, \bibinfo {author} {\bibfnamefont {Eric}\
  \bibnamefont {Ostby}}, \bibinfo {author} {\bibfnamefont {Andre}\ \bibnamefont
  {Petukhov}}, \bibinfo {author} {\bibfnamefont {Harald}\ \bibnamefont
  {Putterman}}, \bibinfo {author} {\bibfnamefont {Chris}\ \bibnamefont
  {Quintana}}, \bibinfo {author} {\bibfnamefont {Pedram}\ \bibnamefont
  {Roushan}}, \bibinfo {author} {\bibfnamefont {Nicholas~C.}\ \bibnamefont
  {Rubin}}, \bibinfo {author} {\bibfnamefont {Daniel}\ \bibnamefont {Sank}},
  \bibinfo {author} {\bibfnamefont {Andrea}\ \bibnamefont {Skolik}}, \bibinfo
  {author} {\bibfnamefont {Vadim}\ \bibnamefont {Smelyanskiy}}, \bibinfo
  {author} {\bibfnamefont {Doug}\ \bibnamefont {Strain}}, \bibinfo {author}
  {\bibfnamefont {Michael}\ \bibnamefont {Streif}}, \bibinfo {author}
  {\bibfnamefont {Marco}\ \bibnamefont {Szalay}}, \bibinfo {author}
  {\bibfnamefont {Amit}\ \bibnamefont {Vainsencher}}, \bibinfo {author}
  {\bibfnamefont {Theodore}\ \bibnamefont {White}}, \bibinfo {author}
  {\bibfnamefont {Z.~Jamie}\ \bibnamefont {Yao}}, \bibinfo {author}
  {\bibfnamefont {Ping}\ \bibnamefont {Yeh}}, \bibinfo {author} {\bibfnamefont
  {Adam}\ \bibnamefont {Zalcman}}, \bibinfo {author} {\bibfnamefont {Leo}\
  \bibnamefont {Zhou}}, \bibinfo {author} {\bibfnamefont {Hartmut}\
  \bibnamefont {Neven}}, \bibinfo {author} {\bibfnamefont {Dave}\ \bibnamefont
  {Bacon}}, \bibinfo {author} {\bibfnamefont {Erik}\ \bibnamefont {Lucero}},
  \bibinfo {author} {\bibfnamefont {Edward}\ \bibnamefont {Farhi}}, \ and\
  \bibinfo {author} {\bibfnamefont {Ryan}\ \bibnamefont {Babbush}},\ }\bibfield
   {title} {\enquote {\bibinfo {title} {{Quantum approximate optimization of
  non-planar graph problems on a planar superconducting processor}},}\ }\href
  {\doibase 10.1038/s41567-020-01105-y} {\bibfield  {journal} {\bibinfo
  {journal} {Nature Physics}\ }\textbf {\bibinfo {volume} {17}},\ \bibinfo
  {pages} {332--336} (\bibinfo {year} {2021})},\ \Eprint
  {http://arxiv.org/abs/2004.04197} {arXiv:2004.04197} \BibitemShut {NoStop}%
\bibitem [{\citenamefont {Gao}\ \emph {et~al.}(2021)\citenamefont {Gao},
  \citenamefont {Kalinowski}, \citenamefont {Chou}, \citenamefont {Lukin},
  \citenamefont {Barak},\ and\ \citenamefont {Choi}}]{Gao2021}%
  \BibitemOpen
  \bibfield  {author} {\bibinfo {author} {\bibfnamefont {Xun}\ \bibnamefont
  {Gao}}, \bibinfo {author} {\bibfnamefont {Marcin}\ \bibnamefont
  {Kalinowski}}, \bibinfo {author} {\bibfnamefont {Chi-Ning}\ \bibnamefont
  {Chou}}, \bibinfo {author} {\bibfnamefont {Mikhail~D.}\ \bibnamefont
  {Lukin}}, \bibinfo {author} {\bibfnamefont {Boaz}\ \bibnamefont {Barak}}, \
  and\ \bibinfo {author} {\bibfnamefont {Soonwon}\ \bibnamefont {Choi}},\
  }\href {\doibase 10.48550/ARXIV.2112.01657} {\enquote {\bibinfo {title}
  {Limitations of linear cross-entropy as a measure for quantum advantage},}\ }
  (\bibinfo {year} {2021})\BibitemShut {NoStop}%
\bibitem [{\citenamefont {Pang}\ \emph {et~al.}(2020)\citenamefont {Pang},
  \citenamefont {Hao}, \citenamefont {Dugad}, \citenamefont {Zhou},\ and\
  \citenamefont {Solomonik}}]{Pang:2020}%
  \BibitemOpen
  \bibfield  {author} {\bibinfo {author} {\bibfnamefont {Yuchen}\ \bibnamefont
  {Pang}}, \bibinfo {author} {\bibfnamefont {Tianyi}\ \bibnamefont {Hao}},
  \bibinfo {author} {\bibfnamefont {Annika}\ \bibnamefont {Dugad}}, \bibinfo
  {author} {\bibfnamefont {Yiqing}\ \bibnamefont {Zhou}}, \ and\ \bibinfo
  {author} {\bibfnamefont {Edgar}\ \bibnamefont {Solomonik}},\ }\bibfield
  {title} {\enquote {\bibinfo {title} {Efficient 2d tensor network simulation
  of quantum systems},}\ }in\ \href {\doibase 10.1109/SC41405.2020.00018}
  {\emph {\bibinfo {booktitle} {SC20: International Conference for High
  Performance Computing, Networking, Storage and Analysis}}}\ (\bibinfo {year}
  {2020})\ pp.\ \bibinfo {pages} {1--14}\BibitemShut {NoStop}%
\bibitem [{\citenamefont {Lin}\ \emph {et~al.}(2021)\citenamefont {Lin},
  \citenamefont {Zaletel},\ and\ \citenamefont {Pollmann}}]{Lin:2021}%
  \BibitemOpen
  \bibfield  {author} {\bibinfo {author} {\bibfnamefont {Sheng-Hsuan}\
  \bibnamefont {Lin}}, \bibinfo {author} {\bibfnamefont {Michael}\ \bibnamefont
  {Zaletel}}, \ and\ \bibinfo {author} {\bibfnamefont {Frank}\ \bibnamefont
  {Pollmann}},\ }\href {\doibase 10.48550/ARXIV.2112.08394} {\enquote {\bibinfo
  {title} {Efficient simulation of dynamics in two-dimensional quantum spin
  systems with isometric tensor networks},}\ } (\bibinfo {year} {2021}),\
  \Eprint {http://arxiv.org/abs/2112.08394} {2112.08394} \BibitemShut {NoStop}%
\bibitem [{\citenamefont {Proctor}\ \emph {et~al.}(2022)\citenamefont
  {Proctor}, \citenamefont {Rudinger}, \citenamefont {Young}, \citenamefont
  {Nielsen},\ and\ \citenamefont {Blume-Kohout}}]{Proctor2020}%
  \BibitemOpen
  \bibfield  {author} {\bibinfo {author} {\bibfnamefont {Timothy}\ \bibnamefont
  {Proctor}}, \bibinfo {author} {\bibfnamefont {Kenneth}\ \bibnamefont
  {Rudinger}}, \bibinfo {author} {\bibfnamefont {Kevin}\ \bibnamefont {Young}},
  \bibinfo {author} {\bibfnamefont {Erik}\ \bibnamefont {Nielsen}}, \ and\
  \bibinfo {author} {\bibfnamefont {Robin}\ \bibnamefont {Blume-Kohout}},\
  }\bibfield  {title} {\enquote {\bibinfo {title} {{Measuring the capabilities
  of quantum computers}},}\ }\href {\doibase 10.1038/s41567-021-01409-7}
  {\bibfield  {journal} {\bibinfo  {journal} {Nature Physics}\ }\textbf
  {\bibinfo {volume} {18}},\ \bibinfo {pages} {75--79} (\bibinfo {year}
  {2022})},\ \Eprint {http://arxiv.org/abs/2008.11294} {arXiv:2008.11294}
  \BibitemShut {NoStop}%
\bibitem [{\citenamefont {Dupont}\ \emph {et~al.}(2022)\citenamefont {Dupont},
  \citenamefont {Didier}, \citenamefont {Hodson}, \citenamefont {Moore},\ and\
  \citenamefont {Reagor}}]{Dupont2022}%
  \BibitemOpen
  \bibfield  {author} {\bibinfo {author} {\bibfnamefont {Maxime}\ \bibnamefont
  {Dupont}}, \bibinfo {author} {\bibfnamefont {Nicolas}\ \bibnamefont
  {Didier}}, \bibinfo {author} {\bibfnamefont {Mark~J.}\ \bibnamefont
  {Hodson}}, \bibinfo {author} {\bibfnamefont {Joel~E.}\ \bibnamefont {Moore}},
  \ and\ \bibinfo {author} {\bibfnamefont {Matthew~J.}\ \bibnamefont
  {Reagor}},\ }\href {\doibase 10.48550/ARXIV.2206.06348} {\enquote {\bibinfo
  {title} {Calibrating the classical hardness of the quantum approximate
  optimization algorithm},}\ } (\bibinfo {year} {2022})\BibitemShut {NoStop}%
\bibitem [{\citenamefont {Sreedhar}\ \emph {et~al.}(2022)\citenamefont
  {Sreedhar}, \citenamefont {Vikstål}, \citenamefont {Svensson}, \citenamefont
  {Ask}, \citenamefont {Johansson},\ and\ \citenamefont
  {García-Álvarez}}]{Sreedhar2022}%
  \BibitemOpen
  \bibfield  {author} {\bibinfo {author} {\bibfnamefont {Rishi}\ \bibnamefont
  {Sreedhar}}, \bibinfo {author} {\bibfnamefont {Pontus}\ \bibnamefont
  {Vikstål}}, \bibinfo {author} {\bibfnamefont {Marika}\ \bibnamefont
  {Svensson}}, \bibinfo {author} {\bibfnamefont {Andreas}\ \bibnamefont {Ask}},
  \bibinfo {author} {\bibfnamefont {Göran}\ \bibnamefont {Johansson}}, \ and\
  \bibinfo {author} {\bibfnamefont {Laura}\ \bibnamefont {García-Álvarez}},\
  }\href {\doibase 10.48550/ARXIV.2207.03404} {\enquote {\bibinfo {title} {The
  quantum approximate optimization algorithm performance with low entanglement
  and high circuit depth},}\ } (\bibinfo {year} {2022})\BibitemShut {NoStop}%
\bibitem [{\citenamefont {Medvidovi{\'{c}}}\ and\ \citenamefont
  {Carleo}(2021)}]{Medvidovi2021}%
  \BibitemOpen
  \bibfield  {author} {\bibinfo {author} {\bibfnamefont {Matija}\ \bibnamefont
  {Medvidovi{\'{c}}}}\ and\ \bibinfo {author} {\bibfnamefont {Giuseppe}\
  \bibnamefont {Carleo}},\ }\bibfield  {title} {\enquote {\bibinfo {title}
  {{Classical variational simulation of the Quantum Approximate Optimization
  Algorithm}},}\ }\href {\doibase 10.1038/s41534-021-00440-z} {\bibfield
  {journal} {\bibinfo  {journal} {npj Quantum Information}\ }\textbf {\bibinfo
  {volume} {7}},\ \bibinfo {pages} {101} (\bibinfo {year} {2021})}\BibitemShut
  {NoStop}%
\bibitem [{\citenamefont {Jónsson}\ \emph {et~al.}(2018)\citenamefont
  {Jónsson}, \citenamefont {Bauer},\ and\ \citenamefont
  {Carleo}}]{Jnsson2018}%
  \BibitemOpen
  \bibfield  {author} {\bibinfo {author} {\bibfnamefont {Bjarni}\ \bibnamefont
  {Jónsson}}, \bibinfo {author} {\bibfnamefont {Bela}\ \bibnamefont {Bauer}},
  \ and\ \bibinfo {author} {\bibfnamefont {Giuseppe}\ \bibnamefont {Carleo}},\
  }\href@noop {} {\enquote {\bibinfo {title} {Neural-network states for the
  classical simulation of quantum computing},}\ } (\bibinfo {year} {2018}),\
  \Eprint {http://arxiv.org/abs/1808.05232} {arXiv:1808.05232 [quant-ph]}
  \BibitemShut {NoStop}%
\bibitem [{\citenamefont {White}(1992)}]{White1992}%
  \BibitemOpen
  \bibfield  {author} {\bibinfo {author} {\bibfnamefont {Steven~R.}\
  \bibnamefont {White}},\ }\bibfield  {title} {\enquote {\bibinfo {title}
  {Density matrix formulation for quantum renormalization groups},}\ }\href
  {\doibase 10.1103/PhysRevLett.69.2863} {\bibfield  {journal} {\bibinfo
  {journal} {Phys. Rev. Lett.}\ }\textbf {\bibinfo {volume} {69}},\ \bibinfo
  {pages} {2863--2866} (\bibinfo {year} {1992})}\BibitemShut {NoStop}%
\bibitem [{\citenamefont {Saberi}\ \emph {et~al.}(2009)\citenamefont {Saberi},
  \citenamefont {Weichselbaum}, \citenamefont {Lamata}, \citenamefont
  {P\'erez-Garc\'{\i}a}, \citenamefont {von Delft},\ and\ \citenamefont
  {Solano}}]{PhysRevA.80.022334}%
  \BibitemOpen
  \bibfield  {author} {\bibinfo {author} {\bibfnamefont {Hamed}\ \bibnamefont
  {Saberi}}, \bibinfo {author} {\bibfnamefont {Andreas}\ \bibnamefont
  {Weichselbaum}}, \bibinfo {author} {\bibfnamefont {Lucas}\ \bibnamefont
  {Lamata}}, \bibinfo {author} {\bibfnamefont {David}\ \bibnamefont
  {P\'erez-Garc\'{\i}a}}, \bibinfo {author} {\bibfnamefont {Jan}\ \bibnamefont
  {von Delft}}, \ and\ \bibinfo {author} {\bibfnamefont {Enrique}\ \bibnamefont
  {Solano}},\ }\bibfield  {title} {\enquote {\bibinfo {title} {Constrained
  optimization of sequentially generated entangled multiqubit states},}\ }\href
  {\doibase 10.1103/PhysRevA.80.022334} {\bibfield  {journal} {\bibinfo
  {journal} {Phys. Rev. A}\ }\textbf {\bibinfo {volume} {80}},\ \bibinfo
  {pages} {022334} (\bibinfo {year} {2009})}\BibitemShut {NoStop}%
\bibitem [{\citenamefont {Stoudenmire}\ and\ \citenamefont
  {White}(2010)}]{Stoudenmire_2010}%
  \BibitemOpen
  \bibfield  {author} {\bibinfo {author} {\bibfnamefont {E~M}\ \bibnamefont
  {Stoudenmire}}\ and\ \bibinfo {author} {\bibfnamefont {Steven~R}\
  \bibnamefont {White}},\ }\bibfield  {title} {\enquote {\bibinfo {title}
  {Minimally entangled typical thermal state algorithms},}\ }\href {\doibase
  10.1088/1367-2630/12/5/055026} {\bibfield  {journal} {\bibinfo  {journal}
  {New Journal of Physics}\ }\textbf {\bibinfo {volume} {12}},\ \bibinfo
  {pages} {055026} (\bibinfo {year} {2010})}\BibitemShut {NoStop}%
\bibitem [{\citenamefont {Hirata}\ \emph {et~al.}(2009)\citenamefont {Hirata},
  \citenamefont {Nakanishi}, \citenamefont {Yamashita},\ and\ \citenamefont
  {Nakashima}}]{Hirata2009}%
  \BibitemOpen
  \bibfield  {author} {\bibinfo {author} {\bibfnamefont {Yuichi}\ \bibnamefont
  {Hirata}}, \bibinfo {author} {\bibfnamefont {Masaki}\ \bibnamefont
  {Nakanishi}}, \bibinfo {author} {\bibfnamefont {Shigeru}\ \bibnamefont
  {Yamashita}}, \ and\ \bibinfo {author} {\bibfnamefont {Yasuhiko}\
  \bibnamefont {Nakashima}},\ }\bibfield  {title} {\enquote {\bibinfo {title}
  {{An Efficient Method to Convert Arbitrary Quantum Circuits to Ones on a
  Linear Nearest Neighbor Architecture}},}\ }in\ \href {\doibase
  10.1109/ICQNM.2009.25} {\emph {\bibinfo {booktitle} {2009 Third International
  Conference on Quantum, Nano and Micro Technologies}}}\ (\bibinfo  {publisher}
  {IEEE},\ \bibinfo {year} {2009})\ pp.\ \bibinfo {pages} {26--33}\BibitemShut
  {NoStop}%
\bibitem [{\citenamefont {Brouwer}\ and\ \citenamefont
  {Beenakker}(1996)}]{Brouwer1996}%
  \BibitemOpen
  \bibfield  {author} {\bibinfo {author} {\bibfnamefont {P.~W.}\ \bibnamefont
  {Brouwer}}\ and\ \bibinfo {author} {\bibfnamefont {C.~W.~J.}\ \bibnamefont
  {Beenakker}},\ }\bibfield  {title} {\enquote {\bibinfo {title} {Diagrammatic
  method of integration over the unitary group, with applications to quantum
  transport in mesoscopic systems},}\ }\href {\doibase 10.1063/1.531667}
  {\bibfield  {journal} {\bibinfo  {journal} {Journal of Mathematical Physics}\
  }\textbf {\bibinfo {volume} {37}},\ \bibinfo {pages} {4904--4934} (\bibinfo
  {year} {1996})}\BibitemShut {NoStop}%
\bibitem [{\citenamefont {Martiel}\ \emph {et~al.}(2021)\citenamefont
  {Martiel}, \citenamefont {Ayral},\ and\ \citenamefont
  {Allouche}}]{Martiel2021}%
  \BibitemOpen
  \bibfield  {author} {\bibinfo {author} {\bibfnamefont {Simon}\ \bibnamefont
  {Martiel}}, \bibinfo {author} {\bibfnamefont {Thomas}\ \bibnamefont {Ayral}},
  \ and\ \bibinfo {author} {\bibfnamefont {Cyril}\ \bibnamefont {Allouche}},\
  }\bibfield  {title} {\enquote {\bibinfo {title} {{Benchmarking quantum
  co-processors in an application-centric, hardware-agnostic and scalable
  way}},}\ }\href {http://arxiv.org/abs/2102.12973} {\ \textbf {\bibinfo
  {volume} {4}} (\bibinfo {year} {2021})},\ \Eprint
  {http://arxiv.org/abs/2102.12973} {arXiv:2102.12973} \BibitemShut {NoStop}%
\bibitem [{\citenamefont {Markov}\ and\ \citenamefont
  {Shi}(2008)}]{Markov2008}%
  \BibitemOpen
  \bibfield  {author} {\bibinfo {author} {\bibfnamefont {Igor~L.}\ \bibnamefont
  {Markov}}\ and\ \bibinfo {author} {\bibfnamefont {Yaoyun.}\ \bibnamefont
  {Shi}},\ }\bibfield  {title} {\enquote {\bibinfo {title} {Simulating quantum
  computation by contracting tensor networks},}\ }\href {\doibase
  10.1137/050644756} {\bibfield  {journal} {\bibinfo  {journal} {SIAM Journal
  on Computing}\ }\textbf {\bibinfo {volume} {38}},\ \bibinfo {pages}
  {963--981} (\bibinfo {year} {2008})},\ \Eprint
  {http://arxiv.org/abs/https://doi.org/10.1137/050644756}
  {https://doi.org/10.1137/050644756} \BibitemShut {NoStop}%
\bibitem [{\citenamefont {Shen}(2001)}]{Shen2001}%
  \BibitemOpen
  \bibfield  {author} {\bibinfo {author} {\bibfnamefont {Jianhong}\
  \bibnamefont {Shen}},\ }\bibfield  {title} {\enquote {\bibinfo {title} {{On
  the singular values of Gaussian random matrices}},}\ }\href {\doibase
  10.1016/S0024-3795(00)00322-0} {\bibfield  {journal} {\bibinfo  {journal}
  {Linear Algebra and its Applications}\ }\textbf {\bibinfo {volume} {326}},\
  \bibinfo {pages} {1--14} (\bibinfo {year} {2001})}\BibitemShut {NoStop}%
\end{thebibliography}%

\end{document}